\documentclass[manuscript]{aastex62}
\usepackage[figuresright]{rotating}

\received{January 1, 2018}
\revised{January 7, 2018}
\accepted{\today}
\submitjournal{ApJ}

\begin{document}

\title{Identifying Galactic Halo Substructure in 6D Phase-space Using $\sim$13,000 LAMOST K Giants}

\correspondingauthor{Xiang-Xiang Xue; Chengqun Yang} \email{xuexx@nao.cas.cn; ycq@bao.ac.cn}

\author{Chengqun Yang}
\affiliation{National Astronomical Observatories, Chinese Academy of Sciences, 20A Datun Road, Chaoyang District, Beijing 100101, P.R.China}
\affiliation{School of Astronomy and Space Science, University of Chinese Academy of Sciences, 19A Yuquan Road, Shijingshan District, Beijing 100049, P.R.China}

\author{Xiang-Xiang Xue}
\affiliation{National Astronomical Observatories, Chinese Academy of Sciences, 20A Datun Road, Chaoyang District, Beijing 100101, P.R.China}

\author{Jing Li}
\affiliation{Physics and Space Science College, China West Normal University, 1 ShiDa Road, Nanchong 637002, P.R.China}
\affiliation{SHAO, Chinese Academy of Sciences, Nandan Road, Shanghai 200030, People’s Republic of China}

\author{Lan Zhang}
\affiliation{National Astronomical Observatories, Chinese Academy of Sciences, 20A Datun Road, Chaoyang District, Beijing 100101, P.R.China}

\author{Chao Liu}
\affiliation{National Astronomical Observatories, Chinese Academy of Sciences, 20A Datun Road, Chaoyang District, Beijing 100101, P.R.China}
\affiliation{School of Astronomy and Space Science, University of Chinese Academy of Sciences, 19A Yuquan Road, Shijingshan District, Beijing 100049, P.R.China}

\author{Gang Zhao}
\affiliation{National Astronomical Observatories, Chinese Academy of Sciences, 20A Datun Road, Chaoyang District, Beijing 100101, P.R.China}
\affiliation{School of Astronomy and Space Science, University of Chinese Academy of Sciences, 19A Yuquan Road, Shijingshan District, Beijing 100049, P.R.China}

\author{Jiang Chang}
\affiliation{National Astronomical Observatories, Chinese Academy of Sciences, 20A Datun Road, Chaoyang District, Beijing 100101, P.R.China}

\author{Hao Tian}
\affiliation{National Astronomical Observatories, Chinese Academy of Sciences, 20A Datun Road, Chaoyang District, Beijing 100101, P.R.China}

\author{Chengdong Li}
\affiliation{National Astronomical Observatories, Chinese Academy of Sciences, 20A Datun Road, Chaoyang District, Beijing 100101, P.R.China}
\affiliation{School of Astronomy and Space Science, University of Chinese Academy of Sciences, 19A Yuquan Road, Shijingshan District, Beijing 100049, P.R.China}



\begin{abstract}
We construct a large halo K-giant sample by combining the positions, distances, radial velocities, and metallicities of over $13,000$ LAMOST DR5 halo K giants with the Gaia DR2 proper motions, which covers a Galactocentric distance range of $5-120$ kpc. Using a position-velocity clustering estimator (the 6Distance), we statistically quantify the presence of position-velocity substructure at high significance: K giants have more close pairs in position-velocity space than a smooth stellar halo. We find that the amount of substructure in the halo increases with increasing distance and metallicity. With a percolation algorithm named friends-of-friends (FoF) to identify groups, we identify members belonging to Sagittarius (Sgr) Streams, Monoceros Ring, Virgo overdensity, Hercules-Aquila Cloud, Orphan Streams and other unknown substructures and find that the Sgr streams account for a large part of grouped stars beyond $20$ kpc and enhance the increase of substructure with distance and metallicity. For the first time, we identify spectroscopic members of Monoceros Ring in the south and north Galactic hemisphere, which presents a rotation of about $185$ km s$^{-1}$ and mean metallicity is $-0.66$ dex.
\end{abstract}

\keywords{Galaxy: evolution --- Galaxy: formation --- Galaxy: halo --- Galaxy: kinematics and dynamics}


\defcitealias{Starkenburg09}{S09}%\citetalias{Starkenburg09}
\defcitealias{Xue11}{X11}
\defcitealias{Xue14}{X14}
\defcitealias{Janesh16}{J16}
\defcitealias{LM10}{LM10}

\section{Introduction}\label{sec:intro}
The hierarchical model of galaxies assembly predicts that a series of accretion and merging events led to the formation of the Milky Way \citep{Searle-Zinn78, WR78, Blumenthal84, BJ05, Springel06}. Such assembly mechanisms are encoded in the stellar members of the Milky Way's halo, which comprises at least two diffuse components, inner- and outer-halo \citep{Carollo07, Carollo10, Carollo12, Beers12, An13, Tissera13, Tissera14}, several streams \citep{Odenkirchen01, GD06}, and numerous overdensities \citep{Belokurov06, Bernard16}. Stellar members of halo streams and overdensities carry information on the merging event that brought these stars into the Galaxy. Therefore, their identification is an important step to understanding galaxy formation. Stars stripped from the merging Galaxy may form structures in the halo in the form of streams, shell or clouds, which can be detected in density space \citep{Ibata01, Newberg02, Majewski03, Belokurov06}, in phase-space \citep{Starkenburg09, Xue11, Janesh16} or in age space \citep{Santucci15, Carollo16, Carollo18}.

Thanks to the wide-field photometric surveys, such as Sloan Digital Sky Survey \citep[SDSS;][]{York00} and Two Micron All Sky Survey \citep[2MASS;][]{Skrutskie06}, many straightforward observational evidences can be found easily from density map of stars. The most prominent and coherent tidal streams are from the Sagittarius dwarf galaxy \citep[hereafter Sgr;][]{Ibata01, Majewski03}, which have been traced entirely around the Milky Way \citep{Majewski03,Sesarl17}. Besides the Sgr streams, many other substructures such as the Virgo Overdensity, the Monoceros ring, the Orphan Stream, Pal 5 and GD-1 were also found from photometric surveys \citep{Odenkirchen01, Newberg02, GD06, Belokurov06}. However, it is difficult to distinguish the stream members from the Galactic field stars using only sky positions and multi-band photometry. Furthermore, the analysis of simulations showed that the halo streams are distributed smoothly in space after a phase-mixing of $\sim \mathrm{10~Gyr}$, but appear clumped in velocity space, especially in inner parts of the galaxy \citep{Helmi99,Helmi03}.

With the development of the spectroscopic surveys, such as Sloan Digital Sky Survey SDSS \citep[SDSS;][]{York00} and the Large Sky Area Multi-Object Fibre Spectroscopic Telescope \citep[LAMOST;][]{Zhao12}, it is possible to obtain 3D positions and radial velocities of numerous stars. However, it was impossible to measure proper motions of distant stars ($\mathrm{>20~kpc}$) with the technology of the day. Many studies have indicated that the Galactic stellar halo indeed possesses detectable substructure in 4D position-velocity space. \citet{Starkenburg09} developed a clustering estimator named 4distance to calculate the ``distance" between two stars in four dimensional position-velocity space of $(l,b,d,rv)$. Combining with friends-of-friends (FoF) algorithm, they identified groups of stars with similar positions and radial velocities from 101 K giants observed by the Spaghetti survey \citep{Morrison00}. Recently, \citet{Xue11} and \citet{Janesh16} adopted 4distance to quantify substructure using much larger samples of halo stars selected from SDSS/SEGUE survey. Furthermore, \citet{Janesh16} has applied FoF to identify grouped stars in 4D position-velocity space associated with Sgr streams, Orphan stream, Cetus Polar stream and other unknown substructure. However, the lack of proper motions is likely to reduce the reliability of identified stream members.

The second data release of $Gaia$ ($Gaia$ DR2) provides most accurate proper motions (good to $\mathrm{0.2~mas~yr^{-1}}$ for $\mathrm{G=17^m}$) and parallaxes (good to $\mathrm{0.1~mas}$ at $\mathrm{G=17^m}$) for more than 1.3 billion sources with $\mathrm{3^m<G<21^m}$ \citep{Gaia18} so far. For majority of $Gaia$ DR2 stars, reliable distance can not be obtained by inverting the parallax, so \citet{BJ18} inferred the distances and their uncertainties of 1.33 billion stars using a weak distance prior that varies smoothly as a function of Galactic longitude and latitude according to a Galaxy model. They pointed out that their approach can infer meaningful distances for stars with negative parallaxes and/or low parallax precision, but will underestimate the distances of distant giants because the distance prior they adopted is dominated by the nearer dwarfs in the model. Therefore, $Gaia$ DR2 parallaxes do not apply to distant giants.

Giants of spectral type K are luminous enough ($\mathrm{-3^m<M_r<1^m}$) to be observed in distant halo, and have been specifically targeted by many wide-field spectroscopic surveys to explore the outer halo of the Galaxy. For example, \citet{Xue14} published a catalog of $\sim$ 6000 halo K giants with distances up to $\mathrm{80~kpc}$ drawn from the Sloan Extension for Galactic Understanding and Exploration \citep[SEGUE;][]{Yanny09}. Recently, the fifth data release (DR5) of LAMOST has published about 9 million spectra, containing about 13,000 halo K giants with good distance estimations (extending to distances of $\mathrm{100~kpc}$; Xue et al. 2019 in preparation), radial velocities, sky positions and metallicities. Hence, in combination with good proper motions published by $Gaia$ DR2, the sample of K giants with LAMOST spectra constitutes by far the largest set of halo stars with 3D positions, 3D velocities and metallicities. This sample enables the attempt at identifying substructures in full phase space.

This paper is organized as follows. In Section \ref{kg}, we simply describe the selection of halo K giants and the estimate of their distances. The methodology of quantifying substructure and group finding approach of friends-of-friends are represented in Section \ref{6d}. We present the results of quantifying substructure in Section \ref{qs} and the identification of substructures in Section \ref{id}. A brief summary is in Section 6.

\section{The Sample}\label{kg}
LAMOST, located in Xinglong station of National Astronomical Observatories of Chinese Academy of Sciences, is a large spectroscopic survey covering -10$^\circ<\delta<+90^\circ$. It can take 4000 low-resolution ($R \sim$ 1800) optical spectra in a single exposure to the magnitude as faint as $V = 17.8^{\rm{m}}$. Exploring the structure and evolution of the Milky Way is one of the major science goals of LAMOST, and the corresponding target selections are designed to fit the scientific motivation \citep{Zhao12,Deng12,LiuC14}. The stellar parameters and radial velocities can be derived by the well-calibrated LAMOST 1D pipeline, which can achieve typical uncertainties of 167 K in effective temperature $T_{\rm{eff}}$, $\mathrm{0.34~dex}$ in surface gravity $\log g$, $\mathrm{0.16~dex}$ in metallicity $\mathrm{[Fe/H]}$, and $\mathrm{5~kms^{-1}}$ in radial velocity $rv$ \citep{Wu11,Wu14}.

\subsection{K Giants in LAMOST DR5}
LAMOST DR5 released about 9 million spectra, of which about 5 million spectra have measurements of stellar parameters and radial velocities. K giants are selected using $T_{\rm{eff}}$ and log $g$ described in \citet{LiuC14}.

The distances of the K giants are determined using a Bayesian method described in \citet{Xue14}, of which the fundamental basis is the color-magnitude diagrams (so-called fiducials) of three globular clusters and one open cluster observed by SDSS. The multi-band photometry of LAMOST K giants is obtained from cross-match with Pan-STARRS1 \citep[PS1;][]{Chambers16} using a match radius of $1\arcsec$. The PS1 magnitudes can be transformed to SDSS system using linear functions of $(g-i)_{P1}$ \citep{Finkbeiner16}, which are derived through common LAMOST K giants with both PS1 and SDSS magnitudes (Xue et al. 2019 in preparation). The extinction is corrected by subtracting the product of $E(B-V)$ from \citet{Schlegel98} and coefficients (3.303 for SDSS $g$ band and 2.285 for SDSS $r$ band) listed in Table 6 of \citet{SF11} from apparent magnitudes. Similar to the Bayesian method of \citet{Xue14}, the best estimates of the distance moduli and their errors can be estimated using the mean and central $\mathrm{68\%}$ interval of the likelihood of the distance moduli. LAMOST $\log g$ is not accurate to discriminate between red clump stars (RC) and red giants, so we avoid assigning distances to giants below the level of the horizontal branch (HB) defined as $\mathrm{(g-r)_0^{HB}= 0.087[Fe/H]^2 + 0.39[Fe/H] + 0.96}$, which is derived by \citet{Xue14} from $\mathrm{[Fe/H]}$ and the $(g-r)_0$ color of the giant branch at the level of HB of eight clusters.

\begin{figure}[htb]
\centering
  \begin{tabular}{@{}c@{}}
    \includegraphics[width=.8\textwidth]{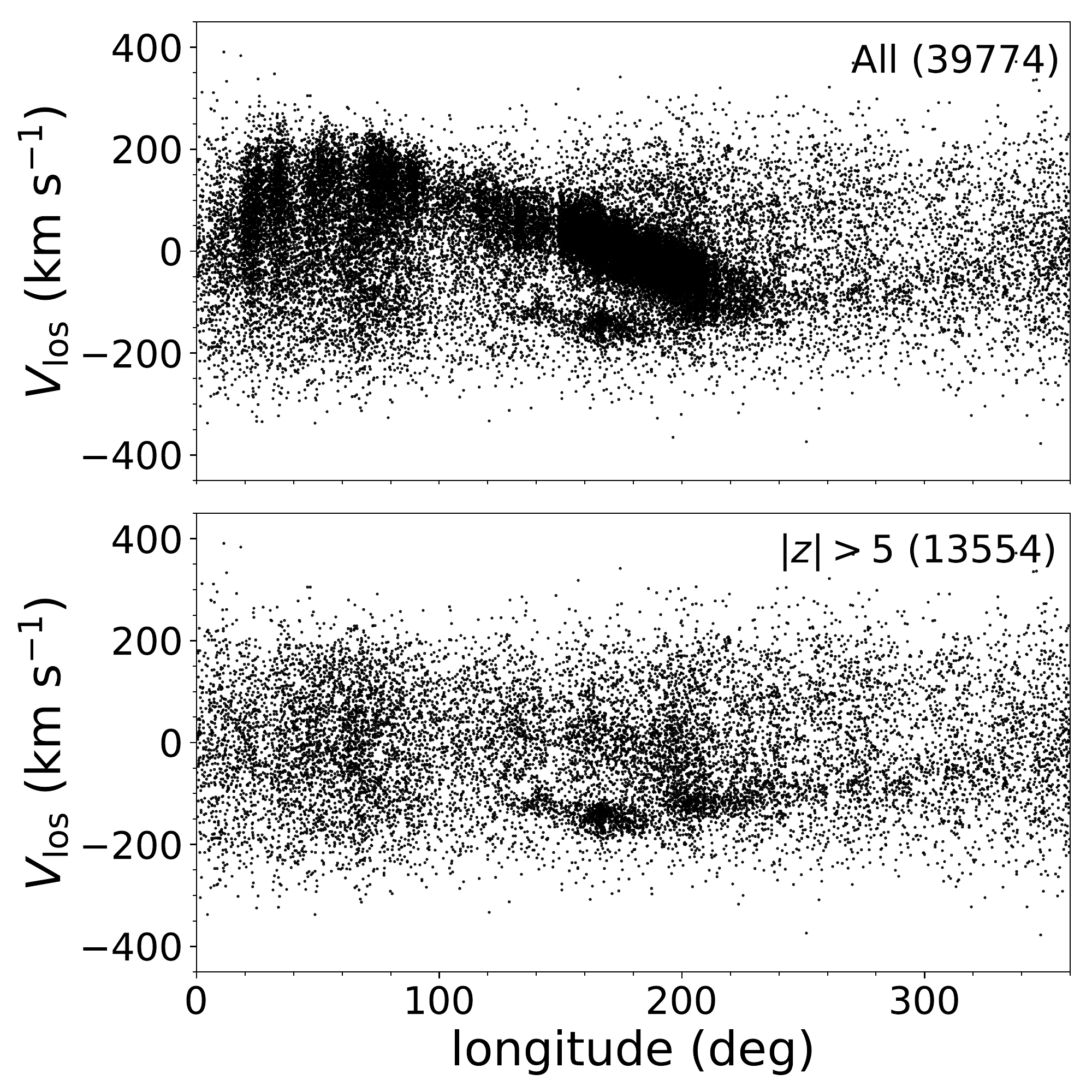}
  \end{tabular}
\caption{Galactic longitude $l$ against the line-of-sight velocity $V_{\rm{los}}$ of LAMOST K giants. The top panel shows the entire sample, and the signature of disk rotation (sin-shaped) is clear here. In the bottom panel, after removing disk stars ($|z| > 5$ kpc), the signature of disk largely disappeared. Some halo substructures are visible.
} \label{all_lv}
\end{figure}

After cross-match with $Gaia$ DR2 with a match radius of $1\arcsec$, there are $\mathrm{39,774}$ LAMOST K giants with sky positions, distances, radial velocities, and proper motions. Figure \ref{all_lv} (upper panel) shows the line-of-sight velocity distribution along with distances of all $\mathrm{39,774}$ K giants, on which the obvious $sin-shape$ indicates a large portion of disk stars.

\subsection{Halo selection}\label{halo selection}
Since we focus on the Galactic halo in this work, we eliminate the K giants within $\mathrm{5~kpc}$ above or below the Galactic disk plane ($|z|\leqslant$ 5 kpc). The right-handed Cartesian coordinate is centered at the Galactic center. The $x$-axis is positive toward the Galactic Center from the Sun, the $y$-axis is along the rotation of the disk, and the $z$-axis towards the North Galactic Pole. The Sun's position is at (-8,0,0) kpc \citep{Reid93}. All velocities are converted to the Galactic standard of rest (GSR) frame by adopting a solar motion of (+10.0,+5.25,+7.17) km s$^{-1}$ \citep{Dehnen98} and the local standard of rest (LSR) velocity of 220 km s$^{-1}$ \citep{Kerr86}. After applying the cut of $|z| > 5$ kpc, the majority of disk stars are eliminated as shown in the lower panel of Figure \ref{all_lv}.

\begin{figure}[htb]
\centering
  \begin{tabular}{@{}c@{}}
    \includegraphics[width=.8\textwidth]{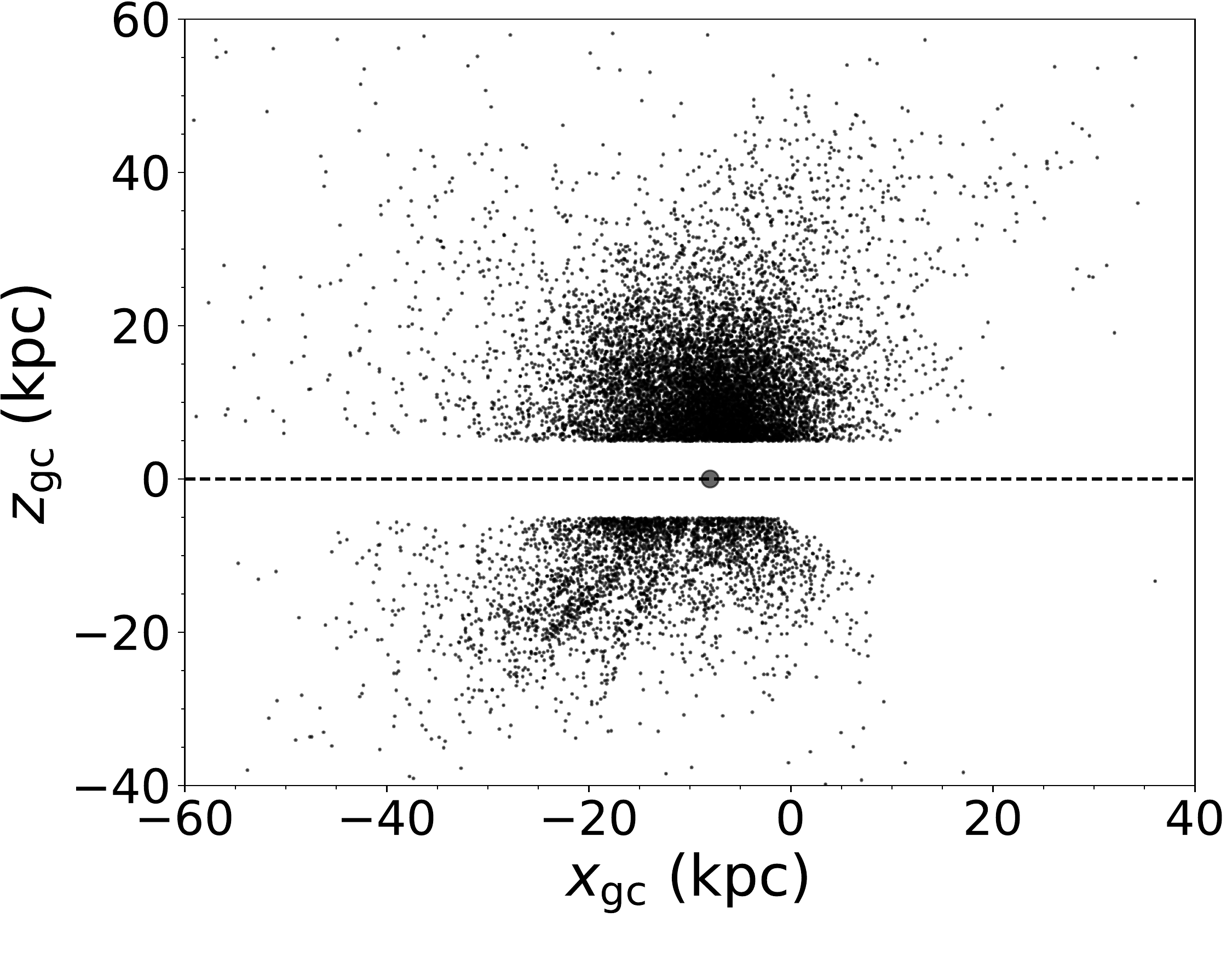}
  \end{tabular}
\caption{The spatial distribution ($x-z$ plane) of $\mathrm{13,554}$ LAMOST halo K giants}.
 \label{all_xz}
\end{figure}

\begin{figure}[htb]
\centering
  \begin{tabular}{@{}cccc@{}}
    \includegraphics[width=.45\textwidth]{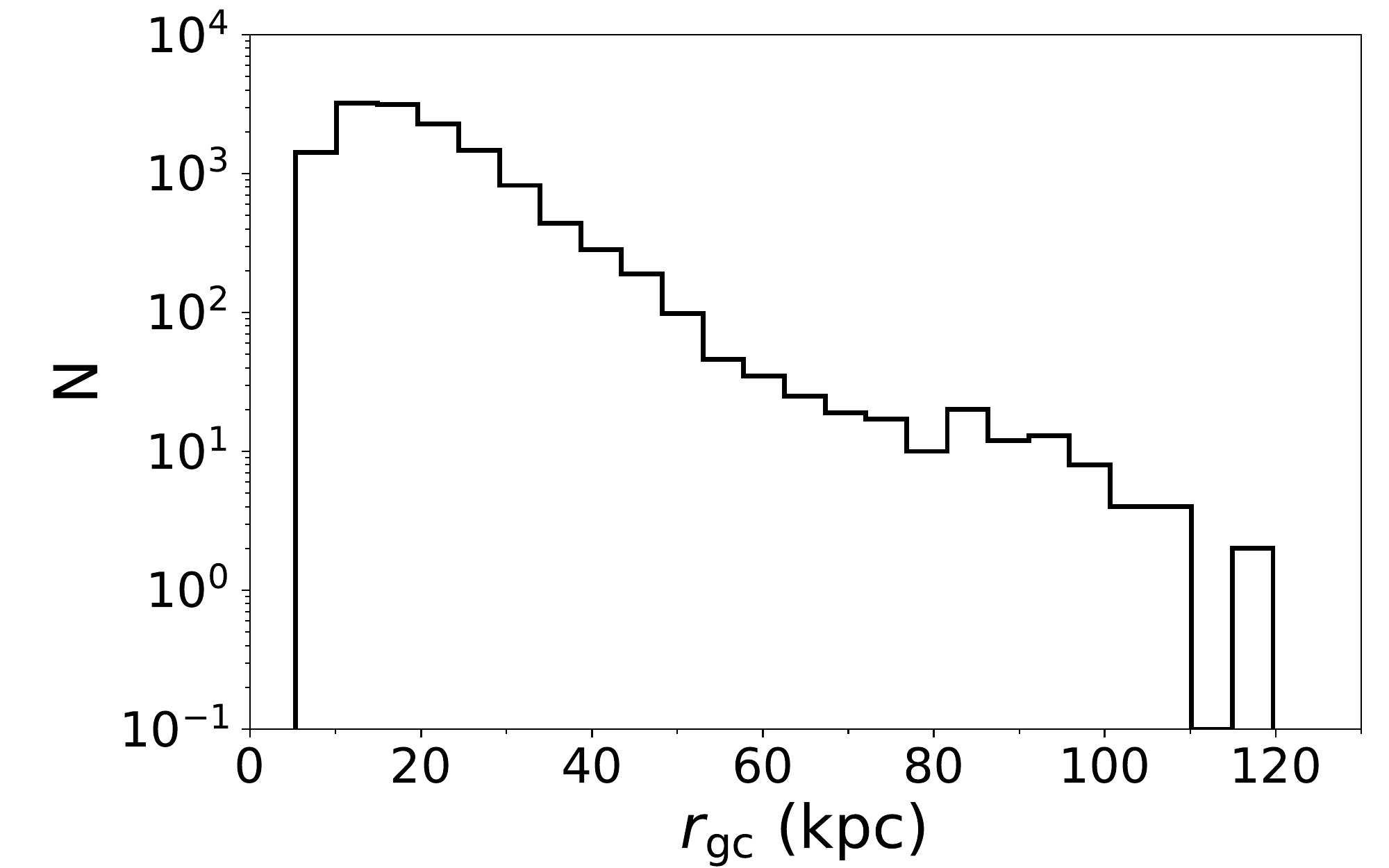} &
    \includegraphics[width=.45\textwidth]{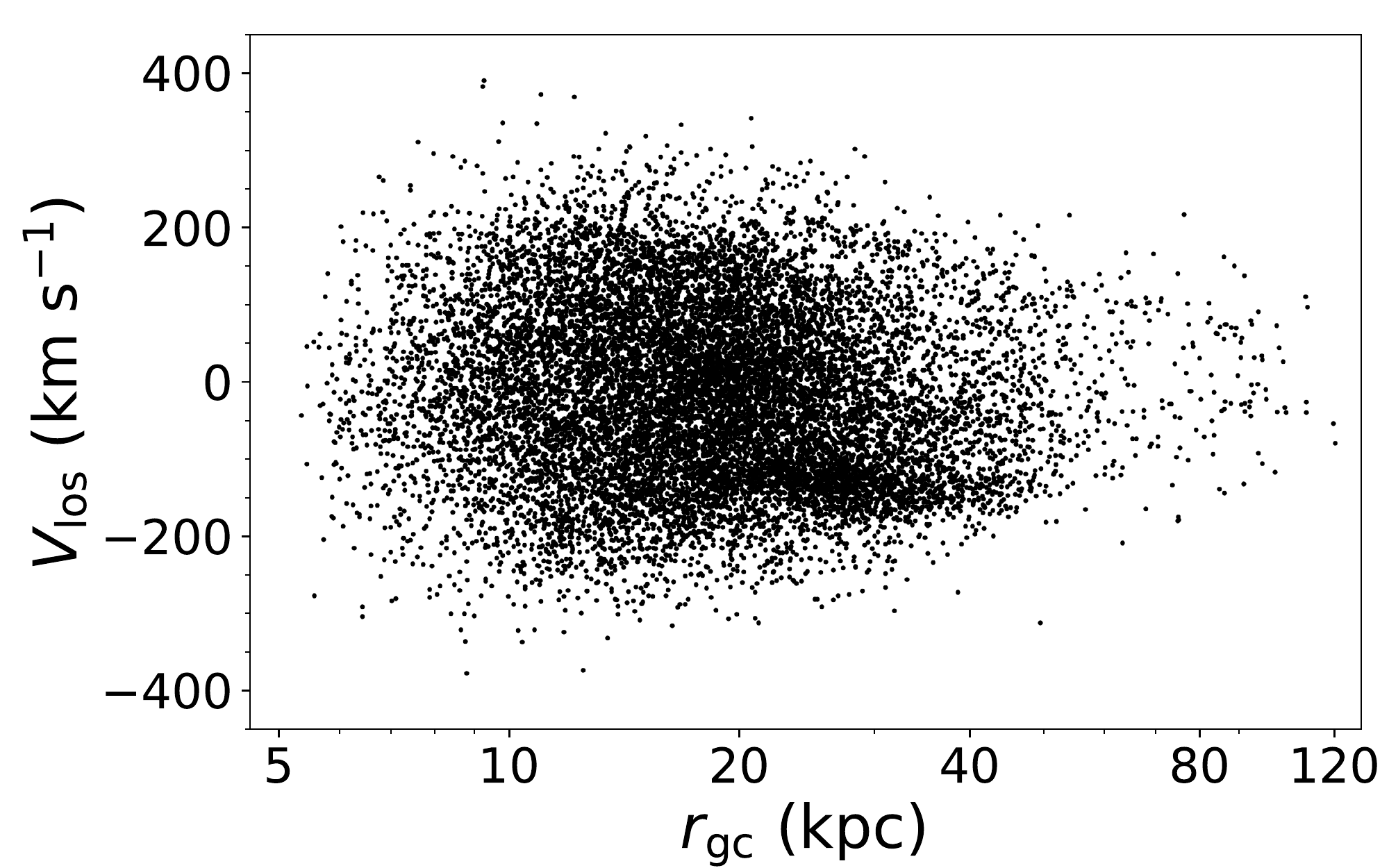} &\\
    \includegraphics[width=.45\textwidth]{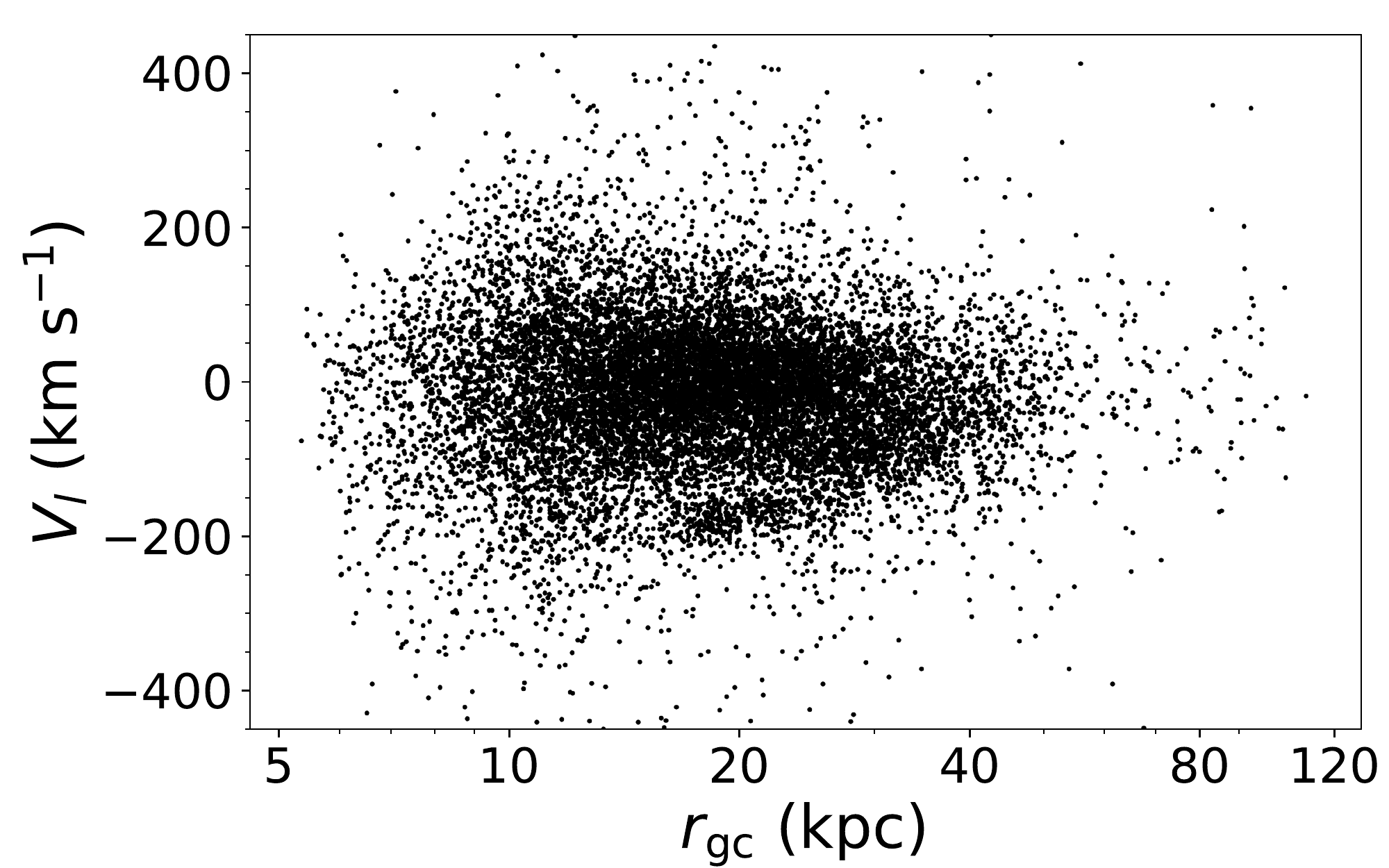} &
    \includegraphics[width=.45\textwidth]{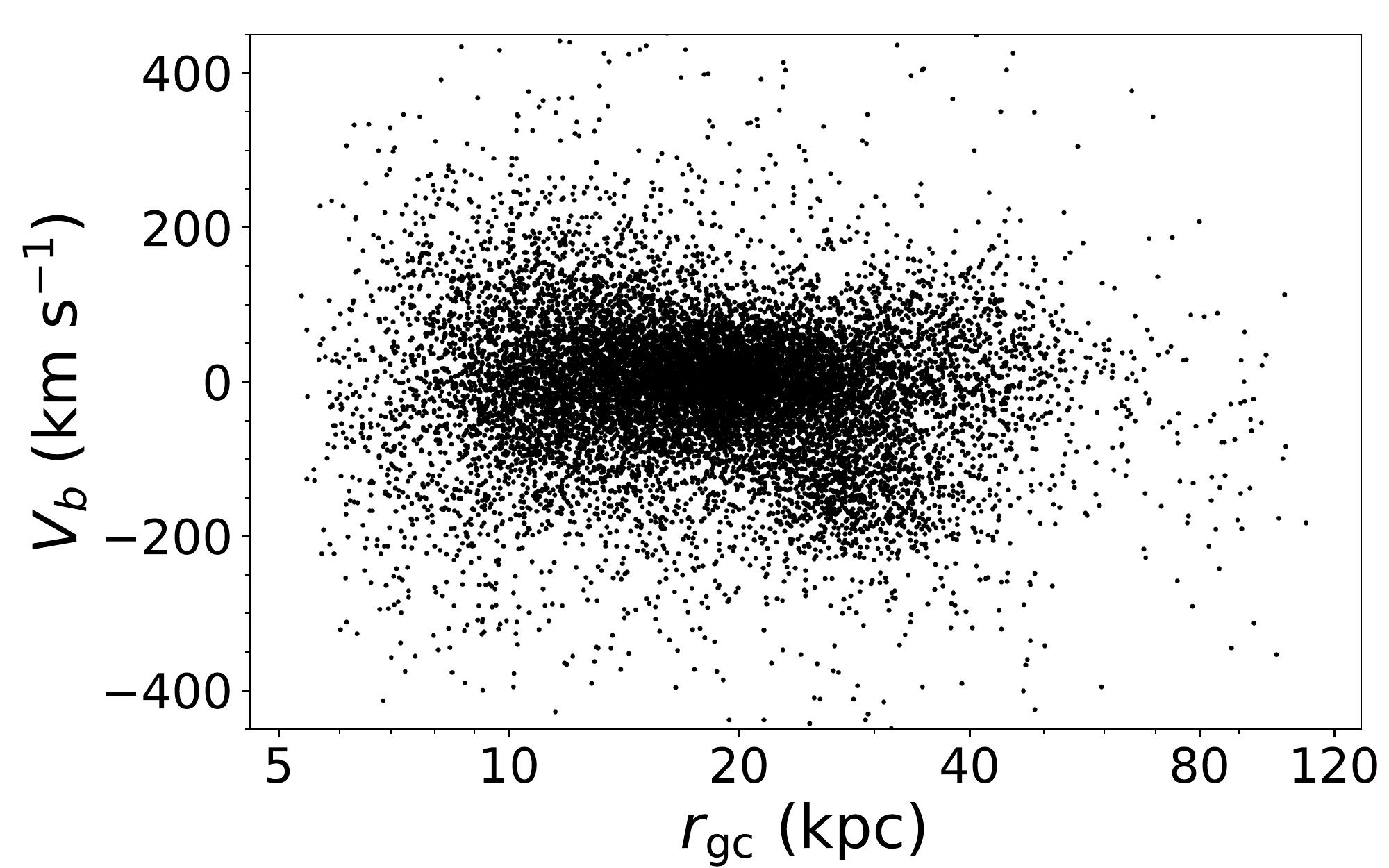} &\\
  \end{tabular}
\caption{The Galactocentric distance distribution and velocity distributions along with Galactocentric distance of LAMOST halo K giants.
} \label{d_v}
\end{figure}

\begin{figure}[htb]
%\centering
  \begin{tabular}{@{}cccc@{}}
    \includegraphics[width=.45\textwidth]{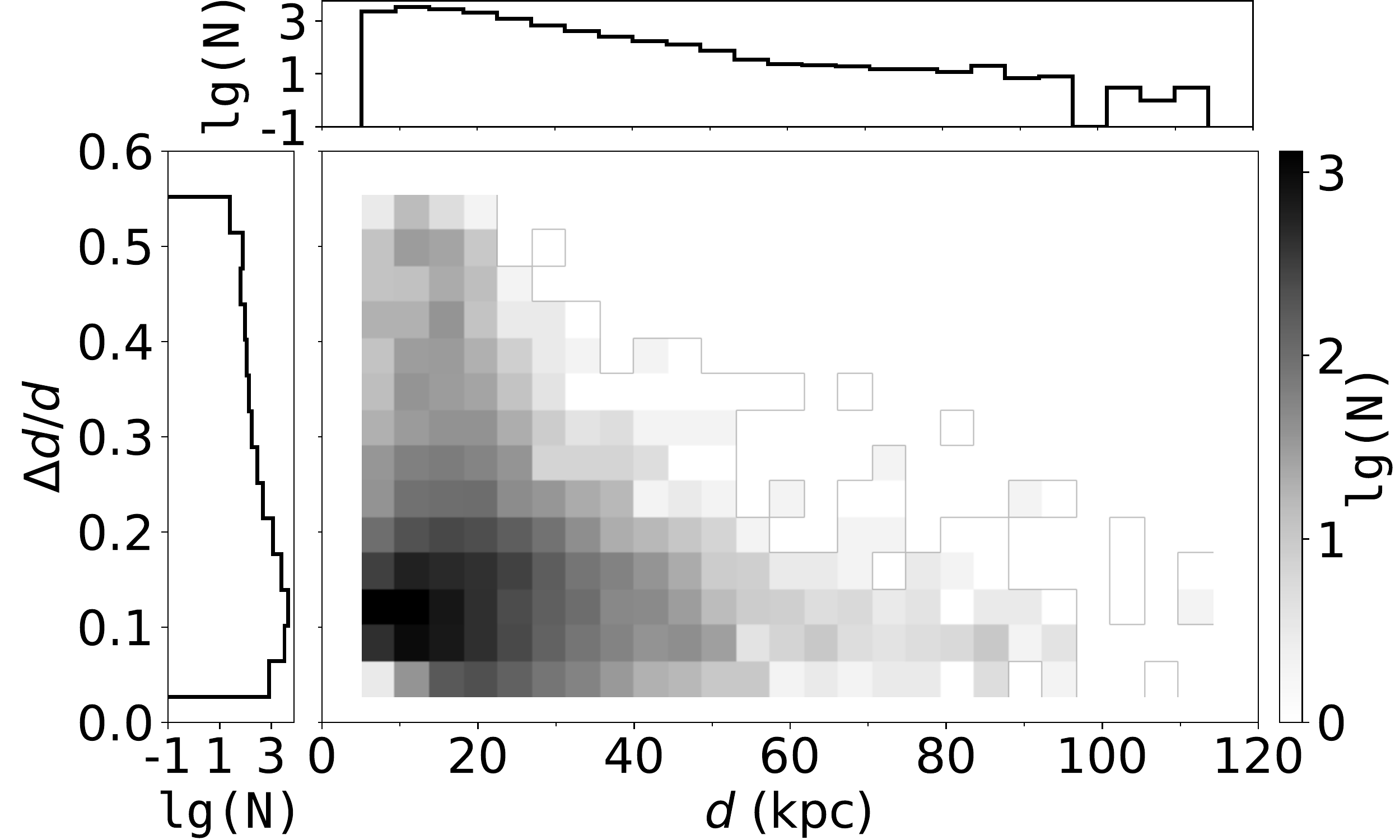} &
    \includegraphics[width=.45\textwidth]{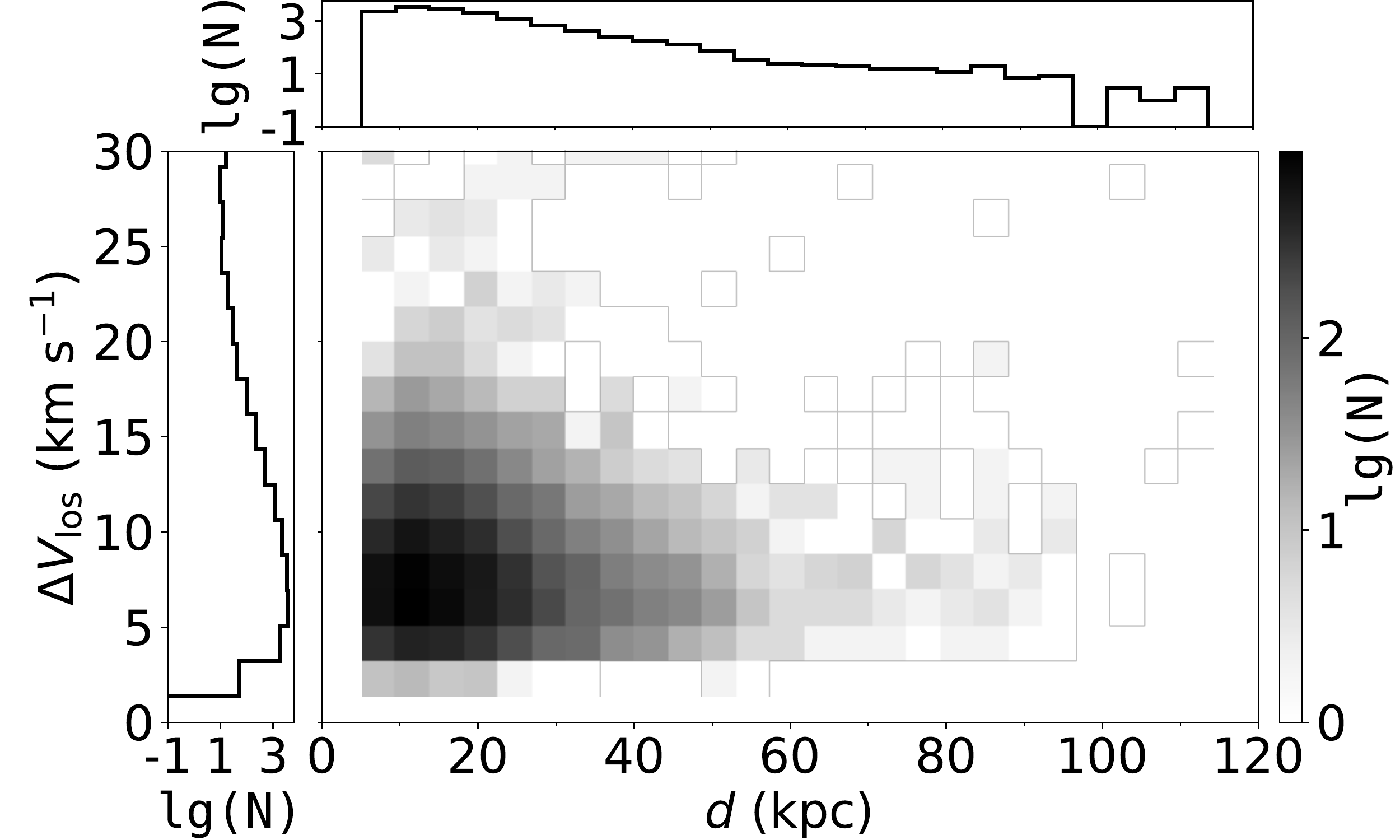} &\\
    \includegraphics[width=.45\textwidth]{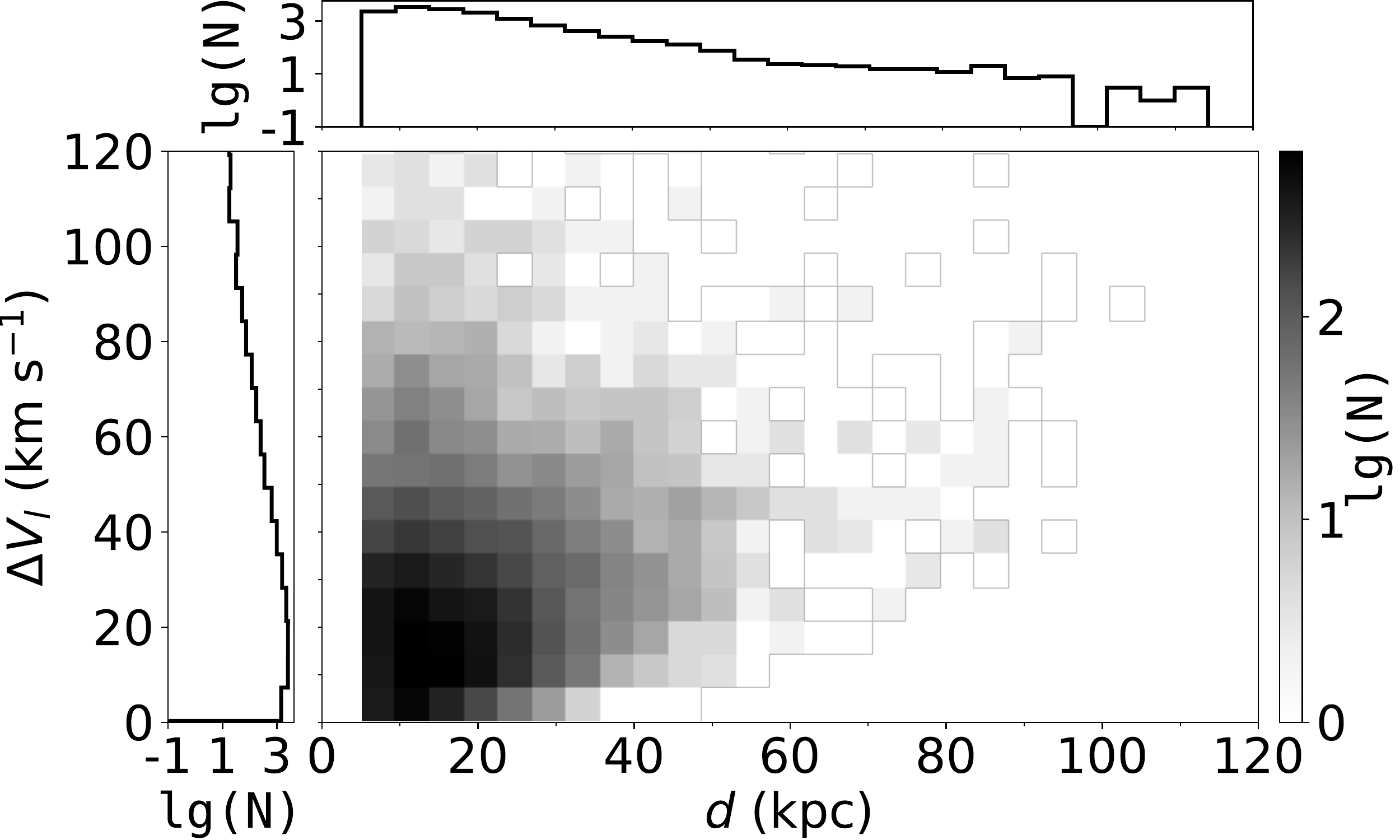} &
    \includegraphics[width=.45\textwidth]{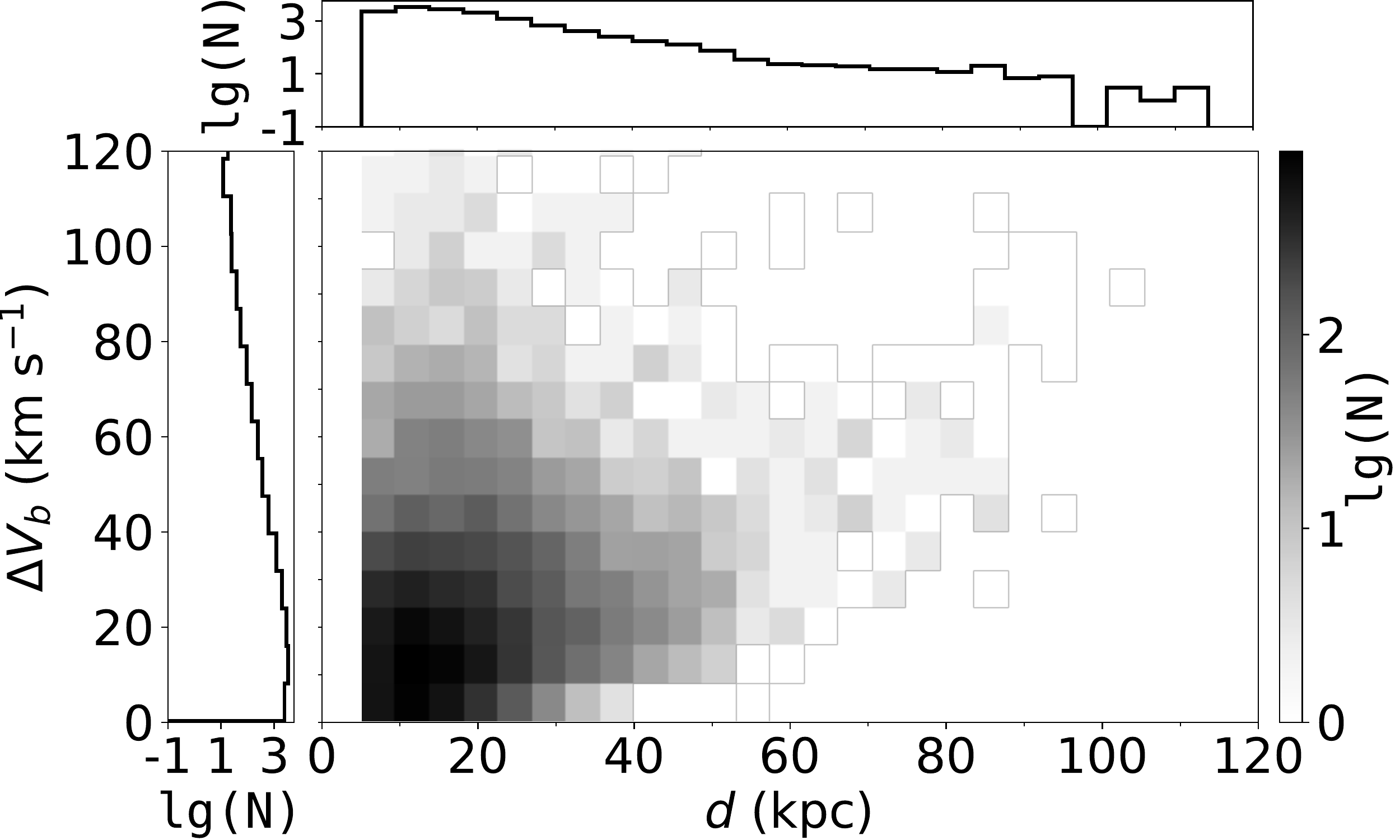} &\\
  \end{tabular}
\caption{The error distributions of distances and velocities along with distances. The distances have a typical error of 13\%. A typical error of 7 km s$^{-1}$ in line-of-sight velocity makes it the most accurate velocity component. The mean errors of the two tangential-velocity components are about 20 km s$^{-1}$, and can spread to $\sim$ 100 km s$^{-1}$.
} \label{d_errs}
\end{figure}

\begin{figure}[htb]
%\centering
  \begin{tabular}{@{}cccc@{}}
    \includegraphics[width=.45\textwidth]{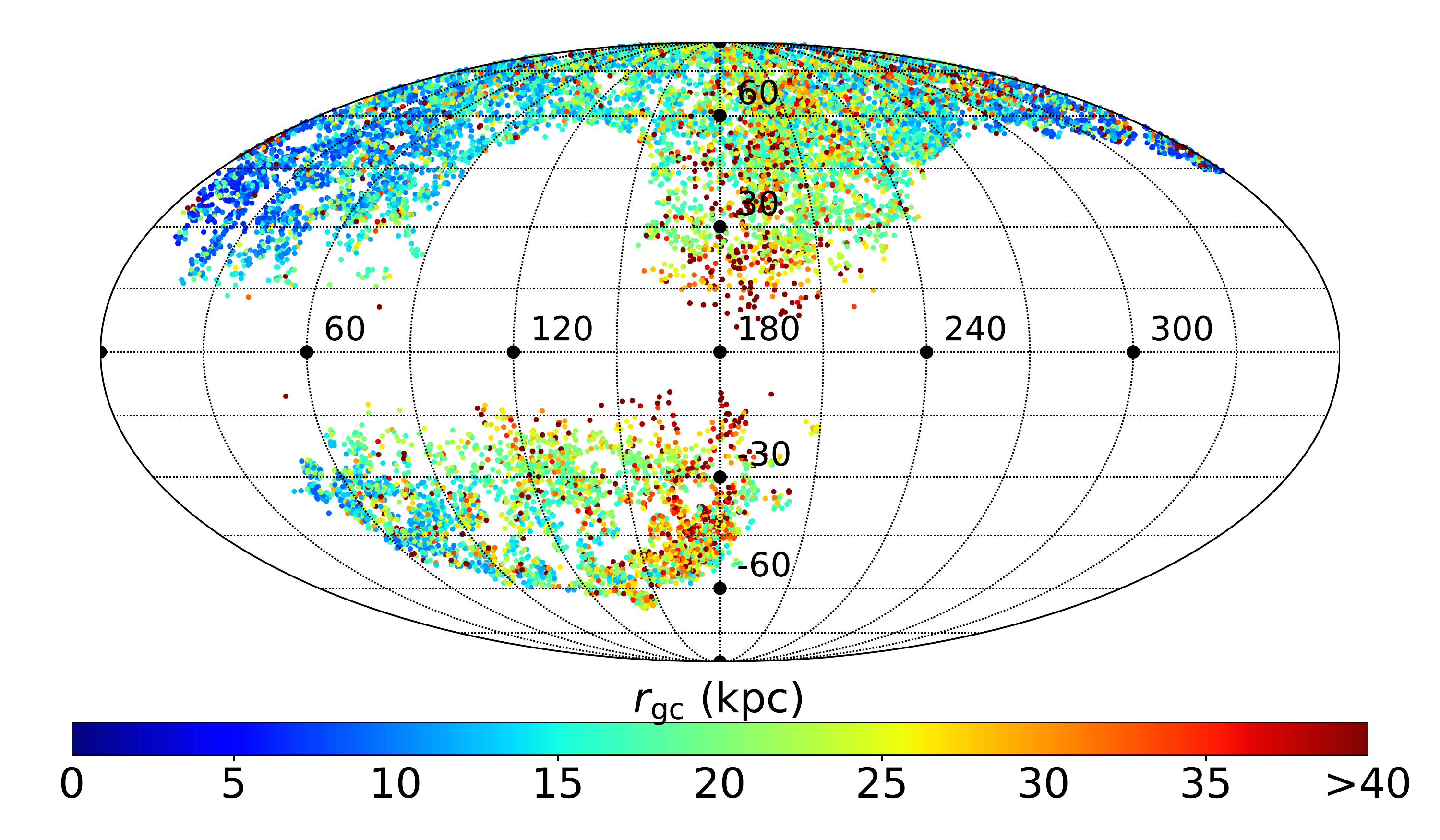} &
    \includegraphics[width=.45\textwidth]{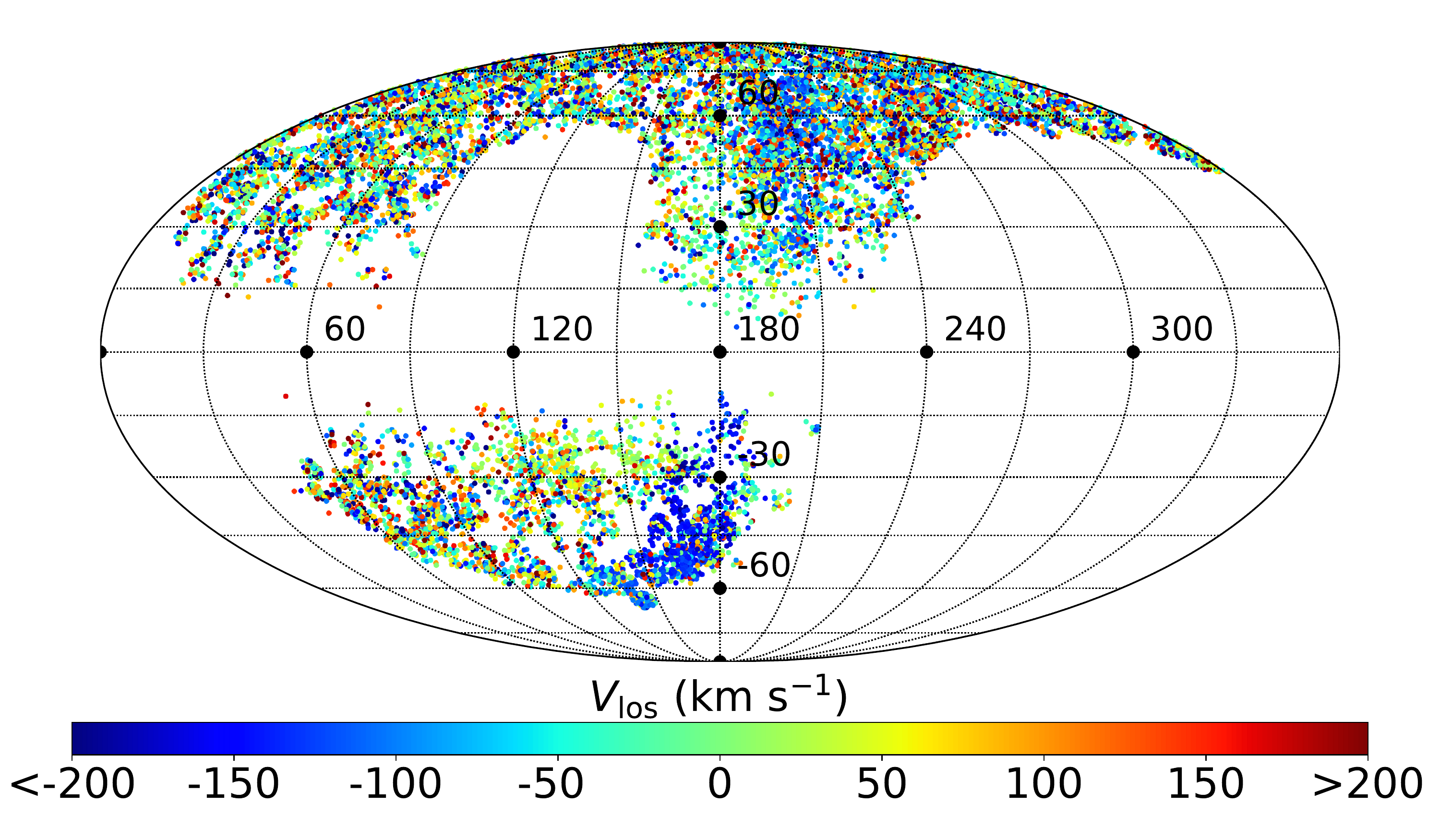} &\\
    \includegraphics[width=.45\textwidth]{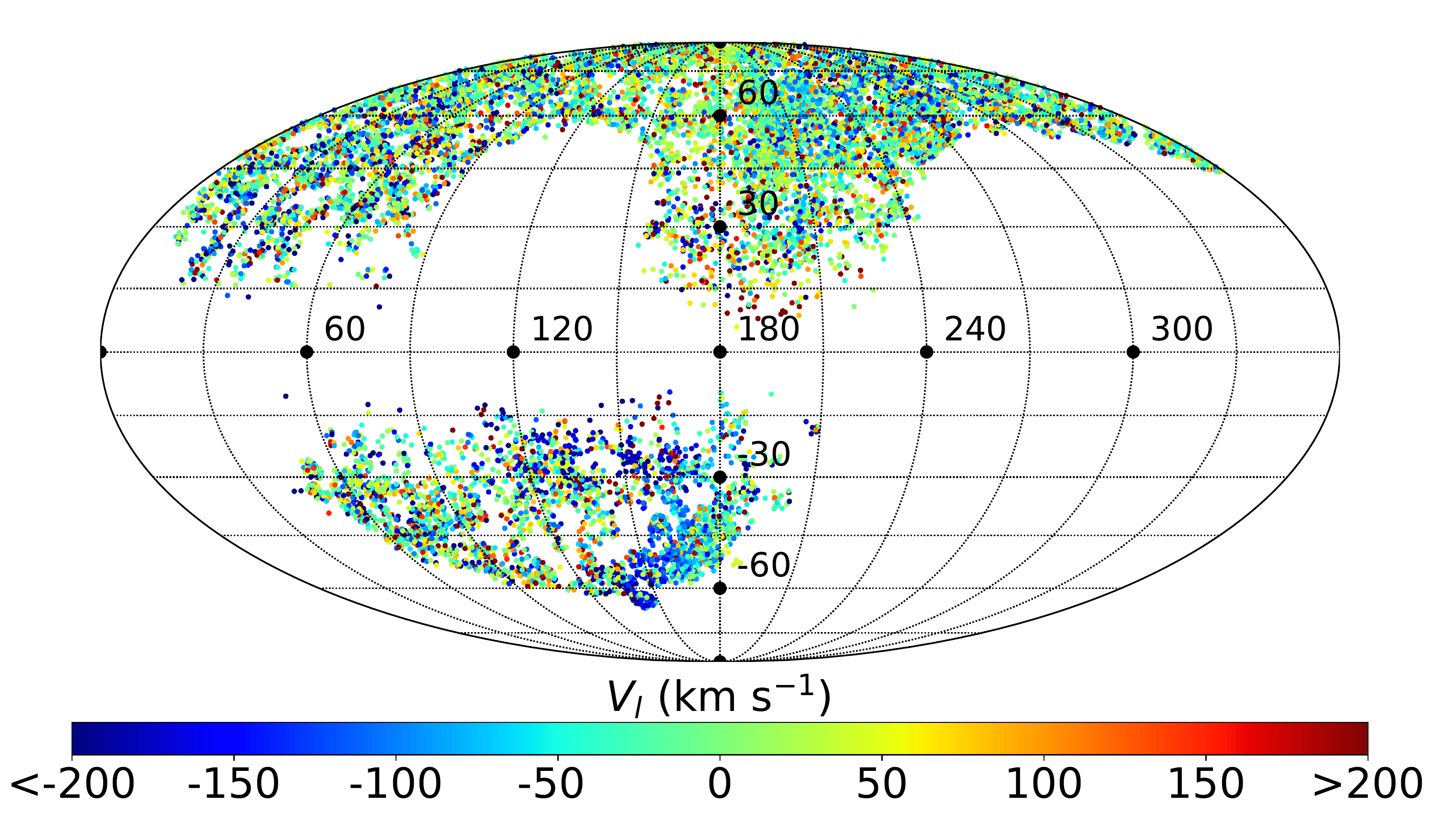} &
    \includegraphics[width=.45\textwidth]{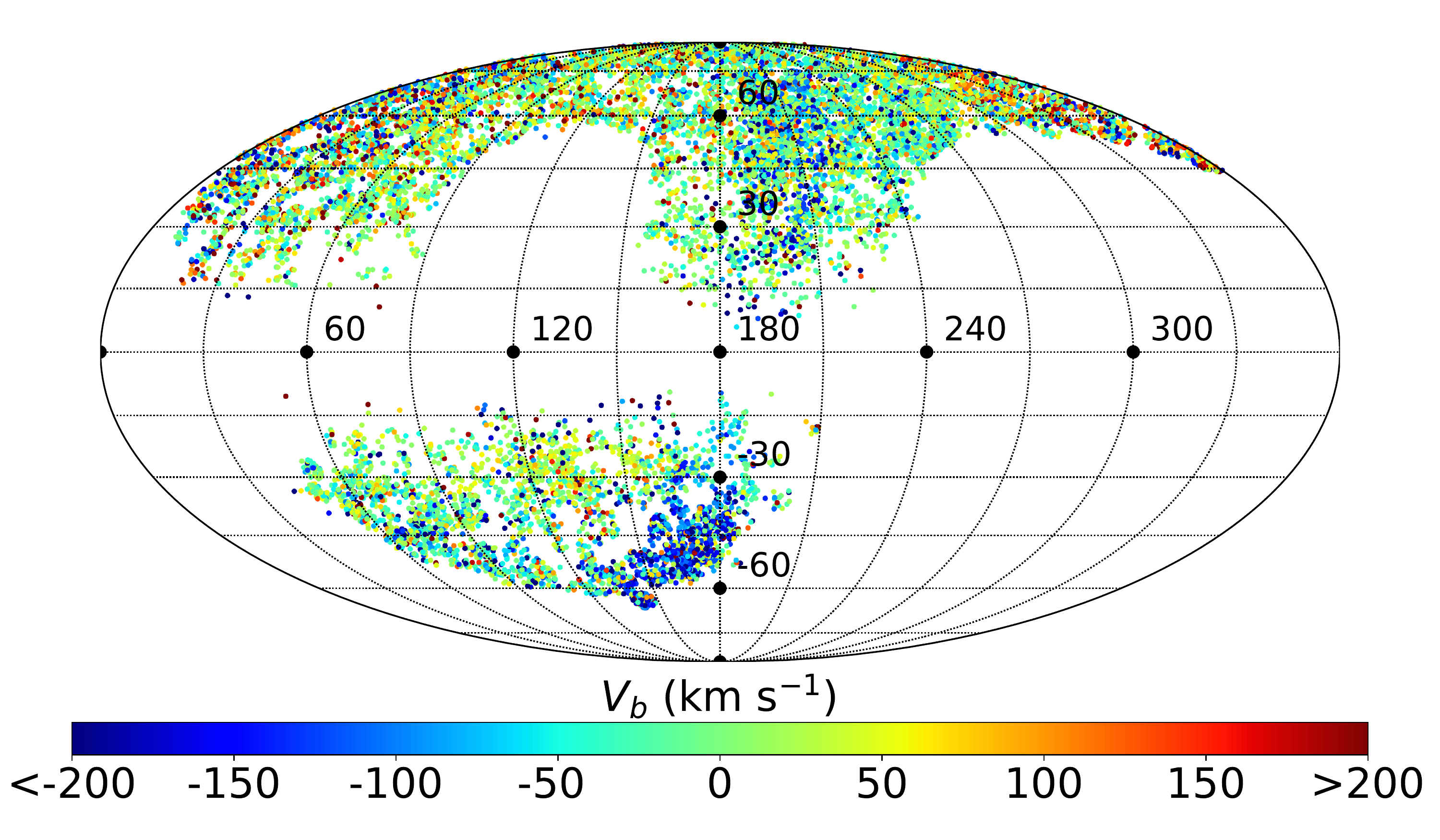} &\\
  \end{tabular}
\caption{Galactic sky coverage for LAMOST halo K giants. Stars are colored according to Galactocentric distance $r_{\rm{gc}}$, line-of-sight velocity $V_{\rm{los}}$, and tangential velocities ($V_{l}$, $V_{b}$). In the region of Sgr streams ($120^\circ < l < 180^\circ$ \& $-60^\circ < b < -30^\circ$ and $180^\circ < l < 200^\circ$ \& $30^\circ < b < 60^\circ$), stars show similar distances and velocities.
} \label{moll}
\end{figure}

Finally, we build a sample of 13,554 halo K giants with 3D positions, 3D velocities, and metallicities. The spatial distribution of the halo K giants in $x-z$ plane is shown in Figure \ref{all_xz}, and the distributions of distances and velocities are shown in Figure \ref{d_v}. The majority of halo K giants in our sample have Galactocentric distances in the range $\mathrm{5-60~kpc}$, with some stars up to $\mathrm{120~kpc}$. The errors of velocities and distances are shown in Figure \ref{d_errs}. The typical errors are 13\% in distance, 7 km s$^{-1}$ in line-of-sight velocity and 20 km s$^{-1}$ in tangential velocities. The sky coverage of the halo K giants with velocity color-coded in Figure \ref{moll} shows that some K giants in the region of Sgr streams have similar velocities, so next we will detect and identify the substructures in position-velocity space from LAMOST halo K giants.

\section{6Distance and Friends-of-friends Algorithm}\label{6d}
We now start quantifying the presence of any kinematic substructure and identifying members of the substructure in 6D phase-space using LAMOST halo K giants. The kinematically cold streams are not strongly phase-mixed, so the adjacent stars in stellar streams are supposed to have similar velocities. Here, we follow \citet{Starkenburg09} and \citet{Janesh16} and develop a statistic that focuses on the incidence of close pairs in $(l,b,d,V_{los},V_l,V_b)$, and then we combine the friends-of-friends algorithm to group stars that are possible in structure.

\subsection{6Distance}
We develop 6Distance from 4distance \citep{Starkenburg09} to calculate a 6D separation of $(l,b,d,V_{los},V_{l},V_{b})$ of any two stars. $(l,b)$ are celestial position in the Galactic coordinate system, $d$ is distance to the Sun, $V_{los}$ is line-of-sight velocity, and $(V_{l},V_{b})$ are tangential velocities along $(l,b)$. All velocities of $(V_{los},V_{l},V_{b})$ are in GSR frame (see Section \ref{halo selection}).

6Distance between two stars $i$ and $j$ is defined as follows:
\begin{equation}
\delta^2_{6d}=\omega_{\theta}\theta^2_{ij}+\omega_{\Delta{d}}(d_i-d_j)^2+
         \omega_{\Delta{V_{l}}}(V_{l,i}-V_{l,j})^2+
         \omega_{\Delta{V_{b}}}(V_{b,i}-V_{b,j})^2+
         \omega_{\Delta{V_{\rm{los}}}}(V_{\rm{los},i}-V_{\rm{los},j})^2,
\label{e_6d}
\end{equation}
where $\theta_{ij}$ is the great circle distance between two stars and calculated by:
\begin{equation}
\cos{\theta_{ij}}= \cos{b_i}\cos{b_j}\cos{(l_i-l_j)}+\sin{b_i}\sin{b_j}.
\end{equation}
The five weights $\omega_{\theta}, \omega_{\Delta{d}}, w_{\Delta V_{\rm{los}}}, w_{\Delta V_l},$ and $w_{\Delta V_b}$ are used to normalize the corresponding components, and define as follows:
\begin{equation}
\begin{array}{l}
\omega_{\theta}     = \frac{1}{\pi^2},\\
\omega_{\Delta{d}}  = \frac{1}{100^2} \frac{(d_{\rm{err}}(i)/d(i))^2+(d_{\rm{err}}(j)/d(j))^2}{2<d_{\rm{err}}/d>^2},\\
\omega_{\Delta{V_*}} = \frac{1}{500^2}\frac{V^2_{\rm{*,err}}(i)+V^2_{\rm{*,err}}(j)}{2<V_{\rm{*,err}}>^2},\\
\label{e_we}
\end{array}
\end{equation}
where $V_*$ stands for $V_l, V_b$, or $V_{\rm{los}}$, $<...>$ denotes the average over all stars. The constants in the weights are the largest angular separation ($\pi$), the largest heliocentric distance separation (100 kpc), and the largest velocity separation (500 km s$^{-1}$), for LAMOST halo K-giant sample. \citet{Starkenburg09} and \citet{Xue11} had pointed out that 4Distance is insensitive to small changes in the weighting factors. We also tried weights ($w_\theta=\frac{1}{<\theta^2>}$, $w_{\Delta d}=\frac{1}{<(\Delta d)^2>}$, $w_{\Delta V_*}=\frac{1}{<(\Delta V_*)^2>}$) defined by \citet{Xue11} and find that different weights affect little to the substructure quantification and identification.

\subsection{The diffuse halo system}\label{diffuse halo}
If position-velocity substructure is present, it is expected that the distribution of $\delta_{6d}$ for the observed sample has more close pairs than the null hypothesis of a diffuse halo system where positions and velocities are uncorrelated. We construct the diffuse halo by only shuffling distances and velocities of our sample, but keeping the angular positions:
\begin{equation}
\delta^2_{6d_{r}}=\omega_{\theta}\theta^2_{ij}+\omega_{\Delta{d}}(d_{i_{r}}-d_{j_{r}})^2+
         \omega_{\Delta{V_{l}}}(V_{l,i_{r}}-V_{l,j_{r}})^2+
         \omega_{\Delta{V_{b}}}(V_{b,i_{r}}-V_{b,j_{r}})^2+
         \omega_{\Delta{V_{\rm{los}}}}(V_{{\rm{los}},i_{r}}-V_{{\rm{los}},j_{r}})^2,
\label{e_6d_s}
\end{equation}
where $\omega_{\theta}, \omega_{\Delta{d}}, w_{\Delta V_{\rm{los}}}, w_{\Delta V_l}, w_{\Delta V_b}$, and the indices $(i,j)$ are exactly the same as in $\delta_{6d}$, but $(i_r,j_r)$ are shuffling indices. The selection function of LAMOST K giants varies with line-of-sight \citep{LiuC17}. However, it is a reasonable assumption that the distance of the stars in the same part of the sky are uncorrelated to the sample selection. Therefore, we do not shuffle the angular positions when constructing the diffuse halo system.

\begin{figure}[htb]
\centering
\includegraphics[scale=0.6]{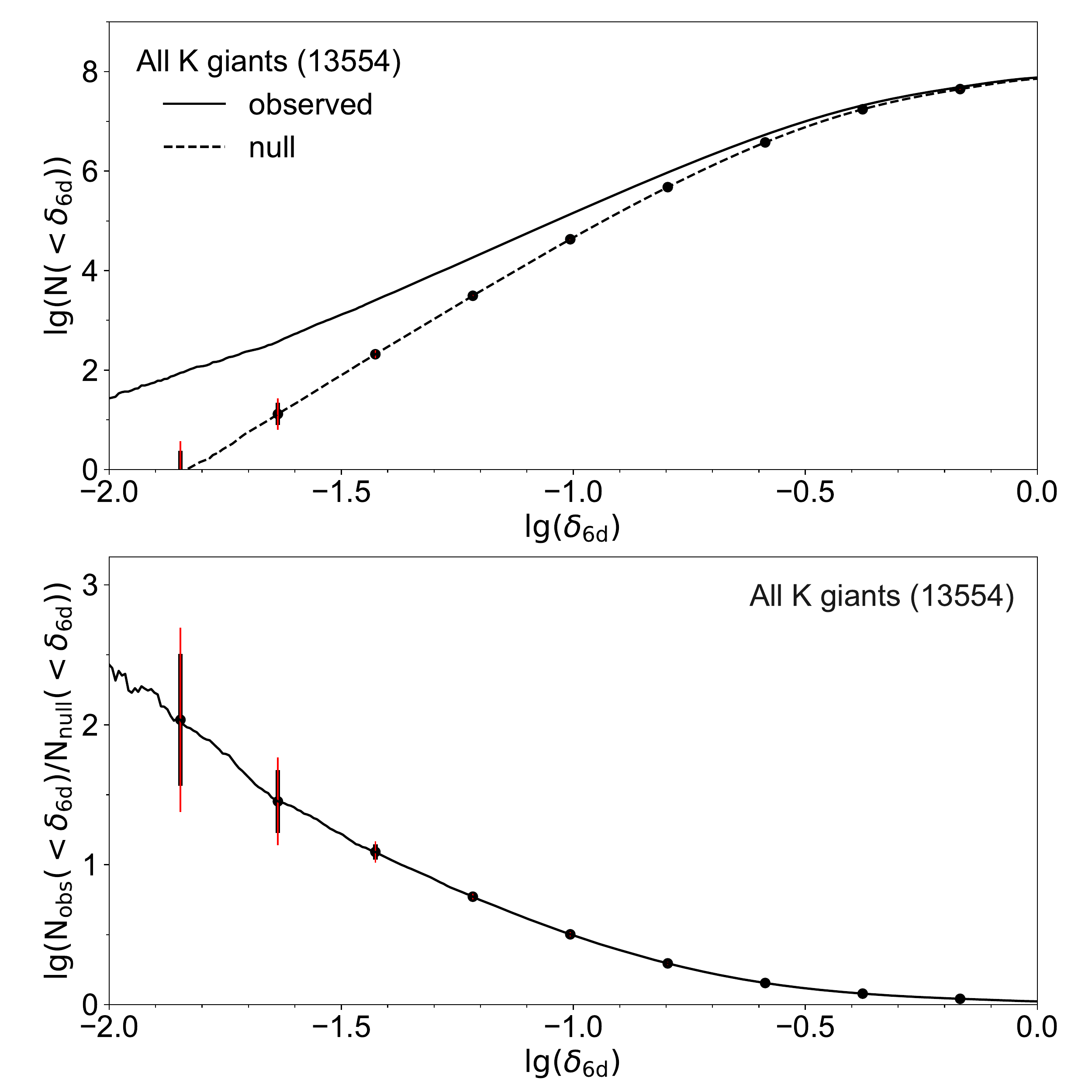}
\caption{Top panel: the close pairs distribution for 13554 halo K giants. $\delta_{6d}$ is the separation of 6D phase-space between any two stars. The solid line is the cumulative distribution of $\delta_{6d}$ for observed sample $N_{\rm{obs}}(<\delta_{6d})$. The dash line is the mean cumulative distribution of $\delta_{6d}$ of 100 Monte Carlo representations of diffuse halo $N_{\rm{null}}(<\delta_{6d})$. The black thick error bars are distribution enclosing 68\% diffuse halo. The red thin error bars enclose 95\% of diffuse halo. Bottom panel: the result of quantifying the halo K giants, $N_{\rm{obs}}(<\delta_{6d})/N_{\rm{null}}(<\delta_{6d})$. Both panels demonstrate the halo K gaints have more close pairs in 6D phase-space than the diffuse halo system.
} \label{6d_all}
\vspace*{\fill}
\end{figure}

Now, we can quantify the degree of substructure in LAMOST halo K giants by comparing the cumulative distribution of $\delta_{6d}$ for halo K-giant sample, $N_{obs}(<\delta_{6d})$, with those of 100 null hypotheses of diffuse halo system $N_{null}(<\delta_{6d})$. Figure \ref{6d_all} shows that $N_{obs}(<\delta_{6d})$ exceeds $N_{null}(<\delta_{6d})$ obviously for small values of $\delta_{6d}$, which means there are more close pairs in halo K giants. Since the null hypotheses of diffuse halo system have the same selection function with LAMOST halo K giants, more close pairs in the halo K giants is unlikely to be a result of selection function. Consequently, LAMOST halo K giants have substructure indeed.

\subsection{Friends-of-Friends Algorithm}\label{fof}
The quantification of substructure is just the first step, and the identification of streams is of particular importance to understand the formation of the Milky Way, such as finding the progenitors of streams, exploring the chemical properties and mass of the progenitors, and constraining the dynamics of the Milky Way.

FoF is a popular percolation algorithm of group finding. It defines groups that contain all stars separated by 6Distance less than a given linking length. \citet{Janesh16} pointed out that FoF algorithm tends to find groups in the region of higher stellar density. As shown in Figure \ref{all_xz}, LAMOST mainly observes northern Galactic hemisphere. \citet{Xu18} found the Galactic halo density profile traced by LAMOST halo K giants shows a single power law with index of $-4\sim-5$. Therefore, we adopt a sky-distance-dependent linking length (i.e., we divide our sample and allocate larger linking lengths for southern Galactic hemisphere and distant sub-samples). Sgr streams are the most prominent, coherent and widely studied tidal streams in the Milky Way, so its a good criterion to test our linking length. We choose linking length for each part by getting enough reliable members of Sgr streams. The reliability of Sgr members is evaluated by positions and velocities of Sgr stream in literature. We will show below (Section \ref{Sec_sgr}) that the Sgr members obtained by our linking lengths are very consistent with simulation \citep[][LM10]{LM10} and observations \citep{Belokurov14, Koposov12}. The details about the sub-samples and linking lengths will be discussed in Section \ref{id}.

Obviously, the method employed here to identify stars in each substructure has an intrinsic uncertainty due to the choice of the linking length and the working coordinates (i.e. position-velocity space in this paper) itself. Therefore, the stellar samples associated to streams or overdensities in this paper suffer of contamination due to the mentioned reasons. However, it is not easy to quantify such contamination exactly. The Sgr streams are very coherent and dense in phase space, so the linking length suitable to identify Sgr streams should be a stringent choice. From the comparison with some known substructure properties in Section \ref{id}, we find the fraction of contamination is not high.

\section{Results on Quantifying Substructure in LAMOST K Giants}\label{qs}
Both observations and simulations found that the Galactic halo is comprised of at least two overlapping components, an inner halo and an outer halo, with different metallcities, spatial distribution, and kinematics. The inner halo is the dominant component at galactocentric distance up to $\sim$ 15-20 kpc and for metallicity $\mathrm{[Fe/H]>-2.0}$ dex, while the outer halo dominates the region beyond 20 kpc and at metallicity $\mathrm{[Fe/H]<-2.0}$ dex \citep{Carollo07, Carollo10, deJong10, Jofre11, Beers12, Kinman12, An13, Hattori13, Kafle13, Tissera13, Tissera14}. The large sample size of LAMOST halo K giants enables us to quantify the substructure in inner halo and outer halo, as well as in different ranges of metallicity, and to test the contribution of Sgr streams.

\subsection{Substructure in Inner and Outer Halo}

As predicted by the hierarchical galaxy formation model, substructures orbiting in outer halo are short after infall and very coherent in space, but substructures in inner halo are long after infall and spatially well-mixed \citep{Helmi08}. Recent studies used main-sequence turnoff stars (MSTO), blue horizontal branch stars (BHB), and K giants to quantify the degree of substructure and found the Galactic halo significantly more structured at larger radii $r_{\rm{gc}} > $20 kpc \citep{Bell08,Xue11, Cooper11, Janesh16, Santucci15, Carollo16}. Many cosmological simulations show a fully phase-mixed inner halo and increasing fraction of the substructure with distance \citep{Bullock01, Napolitano03, BJ05, DH08, Cooper10, Pillepich15, Carollo18}.

\begin{figure}[htb]
\centering
  \begin{tabular}{@{}c@{}}
    \includegraphics[width=.9\textwidth]{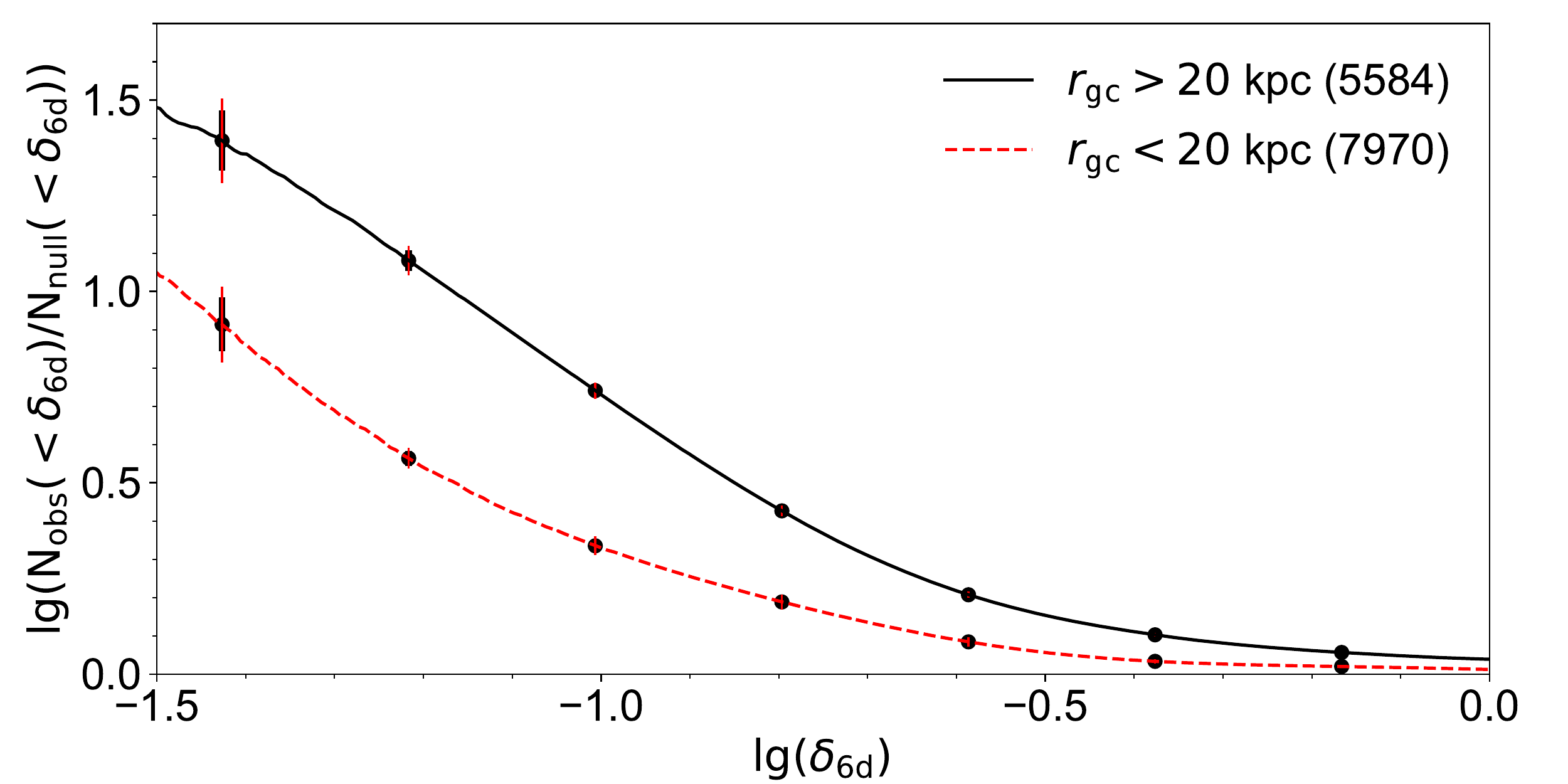}
  \end{tabular}
\caption{The result of quantifying the halo K giants in $r_{\rm{gc}} <$ 20 kpc (red dashed line) and in $r_{\rm{gc}} >$ 20 kpc (black solid line). The substructure signal exists in both regions, while it is stronger in $r_{\rm{gc}} >$ 20 kpc.
} \label{6d_rgc}
\end{figure}

To test it, we divide LAMOST halo K giants into two sub-samples - one with $\mathrm{5 ~kpc<}$ $r_{\mathrm{gc}}$ $\mathrm{<20~kpc}$ and the other with $r_{\mathrm{gc}}$ $\mathrm{>20~kpc}$, and compare the substructure signals in them. Figure \ref{6d_rgc} shows that the sub-sample beyond $\mathrm{20~kpc}$ presents a stronger structure signal, $\sim$ 3 times more than halo stars within $\mathrm{20~kpc}$ at $lg(\delta_{6d})=-1.0$. It means the outer halo is more structured than the inner halo, which is consistent with the previous findings based on observations and simulations.

\subsection{Substructure Dependence on Metallicity}\label{sm}
Covering large range of metallicities makes K giants good representative tracers of Galactic stellar halo. Metallicity of accreted stars can be used to infer the mass of their progenitor according to the mass-metallicity relation \citep{Lee06}. The relation tells us that if a massive dwarf galaxy is accreted, its stellar populations are likely to be metal-rich.

\begin{figure}[htb]
\centering
  \begin{tabular}{@{}c@{}}
    \includegraphics[width=.9\textwidth]{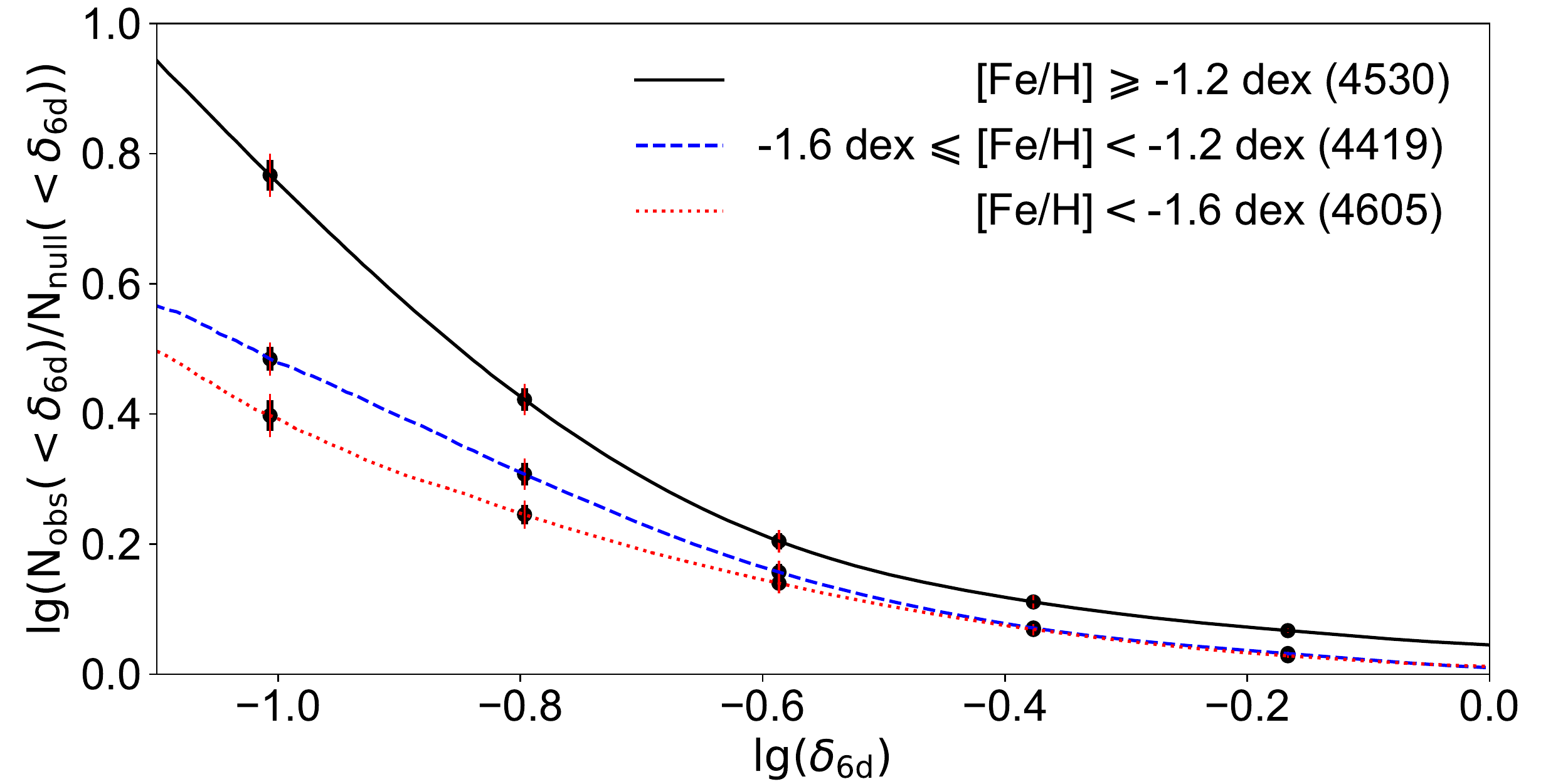}
  \end{tabular}
\caption{The result of quantifying the halo K giants in different metallicity ranges. The substructure signal increases with metallicity.
} \label{6d_feh}
\end{figure}

The meaningful statistics require large enough sample. We divide LAMOST halo K giants into three sub-samples with comparable sizes: one with [Fe/H] $<$ -1.6 dex, one with -1.6 dex $\leqslant$ [Fe/H] $<$ -1.2 dex, and another with [Fe/H] $\geqslant$ -1.2 dex. Figure \ref{6d_feh} shows the substructure measurements of the three sub-samples. The most metal-rich sub-sample has the strongest substructure signal, with $\sim$ 9 times more than diffuse halo at $lg(\delta_{6d})=-1.0$. The sub-sample with intermediate metallicity has $\sim$ 3 times more pairs than diffuse halo at $lg(\delta_{6d})=-1.0$, while the most metal-poor sub-sample shows the weakest substructure signal, with $\sim$ 2.5 more pairs than diffuse halo at $lg(\delta_{6d})=-1.0$. These results suggest that the substructure signal is increasing with metallicity.

\subsection{Contribution of Sgr Streams to the Substructure}
\begin{figure}[htb]
\centering
  \begin{tabular}{@{}c@{}}
    \includegraphics[width=.9\textwidth]{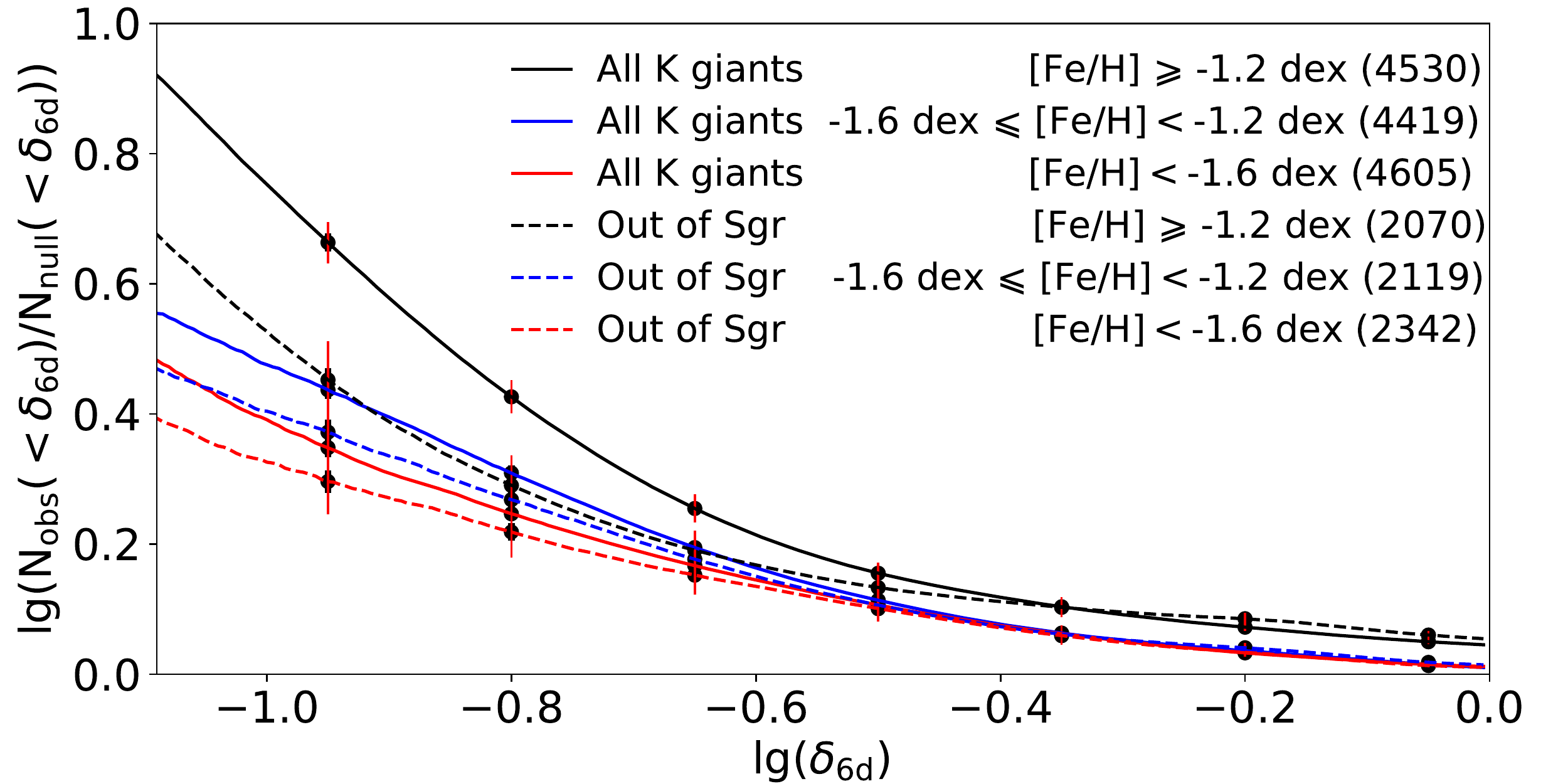}
  \end{tabular}
\caption{The comparison of quantifying the halo K giants before (solid lines) and after (dash lines) removing the Sgr stream (see more detals in the text). The colors represent the different metallicity ranges, which are marked in the legend.
} \label{6d_sgr_feh}
\end{figure}

To study the contribution of the Sgr Stream to the substructure strength, we test the substructure-metallicity relation in two cases, with and without the Sgr Stream stars. In the non-Sgr Stream case, all the K giant stars with $|B|< 12^\circ$ are removed following \citet{Majewski03} and \citet{ Belokurov06}. Figure \ref{6d_sgr_feh} shows the distribution of the relation for the stars with different metallicity ranges in the two cases, solid lines represent the results for all K giant stars and dashed lines for those out of the plane. We can find a clear relation that the substructure strength is higher for the metal-richer sample in both cases, and a significant decrease of the substructure strength with $lg(\delta_{6d})=-1.0$ between the two cases. What's more, the results of the metal-richer stars decrease more than that of the metal-poorer samples, e.g. the substructure strength of the most metal-rich sample decreases from 0.75 down to 0.53. While the strength of the most metal-poor samples decrease from 0.39 down to 0.33. The difference between the two cases indicates that the Sgr Stream significantly enhances this relation, which was also claimed by \citet{Janesh16}. All results above are suggesting that LAMOST is able to provide more help for further substructure investigation in the halo, including the Sgr Stream.

\section{FoF Results}\label{id}
As described in Section \ref{fof}, we divide our sample and allocate different linking lengths for each sub-sample. Specifically, we divide our sample into 7 sub-samples according to the sky coverage and distance, as shown in Table \ref{parts}. Note that there are some overlapped regions between the 7 sub-samples. The overlapped regions of sub-samples will produce common grouped members in the result of FoF groups. We will remove the common members from distant sub-sample groups to make sure the grouped members are unduplicated. By comparing our Sgr groups with the known properties of Sgr stream (e.g., distance, position, and velocities), we determine the linking length for each sub-sample. The specific physical sizes of each component corresponding to the linking lengths can be found in Table \ref{lk}. The physical size is assuming two stars have 5 identical components of 6 phase-space, then calculating the difference component at a given linking lengths. For example, if two stars have identical values of $l,b,d,V_{l},V_{b}$, a difference of 25 km s$^{-1}$ in $V_{\rm{los}}$ would produce a $\delta_{6d}$ of 0.05.

\begin{figure}[htb]
%\centering
  \begin{tabular}{@{}c@{}}
  \includegraphics[width=.95\textwidth]{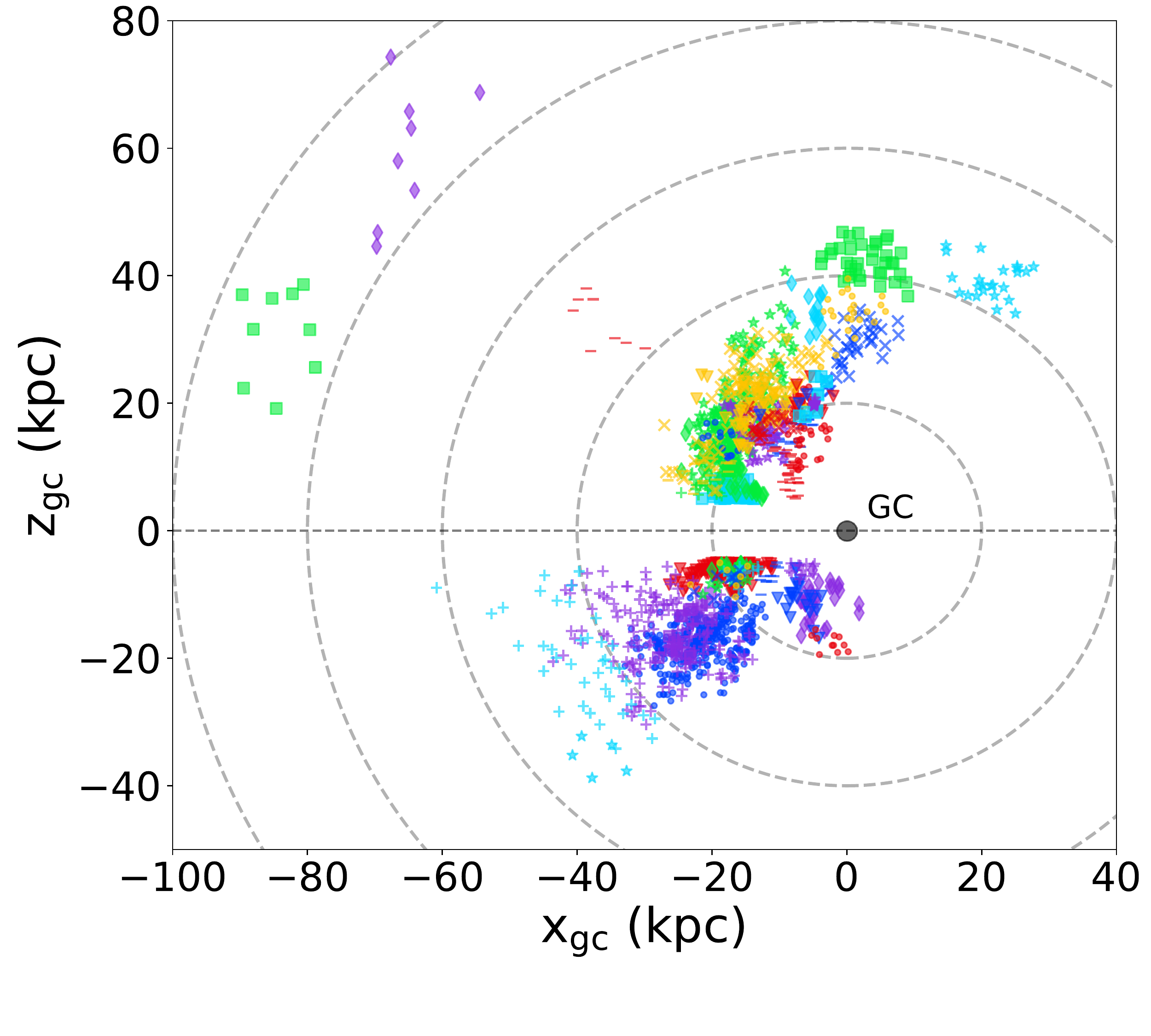}
  \end{tabular}
\caption{The $x-z$ plane distribution of 43 groups (1867 stars) identified from LAMOST halo K giants. We combine different colors and symbols to represent different groups. Dashed lines are every 20 kpc in $r_{\rm{gc}}$. Because of the limited number of colors and symbols, some different groups are shown with same color and symbol.
} \label{all_groups}
\end{figure}

\begin{figure}[htb]
%\centering
  \begin{tabular}{@{\hspace{-0.75cm}}c@{}}
  \includegraphics[width=\textwidth]{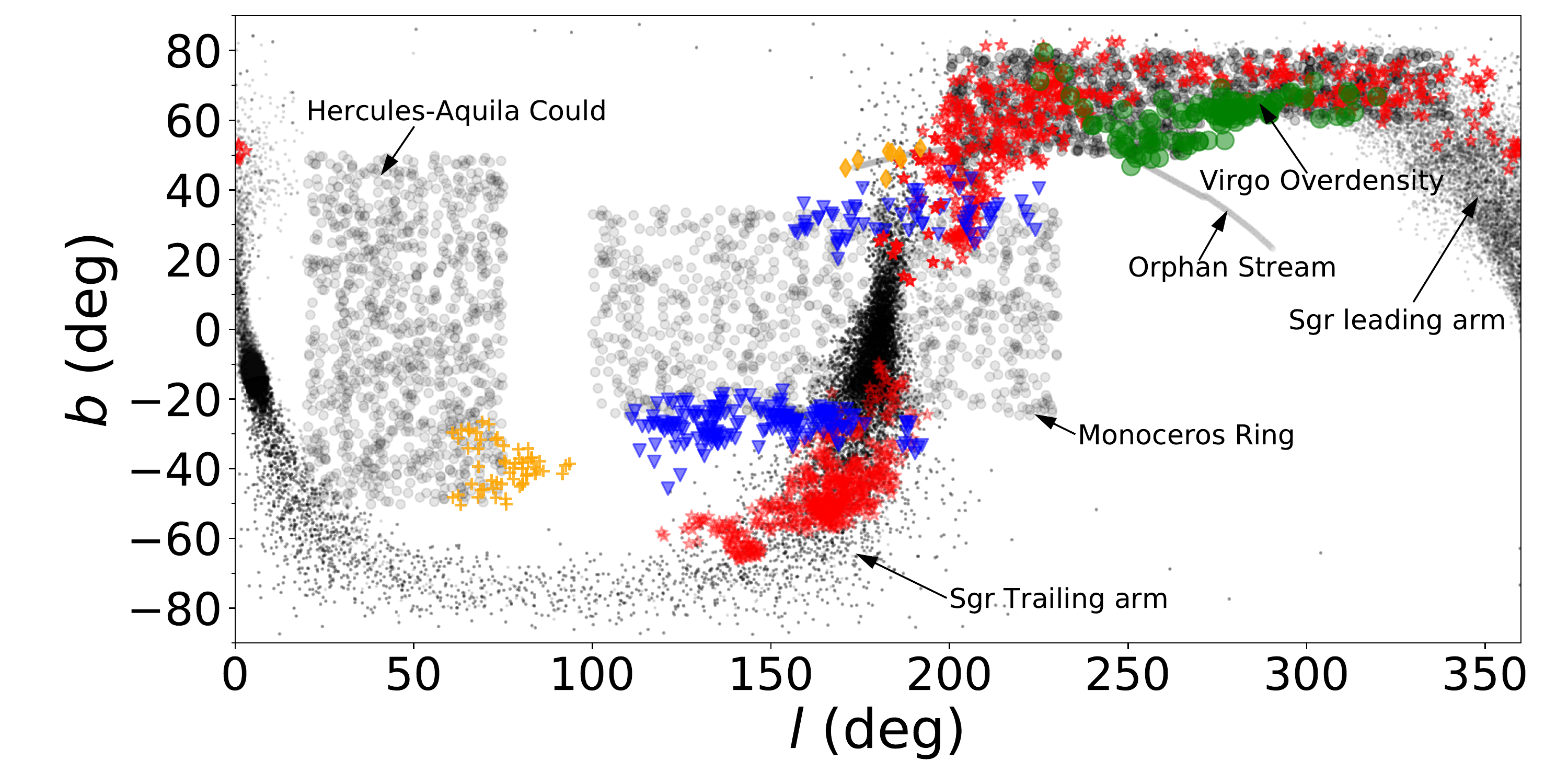}
  \end{tabular}
\caption{The sky coverage of five known substructures identified from LAMOST halo K giants: Sgr Stream (red stars), Monoceros Ring (blue triangles), Virgo Overdensity (green circles), Hercules-Aquila Cloud (orange pluses), and Orphan Stream (orange diamonds). The background (gray dots/filled circles) is literature sky coverage of these known substructure: Sgr leading arm and Sgr trailing arm \citep{LM10}, Monoceros \citep{Slater14}, Virgo Overdensity \citep{Bonaca12}, Hercules-Aquila Could \citep{Belokurov07}, and Orphan stream \citep{Newberg10}.
} \label{known_lb}
\end{figure}

Finally, we identify 25 groups (1517 K giants), associated to 5 known substructures: Sgr stream \citep{Ibata01, Majewski03}, Monoceros Ring \citep{Newberg02}, Virgo Overdensity \citep{Newberg02}, Hercules-Aquila Cloud \citep{Belokurov07}, and Orphan Stream \citep{Grillmair06, Belokurov07}. Besides, 18 groups (350 K giants) can not be linked to any known substructure, so they may relate to some unknown substructures. In total, 1867 grouped stars are identified as shown in Figure \ref{all_groups}, and the known substructures' sky distribution is shown in Figure \ref{known_lb}. The corresponding properties of known substructures and unknown groups are listed in Table \ref{t_known} and Table \ref{t_unknown}, respectively.

\subsection{Attributing Groups to Sgr Stream}\label{Sec_sgr}
Sgr stream is the most prominent stellar stream, and it has become an important tool for studying the Milky Way halo. In Section \ref{kg}, the spatial and velocity distributions have shown the exist of Sgr stream in the sample of LAMOST K giants (see Figure \ref{all_lv} and Figure \ref{moll}). In this section, we link the FoF groups to Sgr streams by comparing them with models (\citetalias{LM10}) and observations \citep{Belokurov14, Koposov12}.

\begin{figure}[htb]
%\centering
  \begin{tabular}{@{}cccc@{}}
    \includegraphics[width=.45\textwidth]{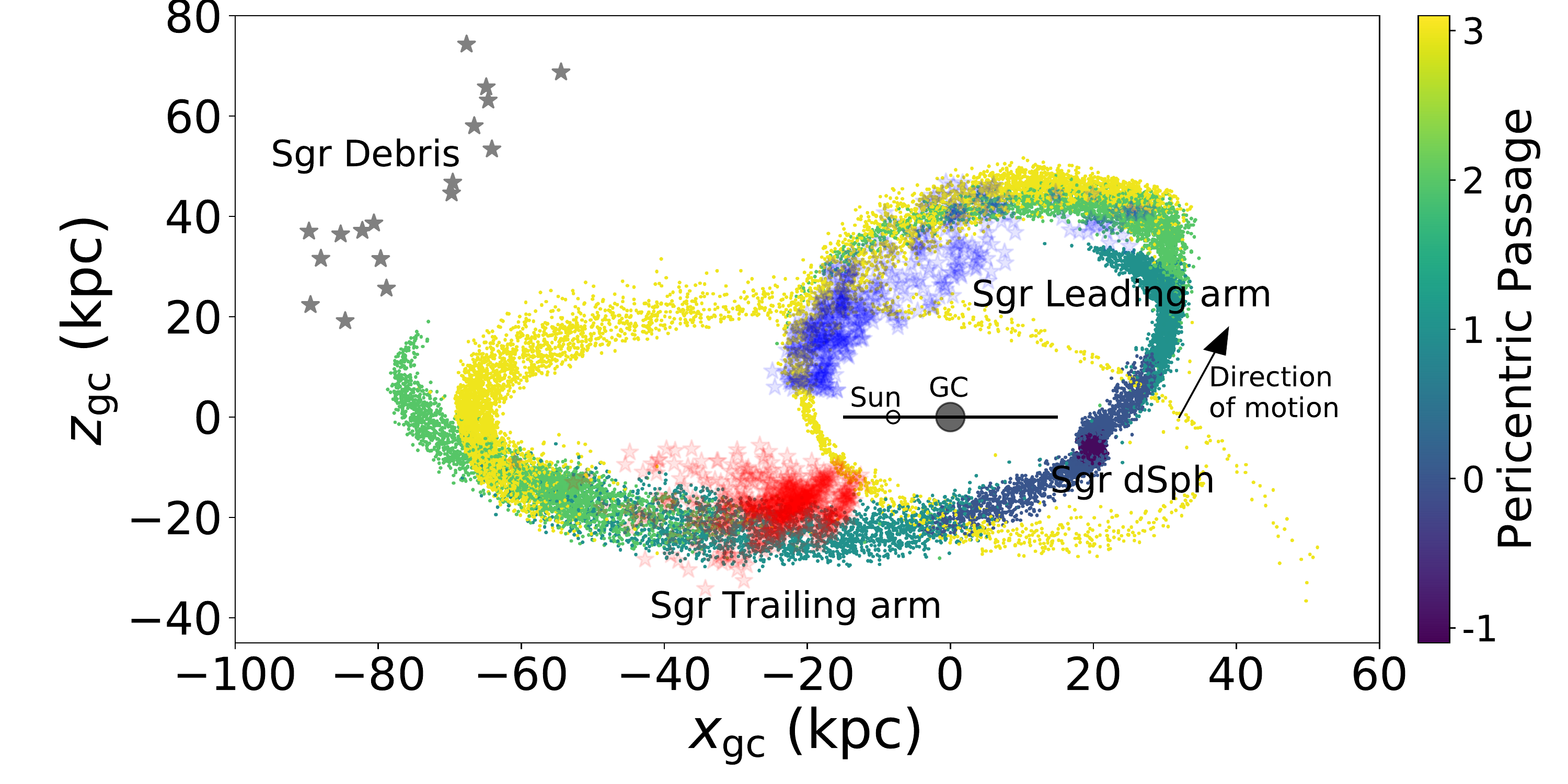} &
    \includegraphics[width=.45\textwidth]{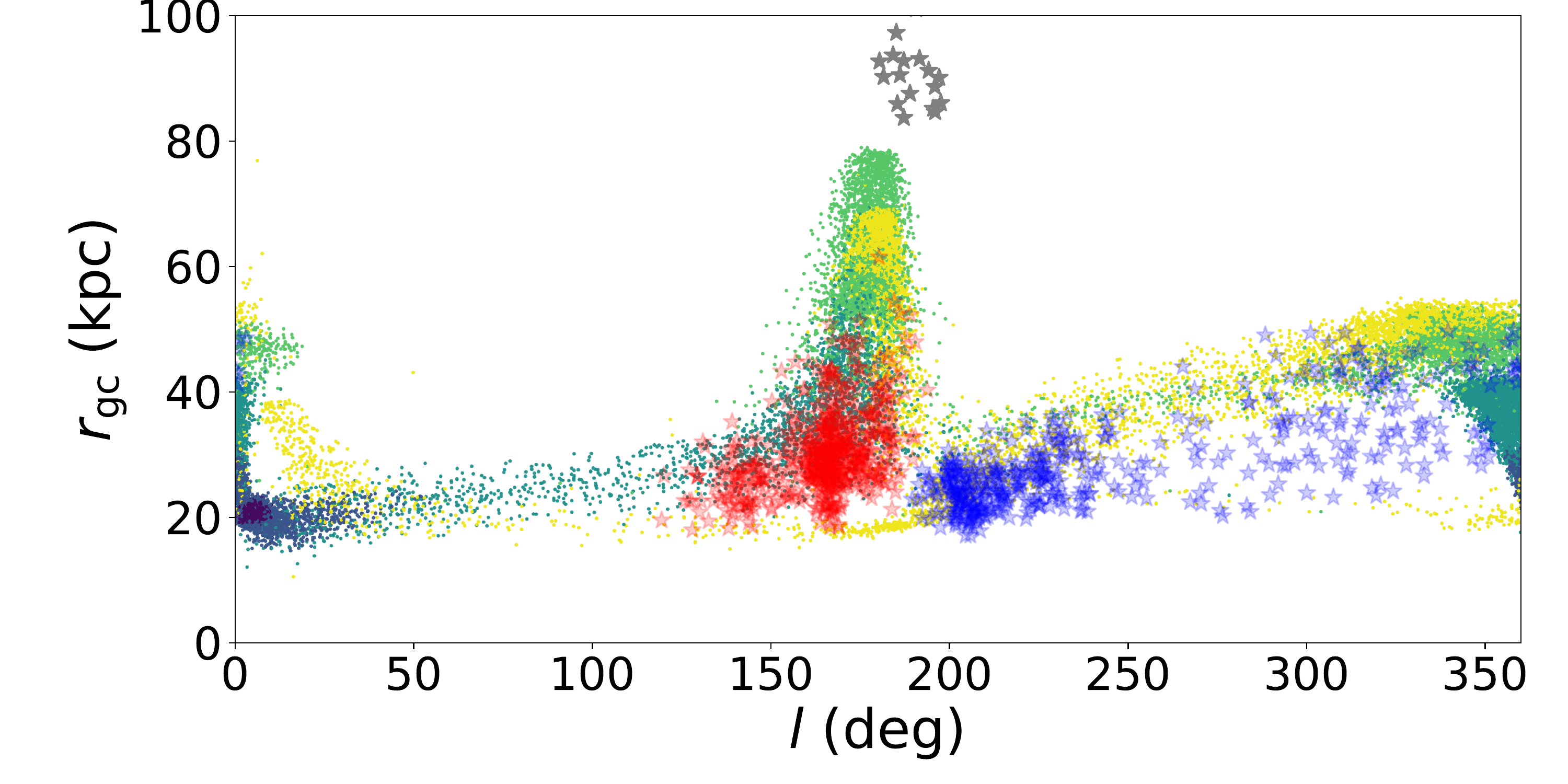} &\\
    \includegraphics[width=.45\textwidth]{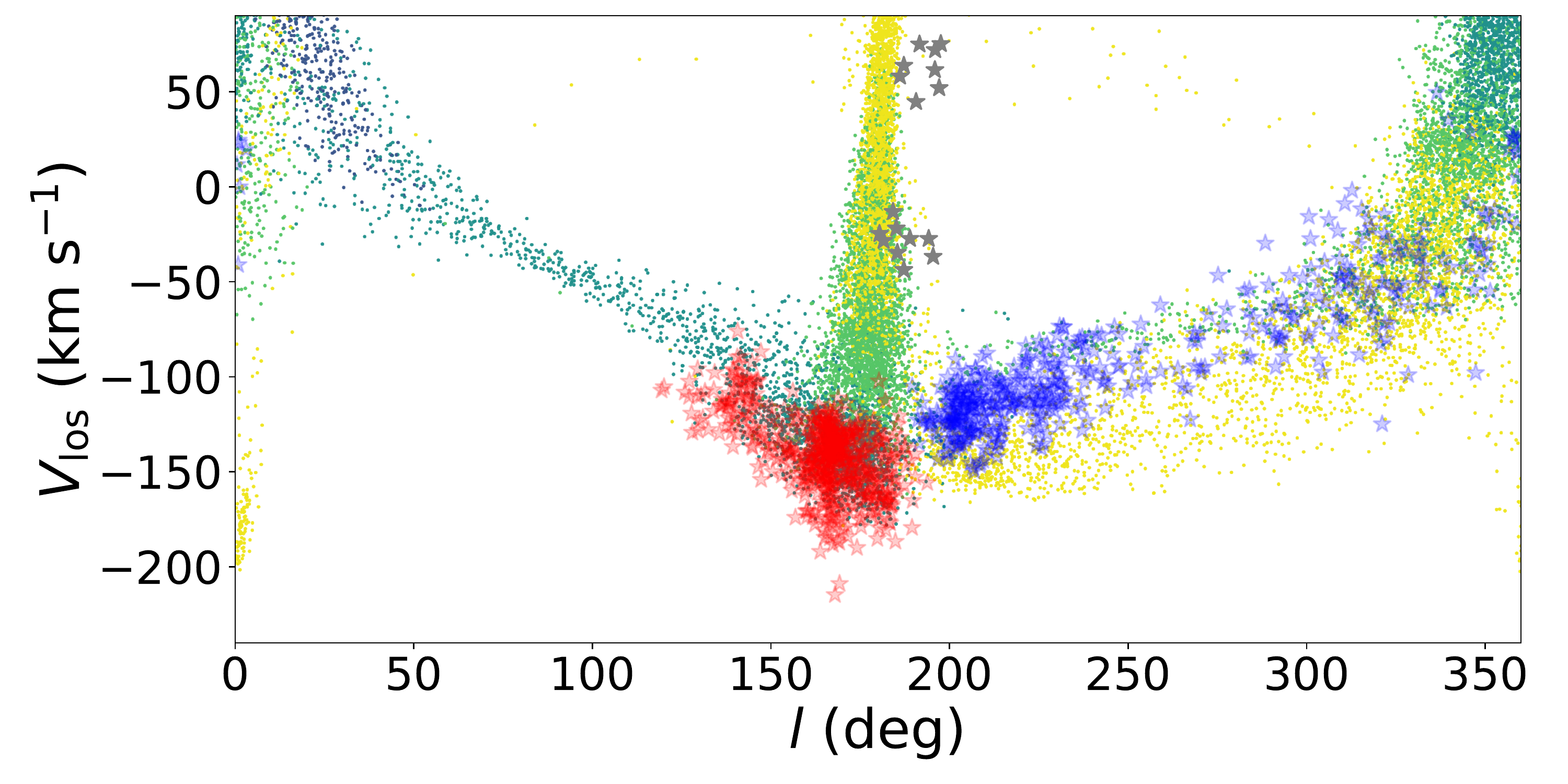} &
    \includegraphics[width=.45\textwidth]{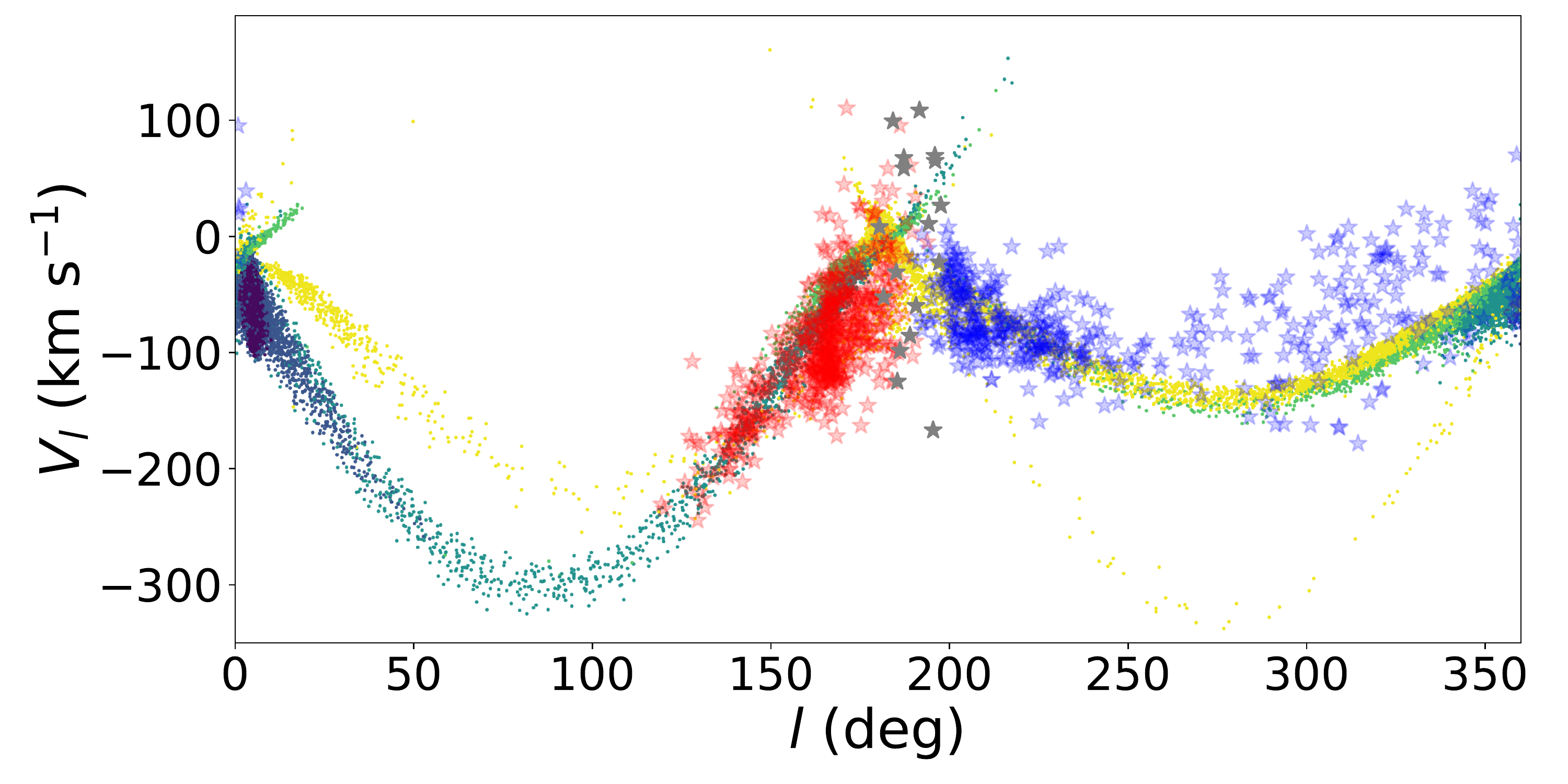} &\\
    \includegraphics[width=.45\textwidth]{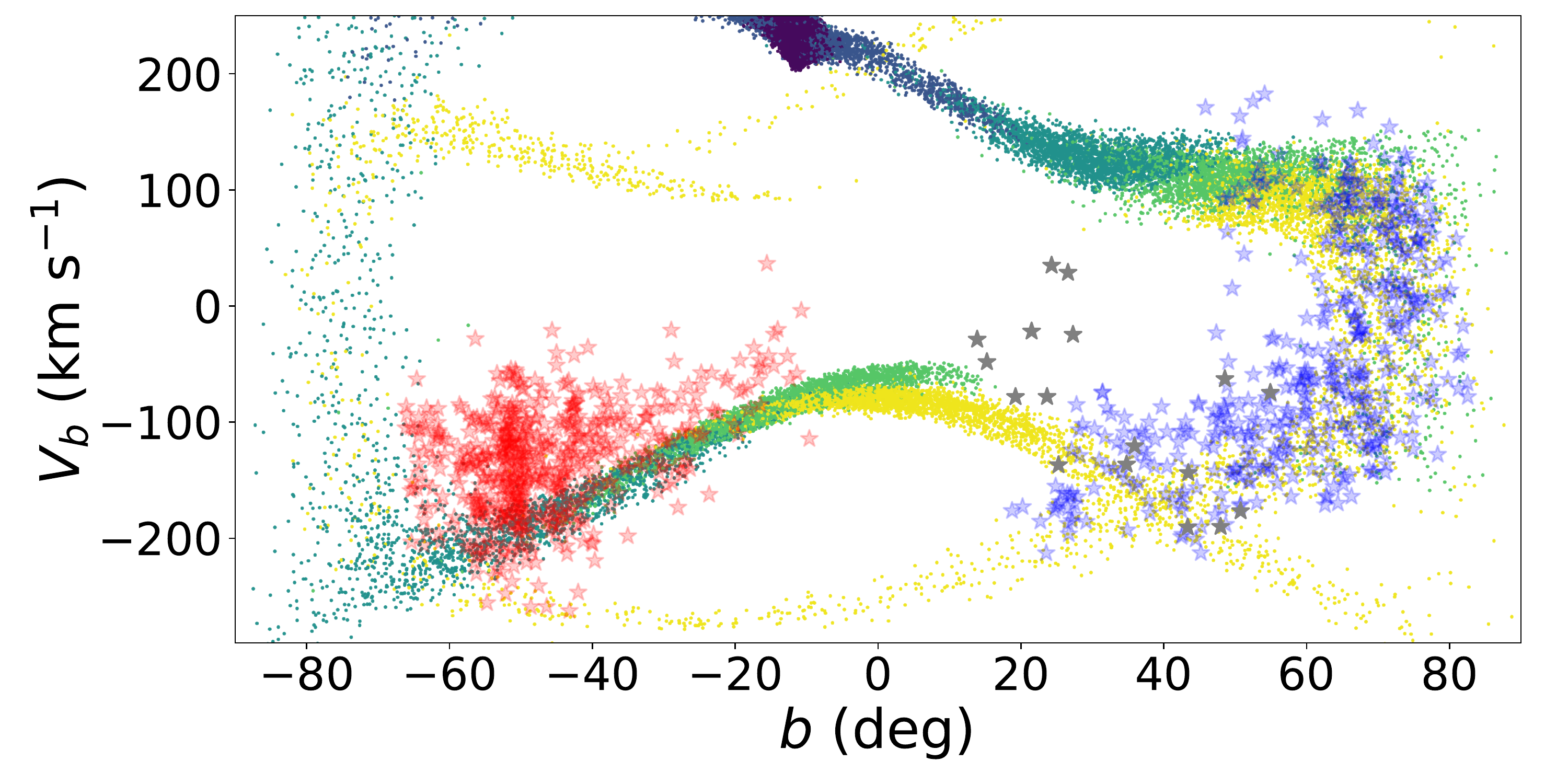} &
    \includegraphics[width=.45\textwidth]{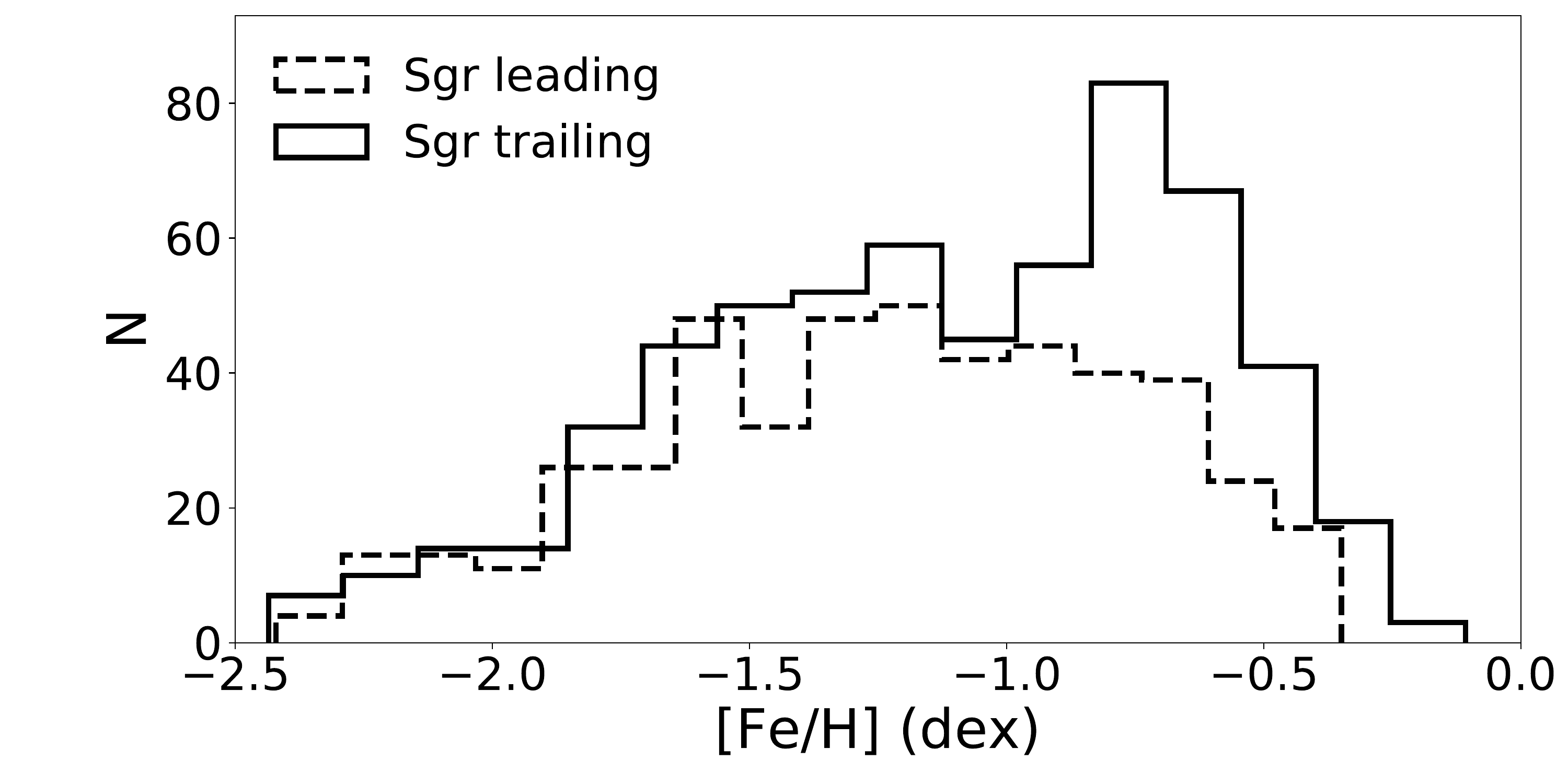} &\\
  \end{tabular}
\caption{The comparison of Sgr members identified in LAMOST halo K giants with \citetalias{LM10} model. There are 8 FoF groups relate to Sgr leading arm (blue star symbols), 3 FoF groups attributed to Sgr trailing arm (red star symbols), and 2 FoF groups belonging to Sgr debris (grey star symbols). The color dots are from LM10 model, which is color-coded by the pericentric passage (values of -1 indicate debris which is still bound at the present day, values of 0 indicate debris stripped on the most recent pericentric passage of Sgr, and values of 1 indicate debris stripped on the previous pericentric passage, see \citetalias{LM10} for details). Comparing with \citetalias{LM10} model, Sgr streams traced by LAMOST K gaints are more diffuse, and locate closer. The mean [Fe/H] values of Sgr leading and trailing arms are -1.24 dex and -1.12 dex, and their dispersion are 0.47 dex and 0.49 dex, which consistent with the prediction of the model that the leading arm is composed by more metal-poor stars due to the origin from the periphery of Sgr dwarf galaxy.
} \label{sgr}
\end{figure}

Comparing with the five most recent pericentric passages of \citetalias{LM10} model, we find 11 groups match well with \citetalias{LM10} model (see Figure \ref{sgr}), of which 8 groups belong to Sgr leading arm (blue stars), and 3 groups belong to Sgr trailing arm (red stars). Figure \ref{sgr} shows Sgr streams traced by K giants have larger dispersion (even offset) in distance and tangential velocities than \citetalias{LM10} model. Larger dispersion may be caused by the errors of distances. Unlike ``standard candles" (e.g., BHB with distances good to 5\%, RR Lyrae stars with distances good to 3\% and red clump stars with distances better than 10\% ), K giants have a typical error of about 15\% because their intrinsic luminosities vary by two orders of magnitude with color and depend on metallicity and age. Given that tidal stripping generally eats away a satellite from the outside, so more recent pericentric passages means smaller mean internal (to the dwarf galaxy) orbital radii before they were unbound, and having higher metallicity \citep{Majewski13, Hasselquist17}. Figure \ref{sgr} shows the Sgr trailing members are located in more recent pericentic passages than Sgr leading members, and the mean [Fe/H] value of the Sgr trailing members are indeed higher than that of the Sgr leading members. Besides the matched groups with \citetalias{LM10} model, there are 2 groups beyond distance range of \citetalias{LM10} model shown as the grey stars in Figure \ref{sgr}. However, they match well with Sgr debris found by \citet{Belokurov14}.

\begin{figure}[htb]
\centering
  \begin{tabular}{@{}cccc@{}}
    \includegraphics[width=1.0\textwidth]{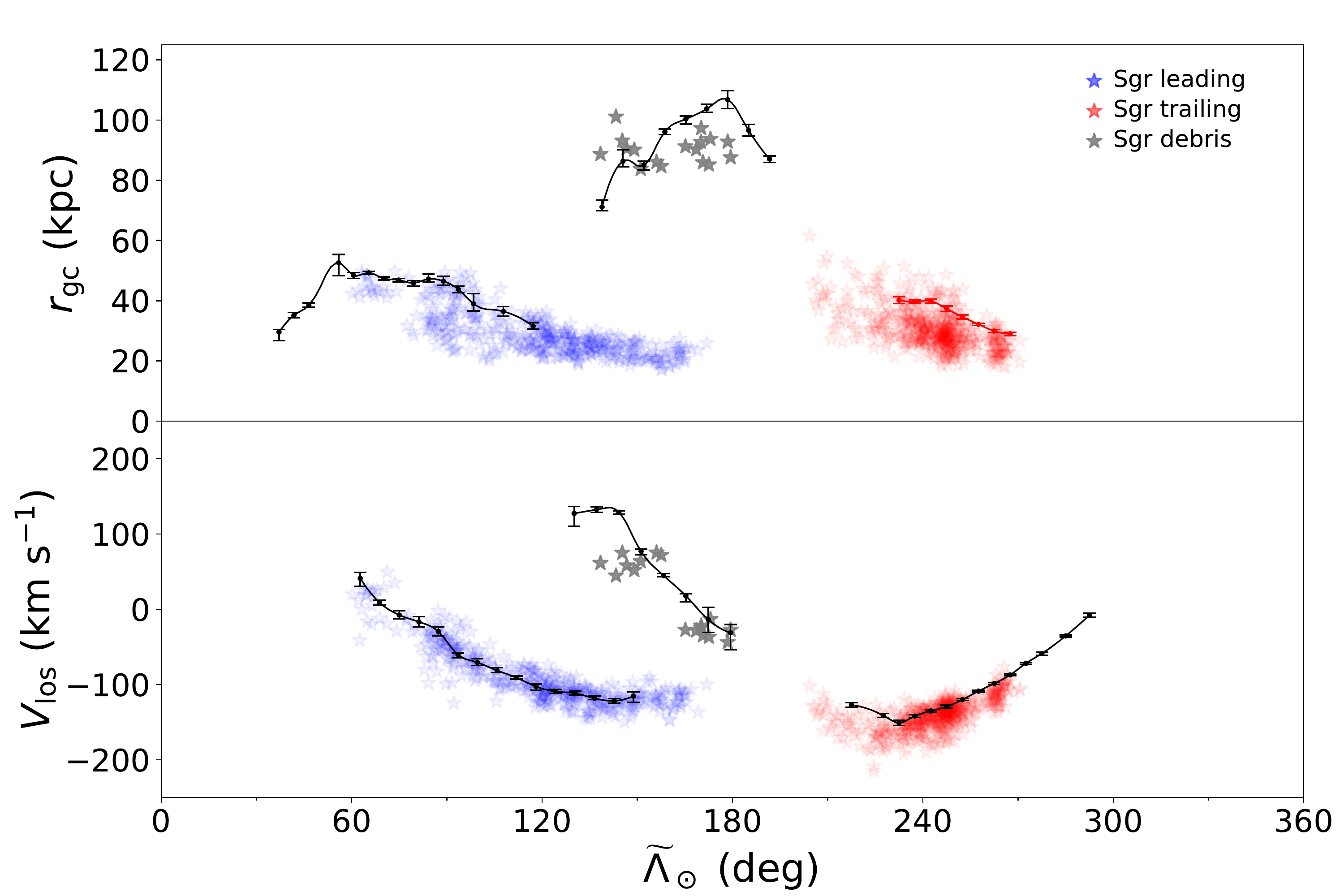}
  \end{tabular}
\caption{Comparison with literature Sgr streams traced by SDSS BHB and RC in $r_{\rm{gc}}$-$\widetilde{\Lambda}_\odot$ plane and $V_{\rm{los}}$-$\widetilde{\Lambda}_\odot$ plane. $\widetilde{\Lambda}_\odot$ is same as the definition in \citet{Belokurov14}. Blue, red, and grey star symbols represent the members of Sgr leading arm, trailing arm, and debris identified in LAMOST halo K giants. Black dots with error bars are from tables 1-5 of \citet{Belokurov14}, and the red dots with error bars are from table 2 of \citet{Koposov12}, which has been increased by 0.35 mag to correct for the reddening towards the progenitor. The Sgr streams traced by LAMOST K giants locate closer than both the BHB stars of \citet{Belokurov14} and the RC stars of \citet{Koposov12}.
} \label{sgr_BK}
\end{figure}

\citet{Koposov12} and \citet{Belokurov14} traced the Sgr streams using red clump stars (RC), blue horizontal branch stars (BHB), main-sequence turn-off stars (MSTO), and red giants (RGB) drown from SDSS. Figure \ref{sgr_BK} shows that the members associated with Sgr streams match well in line-of-sight velocity with tracks found by \citet{Belokurov14} using SDSS giant stars, but locate closer than the tracks traced by BHB stars and RC stars.

\subsection{Attributing Groups to Monoceros Ring}
Monoceros Ring is a large overdensity firstly discovered by \citet{Newberg02}, and subsequent studies have shown it is a ring-like low latitude structure and could potentially encircle the entire galaxy \citep{Yanny03, Ibata03, Rocha-Pinto03}. \citet{Yanny03} traced the structure from $l =180^\circ$ to $227^\circ$ with SDSS faint turnoff stars ($(g-r)_0=0.2$, $g_0$=19.45). They found the substructure extends 5 kpc above and below the plane of the Galaxy, and stars at southern portion is about 2 kpc farther than those of northern portion. \citet{Rocha-Pinto03} used M giants from 2MASS, and they found the structure both in the north and south hemispheres and spanned at least 100$^\circ$. \citet{Ibata03} detected the structure from colour-magnitude diagram in many lower latitude ($|b|<50^\circ$) fields of Isaac Newton Telescope Wide Field Survey. Their structure from $(V-i)_0 \sim 0.45$, $V_0\sim0.9$ curved to $(V-i)_0 \sim 1.0$, $V_0\sim21.45$ in colour-magnitude diagram, which was also seen in the SDSS Monoceros fields \citep{Newberg02}. \citet{Slater14} found the structure stretching from 100$^\circ$ to 230$^\circ$ in Galactic longitude, and covering from -30$^\circ$ to 35$^\circ$ in Galactic latitude using Pan-STARRS1 survey. \citet{LiJ12} identified the structure from SEGUE spectroscopy in northern Galactic hemisphere, and found a good match with the disrupting dwarf galaxy model by \citet{Penarrubia05}. At present, there is little consensus on the origin of the Monoceros Ring. Some studies attributed Monoceros Ring to the accretion debris from satellite \citep{Yanny03, Martin04}. While some studies argued that it may be parts of the flare or warp of the disk \citep{Momany06, Moitinho06, CC09, HL11}. Rencently, \citet{Xu15} discovered an oscillating asymmetry in the disk in the direction of the anticenter, and associated the third oscillation line with the Monoceros Ring.

\begin{figure}[htb]
\centering
  \begin{tabular}{@{\hspace{-0.5cm}}c@{}}
  \includegraphics[width=0.8\textwidth]{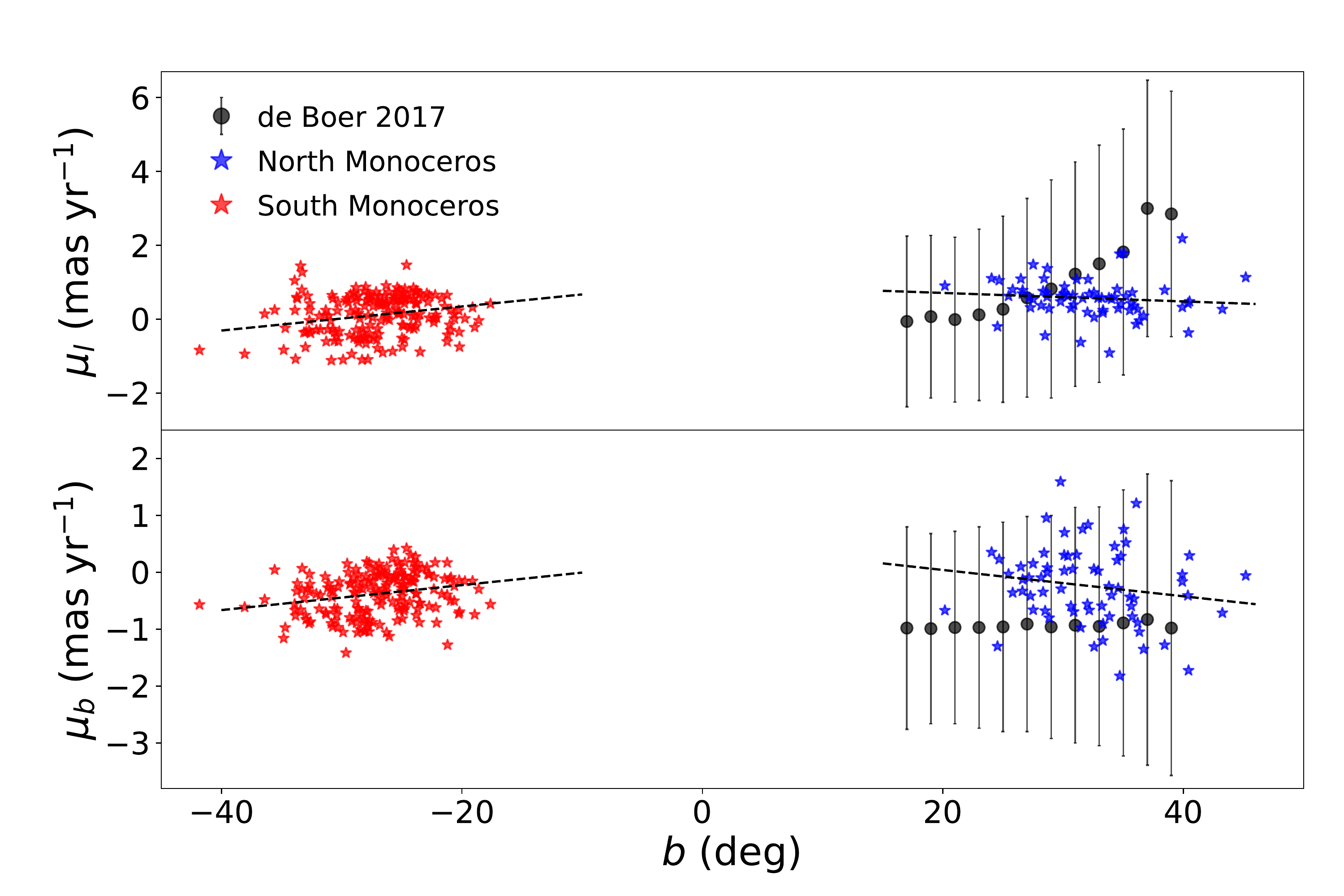}
  \end{tabular}
  \begin{tabular}{@{\hspace{1.0cm}}cccc@{}}
    \includegraphics[width=.45\textwidth]{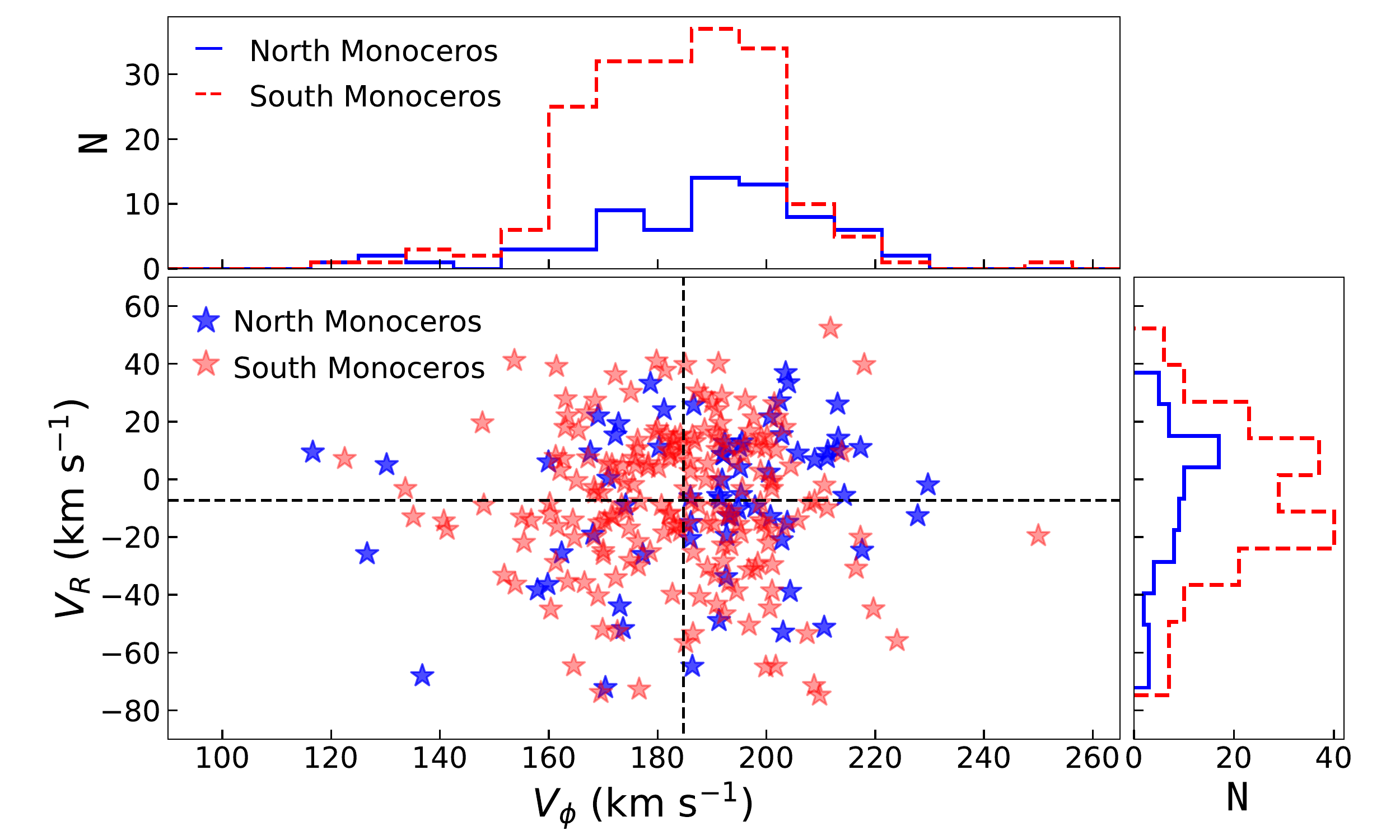} &
    \includegraphics[width=.45\textwidth]{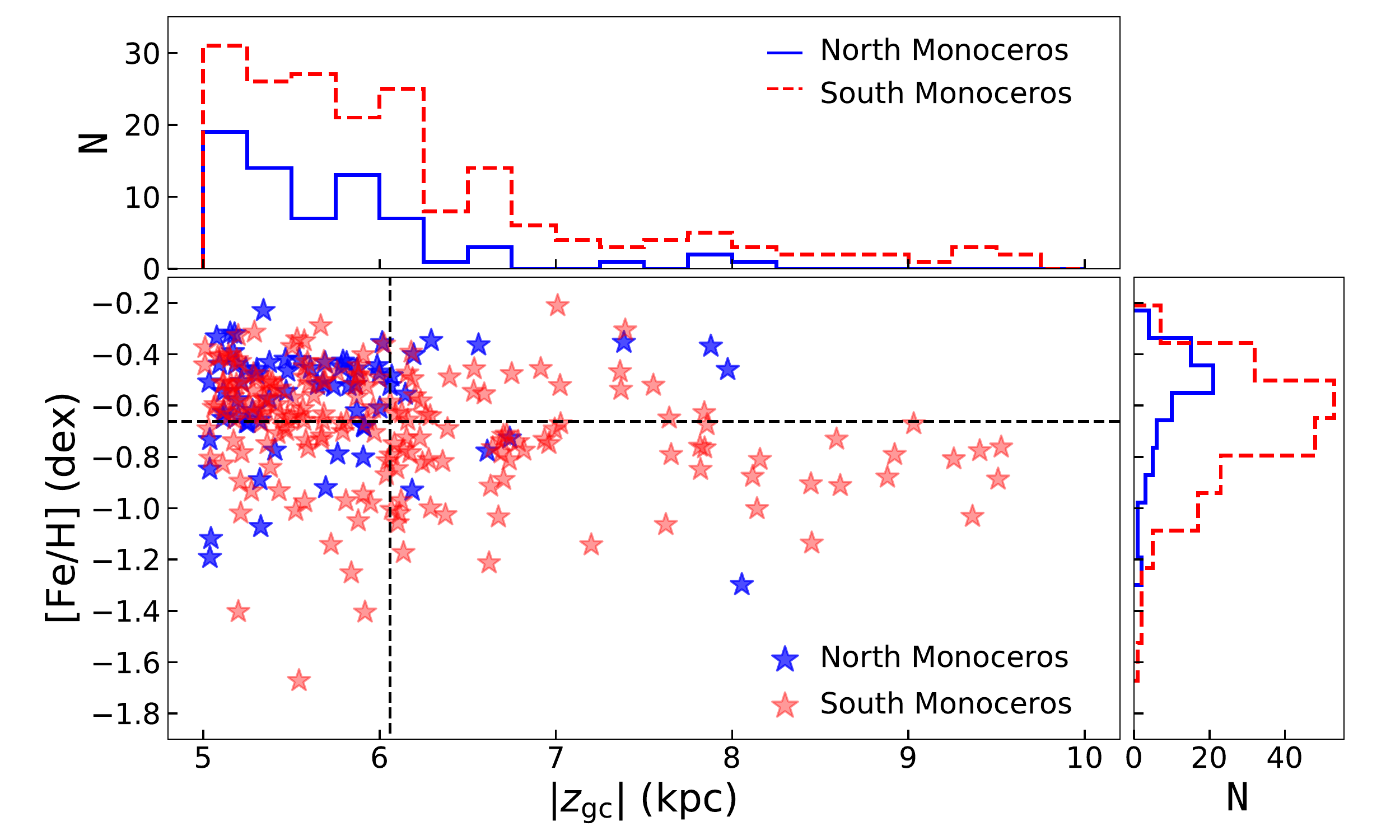} &\\
  \end{tabular}
\caption{Top two panels: the comparison of the Monoceros Ring identified in LAMOST K giants with SDSS-Gaia stars from \citet{Boer18} in $\mu_{l}$-$b$ plane and $\mu_{b}$-$b$ plane. The proper motions of the Monoceros Ring identified in LAMOST K giants (blue and red star symbols) are from Gaia DR2. Black dashed lines are obtained by linear fitting of red and blue star symbols. The mode of proper motions (black dots) and dispersions (black vertical lines) presented in \citet{Boer18} are measured from the comparison between the positions of the source in SDSS and Gaia DR1. For north Monoceros Ring, the proper motion dispersions in this work are smaller than those of \citet{Boer18}. The proper motions of Monoceros ring traced by LAMOST K giants show slight gradient along Galactic longitude. Lower left panel: the distribution of Monoceros Ring traced by LAMOST K giants in $V_{\phi}$-$V_R$ plane ($<V_{\phi}>$=185 km s$^{-1}$, $<V_R>$= -7 km s$^{-1}$). Lower right panel: the distribution of Monoceros Ring traced by LAMOST K giants in $|z_{\rm{gc}}|$-[Fe/H] plane. The mean [Fe/H] is -0.66 dex, and the dispersion of the [Fe/H] is 0.23 dex. About 85\% of the Monoceros members locate in the region of 5 kpc $<|z|<$ 7 kpc. The mean rotation velocity and metallicity may reflect the contamination from other Galactic components, the thick disk in particular.
} \label{mon_pm}
\end{figure}

\begin{figure}[htb]
\centering
  \begin{tabular}{@{\hspace{0cm}}c@{}}
  \includegraphics[width=0.80\textwidth]{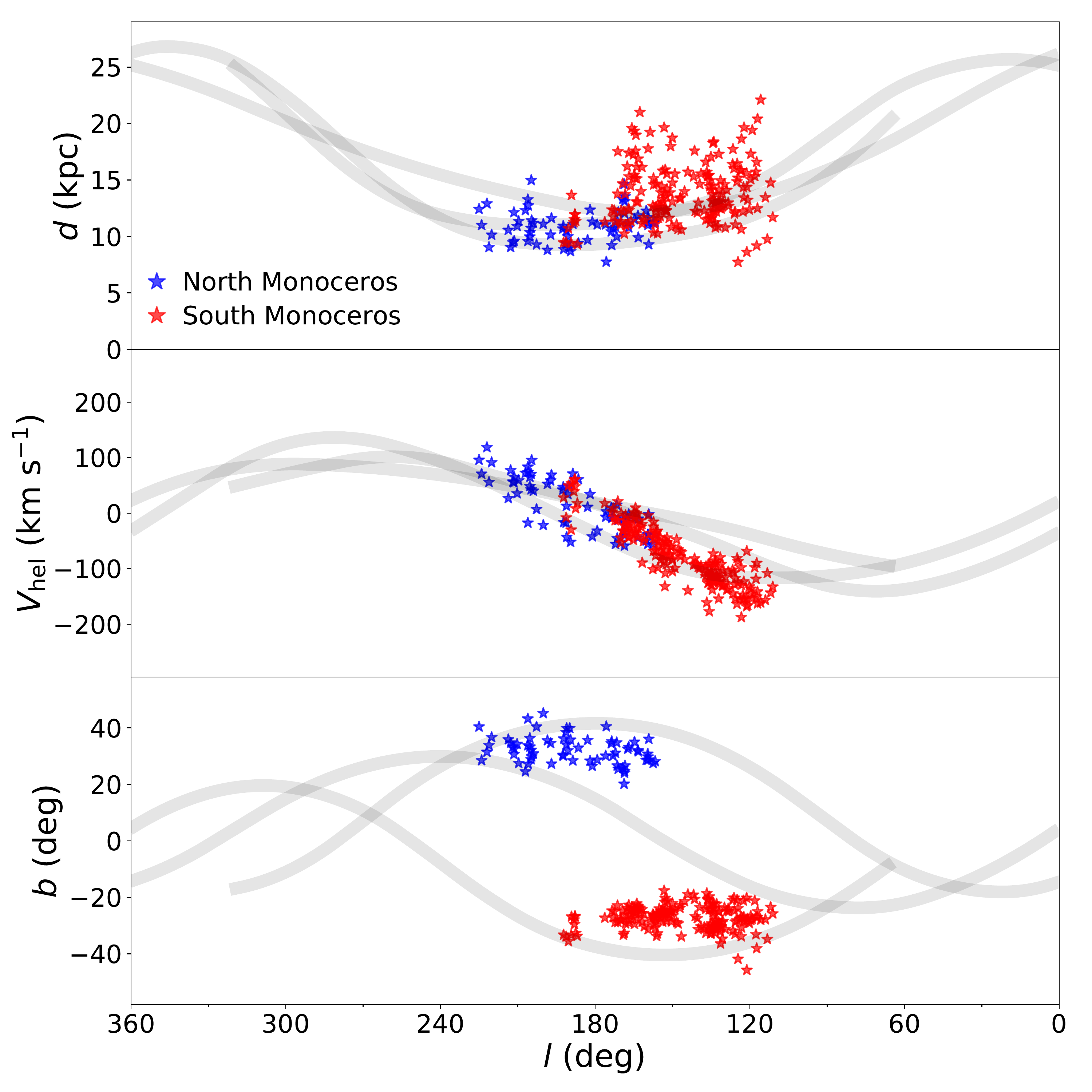}
  \end{tabular}
\caption{The comparison of the Monoceros Ring identified in LAMOST K giants (4 groups) with the simulation of \citet{Penarrubia05} in heliocentric distance $d$, heliocentric radial velocity $V_{hel}$, and sky coverage ($l,b$). The blue and red star symbols are the Monoceros Ring traced by LAMOST K giants in northern Galactic hemisphere and southern Galactic hemisphere, respectively. The grey bands represent the rough tracks of the simulation from \citet{Penarrubia05}, which indicate a model orbit of a disrupting dwarf. The Monoceros Ring in north sky traced by LAMOST K giants matches very well with the simulation. The south part of Monoceros ring identified in LAMOST K giants are consistent with simulation in $V_{hel}$, but slighly off simulation tracks in distance and sky coverage.
} \label{mon_PE}
\end{figure}

In this work, we identify spectroscopic members of Monoceros Ring in both northern and southern Galactic hemisphere for the first time. There are four groups belonging to Monoceros Ring, of which one group is in the northern Galactic hemisphere and the other three groups are in the southern Galactic hemisphere. The members of Monoceros Ring mostly locate at 5-7 kpc from Galactic disk plane and show a mean rotation of $\sim$ 185 km s$^{-1}$, and mean metallicity is -0.66 dex (see Figure \ref{mon_pm}). Comparing with \citet{Boer18}, Gaia DR2 proper motions of Monoceros Ring show smaller dispersion and slighter gradients with Galactic latitude than SDSS-Gaia-DR1 shown as Figure \ref{mon_pm}. We compare the Monoceros members with the simulation by \citet{Penarrubia05}, which modeled the Monoceros Ring as the result of a disrupted dwarf galaxy. Figure \ref{mon_PE} shows the Monoceros members are consistent with the model in the northern Galactic hemisphere, but more distant and lower in Galactic latitude than the model in the southern Galactic hemisphere. The values of mean rotation velocity and metallicity for the Ring may reflect the contamination from other Galactic components, the thick disk in particular. Thus, the data-model inconsistency is likely caused by different origin mechanisms, or by contamination from other Galactic components.

\subsection{Attributing Groups to Virgo Overdensity}
\begin{figure}[htb]
%\centering
  \begin{tabular}{@{}cccc@{}}
    \includegraphics[width=.45\textwidth]{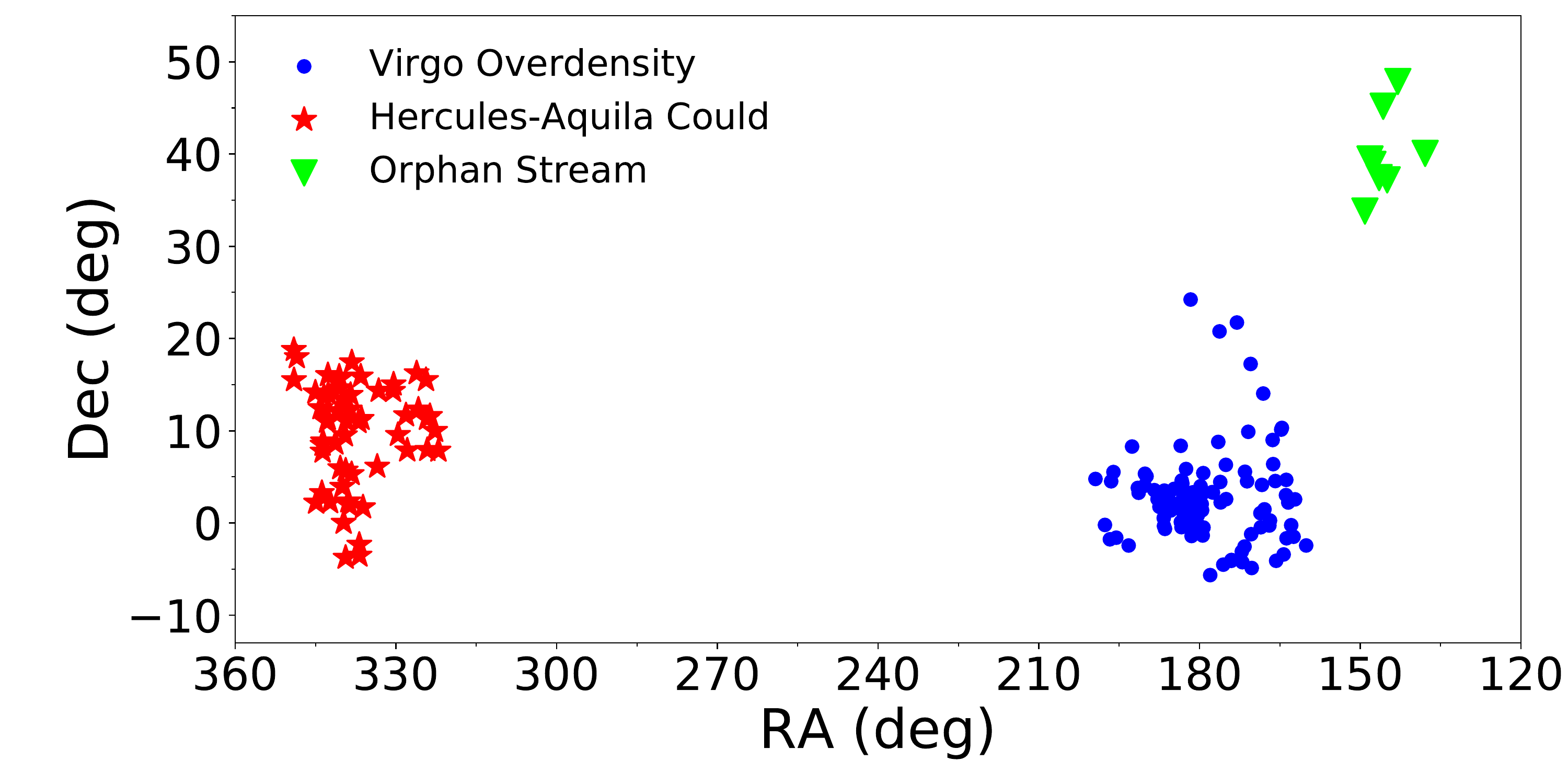} &
    \includegraphics[width=.45\textwidth]{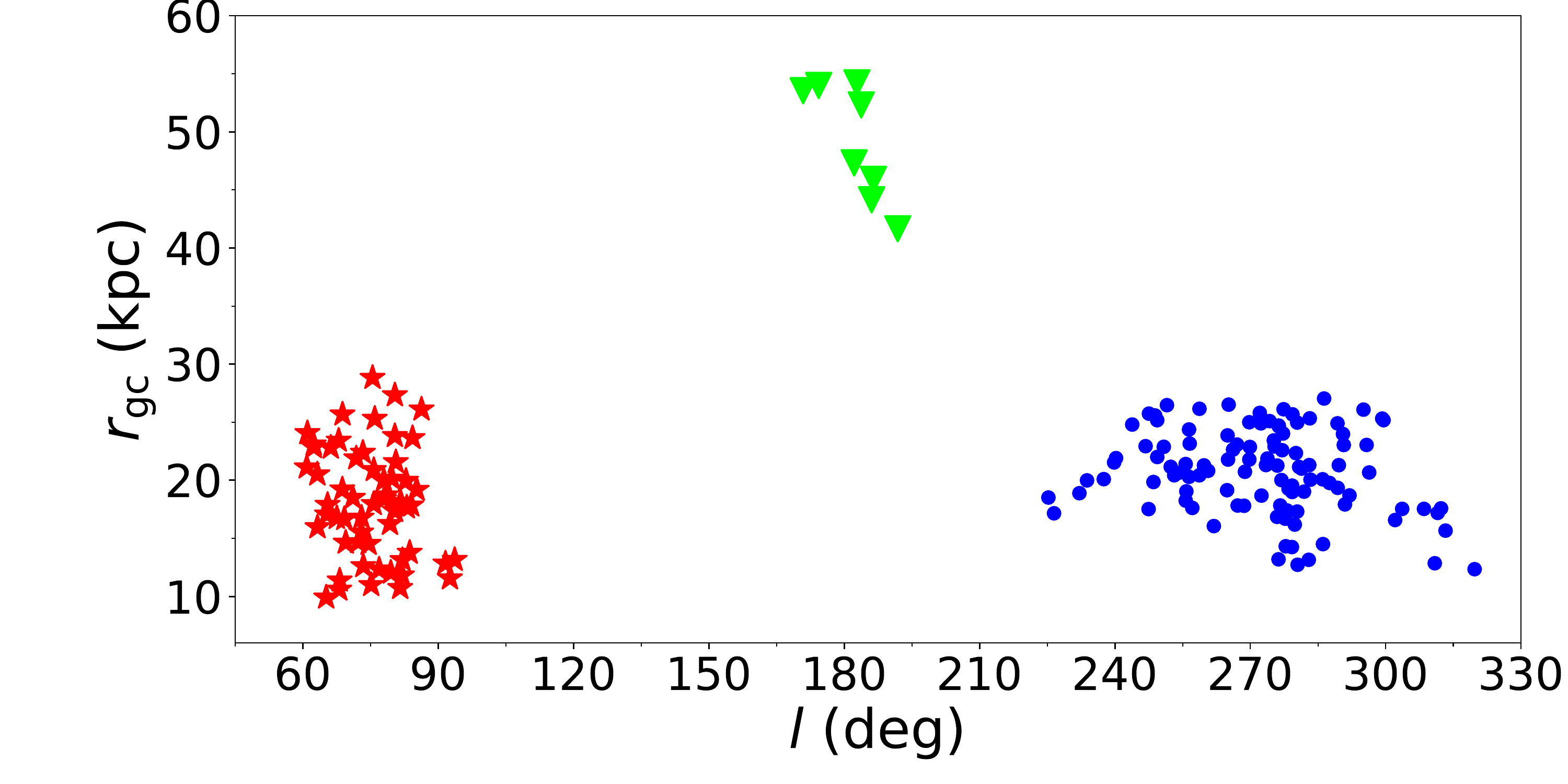} &\\
    \includegraphics[width=.45\textwidth]{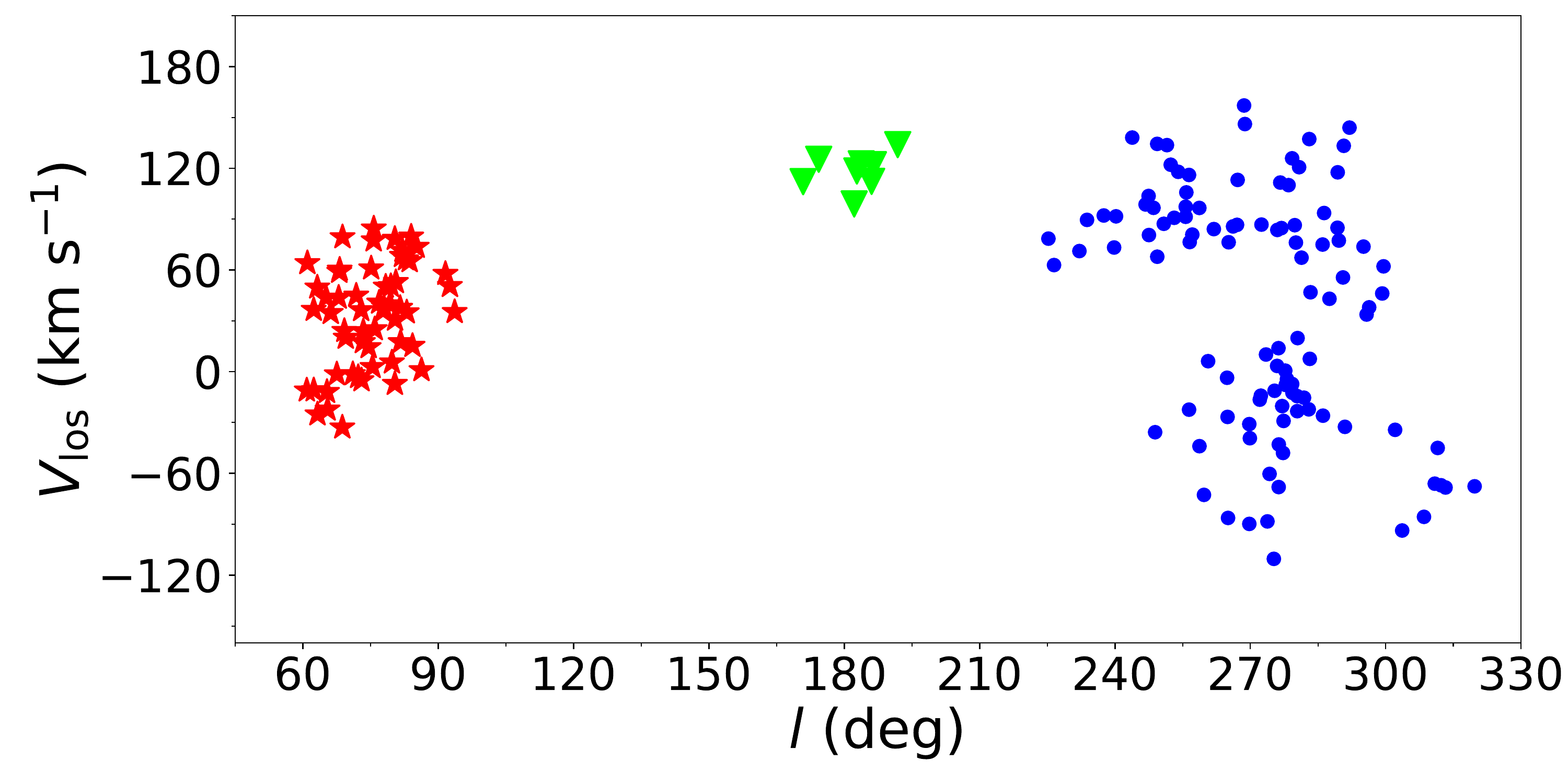} &
    \includegraphics[width=.45\textwidth]{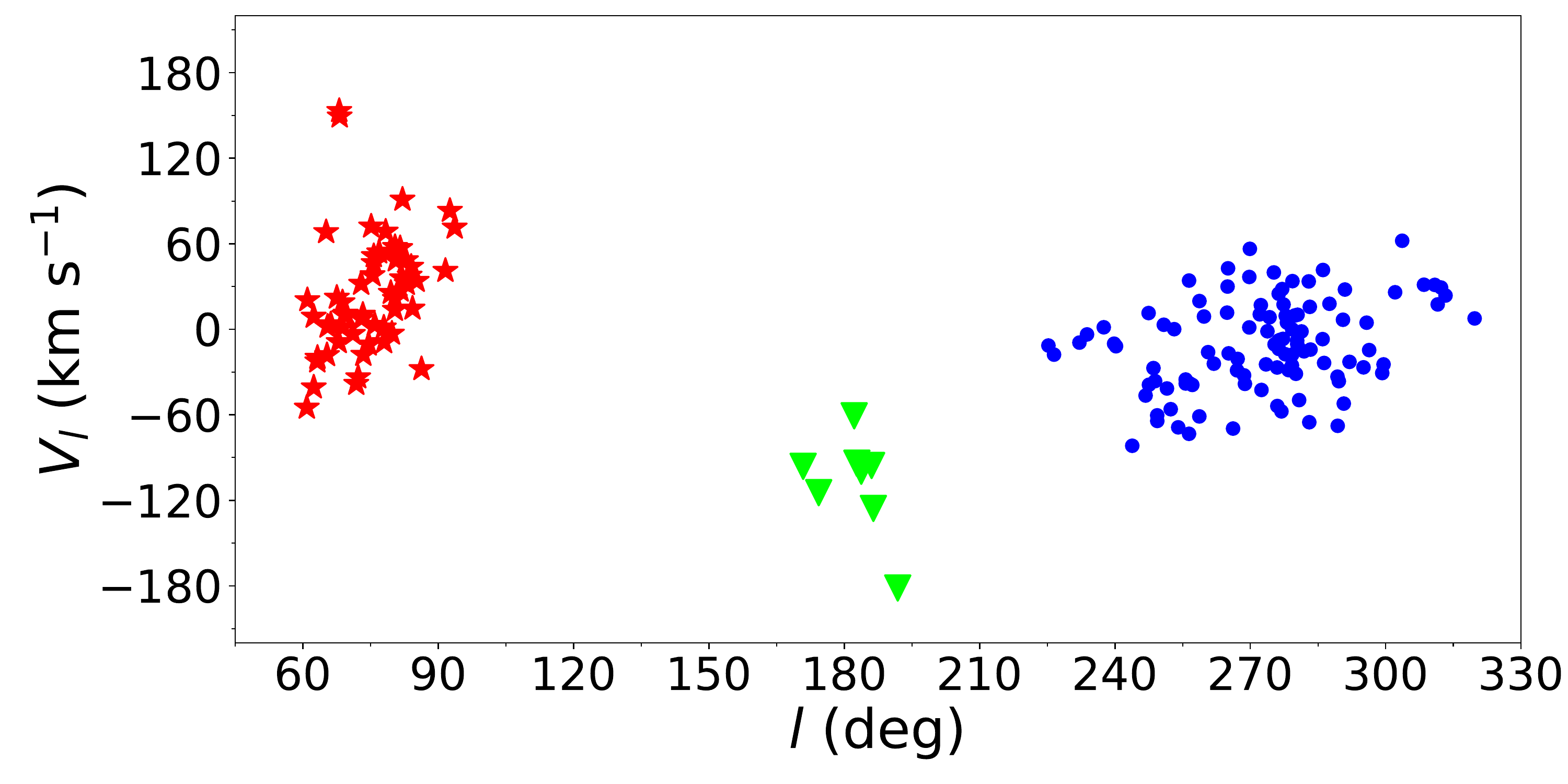} &\\
    \includegraphics[width=.45\textwidth]{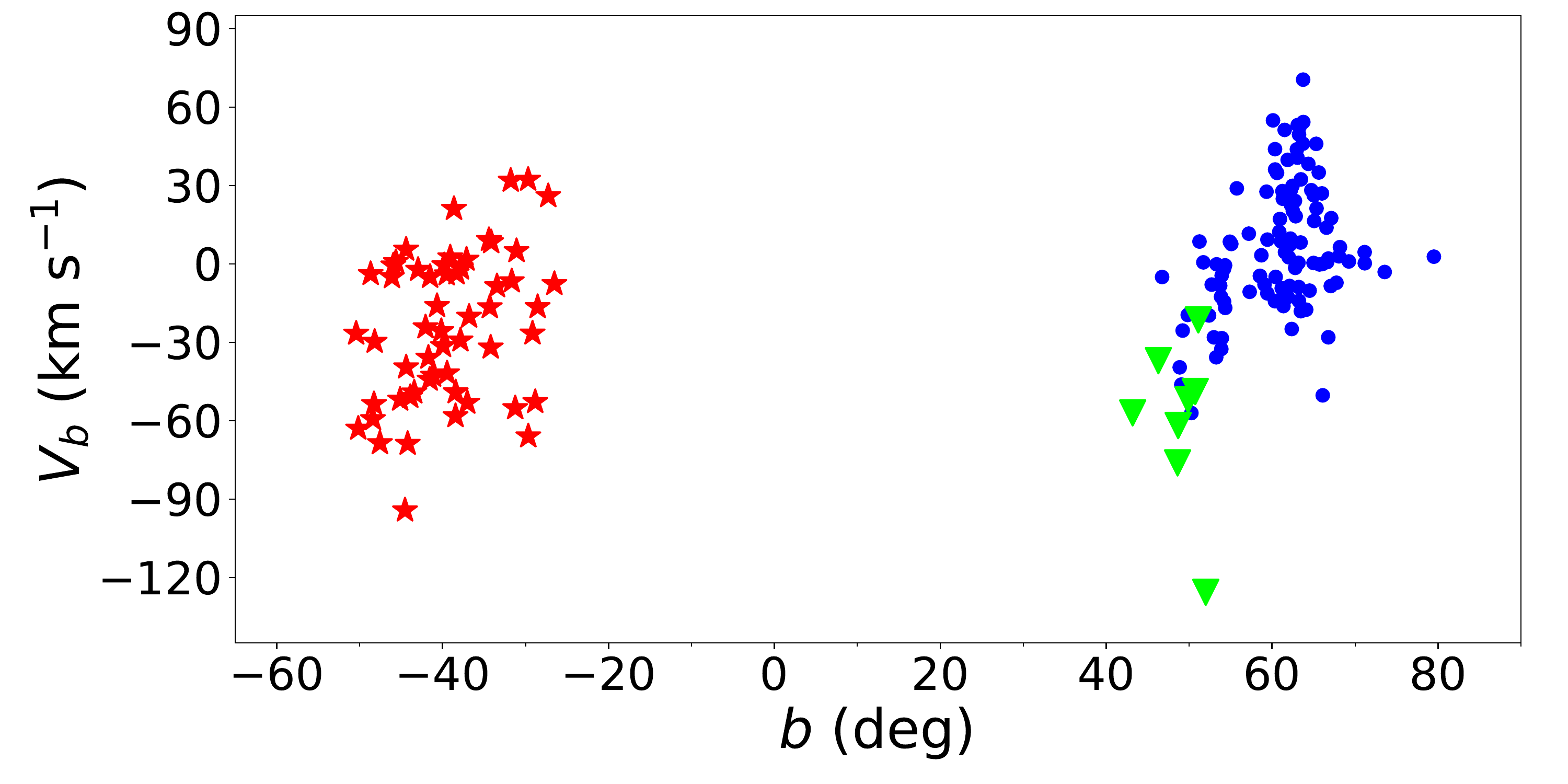} &
    \includegraphics[width=.45\textwidth]{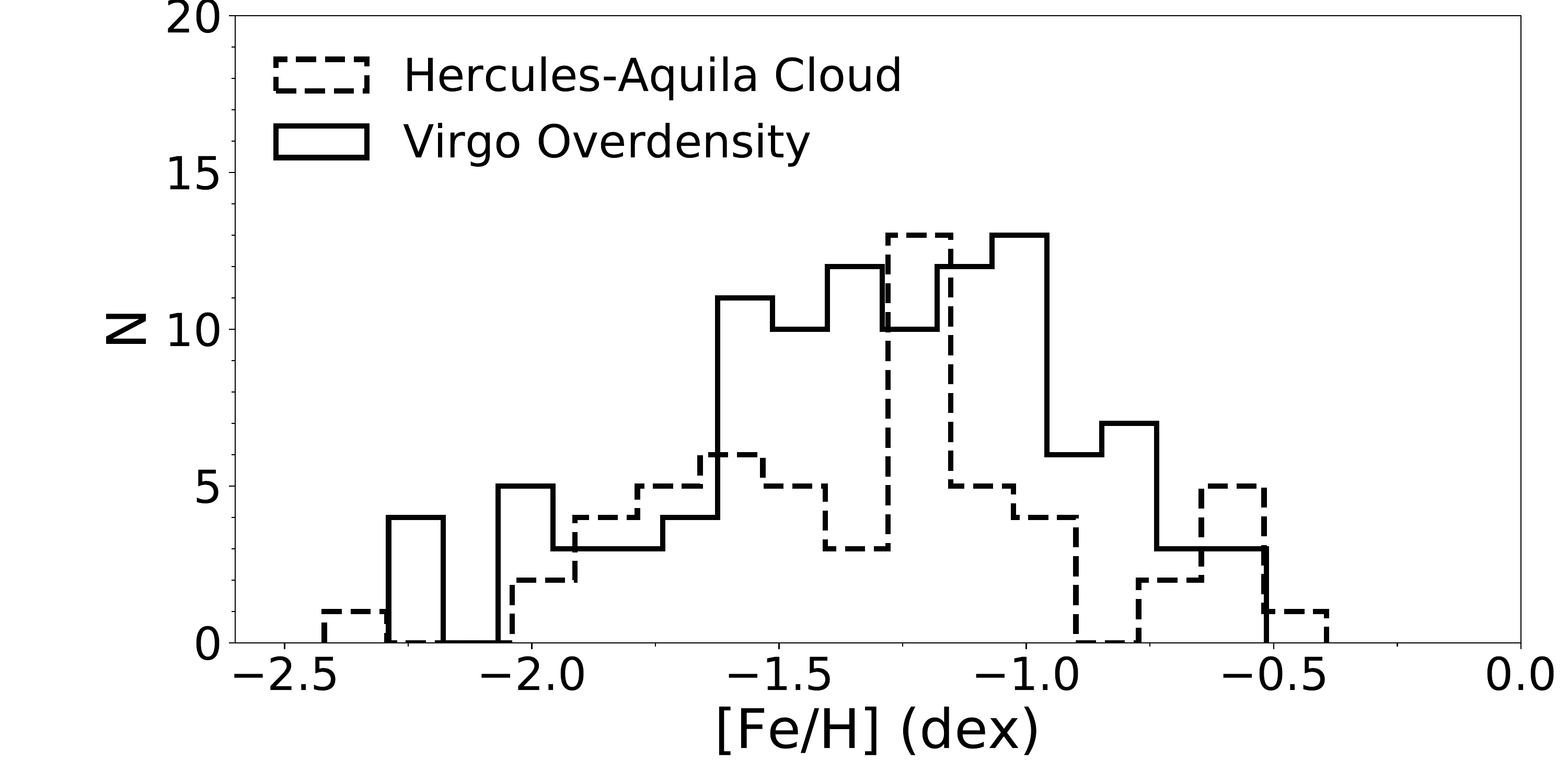} &\\
  \end{tabular}
\caption{The position and velocity distributions of Virgo Overdensity (blue dots), Hercules-Aquila Cloud (red star symbols), and Orphan Stream (green triangles) identified in LAMOST K giants, and the metallicity distributions of the Virgo Overdensity, the Hercules-Aquila Cloud.  There are 4 FoF groups relate to Virgo Overdensity, 3 FoF groups relate to Hercules-Aquila Cloud, and 1 FoF group relate to Orphan Stream. The mean [Fe/H] values of Hercules-Aquila Cloud and Virgo Overdensity are both -1.31 dex, and their dispersion are 0.42 dex and 0.40 dex, and the metallicity distribution of Hercules-Aquila Cloud shows three peaks at -1.6 dex, -1.2 dex, and -0.6 dex.
} \label{other_known}
\end{figure}

The substructure in Virgo constellation is very complex, and its nature is still uncertain. Because of a much higher stellar density exhibited by turnoff stars from SDSS, this region has become known as the ``Virgo Overdensity" \citep{Newberg02}. Virgo Overdensity is located 10$\sim$20 kpc away from the Sun, over 1000 deg$^2$ \citep{Newberg07, Juric08, Bonaca12, Duffau14}. We find four groups located in Virgo Overdensity. As shown in Figure \ref{other_known}, the Virgo Overdensity members has a mean metallicity of -1.31 dex, and heliocentric distance from 13 to 25 kpc.%, and mean line-of-sight velocity $V_{\rm{los}}=$ 36.4 km s$^{-1}$.

\subsection{Attributing Groups to Hercules-Aquila Cloud}
Hercules-Aquila Cloud was found as an overdensity using MSTO in SDSS DR5 by \citet{Belokurov07}. They suggested that this cloud covers a huge area of sky, centered on Galactic longitude $l \sim 40^\circ$, Galactic latitude $b$ from $- 50^\circ$ to $+ 50^\circ$, and line-of-sight velocity $V_{\rm{los}} \sim 180\ \rm{km\ s^{-1}}$. Subsequently, \citet{Sesarl10} used RR Lyrae from SDSS found the Hercules-Aquila Cloud contained at least 1.6 times stellar density of the halo at heliocentric distance of 15 to 25 kpc, and its mean metallicity is similar to Galactic halo. \citet{Watkins09} found the heliocentric distance of RR Lyrae stars in Hercules-Aquila Cloud is 21.9$\pm$12.1 kpc, and metallicity is -1.43$\pm$0.36 dex. Their study additionally presented a estimate of velocity of Hercules-Aquila Cloud. In their bottom panel of Figure 17, the mean velocity of MSTO stars in Hercules-Aquila Cloud is centered around $V_{\rm{los}}=$ 25km s$^{-1}$. \citet{Simion14} mapped the Hercules-Aquila Cloud using RR Lyrae from the Catalina Sky Survey \citep{Drake14}. They found this substructure is more prominent in the southern Galactic hemisphere than in the north, peaking at a heliocentric distance of 18 kpc. We find three groups are asscoiated with Hercules-Aquila Could in the halo K giants. As shown in Figure \ref{other_known}, the Hercules-Aquila Cloud members has a mean metallicity of -1.31 dex, heliocentric distance from 12 to 21 kpc, and mean line-of-sight velocity $V_{\rm{los}}=$ 33.37 km s$^{-1}$.

\subsection{Attributing Groups to Orphan Stream}
Orphan stream is a roughly $1 \sim 2^\circ$ wide stellar stream, and its progenitor has not been identified yet. It runs from ($165^\circ,-17^\circ$) to ($143^\circ,48^\circ$) in equatorial coordinate \citep{Grillmair06, Belokurov07}. In addition, it was also traced with RR Lyrae stars by \citet{Sesarl13}. They found that the most distant parts of Orphan stream is $40 \sim 50$ kpc from the Sun, mean [Fe/H] value of -2.1 dex, and line-of-sight velocity $V_{\rm{los}} \sim 100$ km s$^{-1}$. We find a group matches well with all these conditions. As shown in Figure \ref{other_known}, its mean [Fe/H] is -2.02 dex, and mean line-of-sight velocity $V_{\rm{los}}=$ 118 km s$^{-1}$.

\subsection{Unknown Groups}
\begin{figure}[htb]
%\centering
  \begin{tabular}{@{}cccc@{}}
    \includegraphics[width=.45\textwidth]{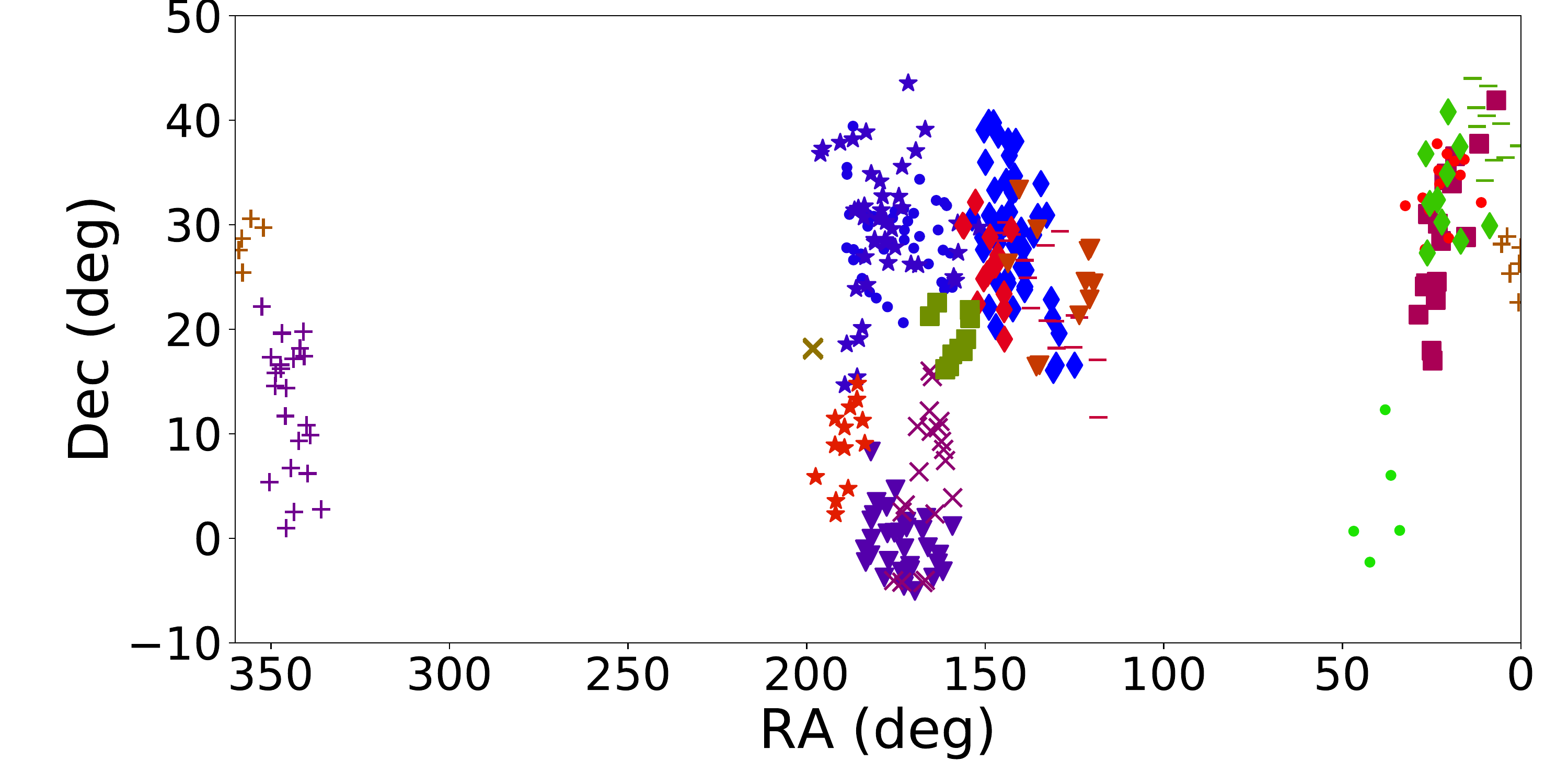} &
    \includegraphics[width=.45\textwidth]{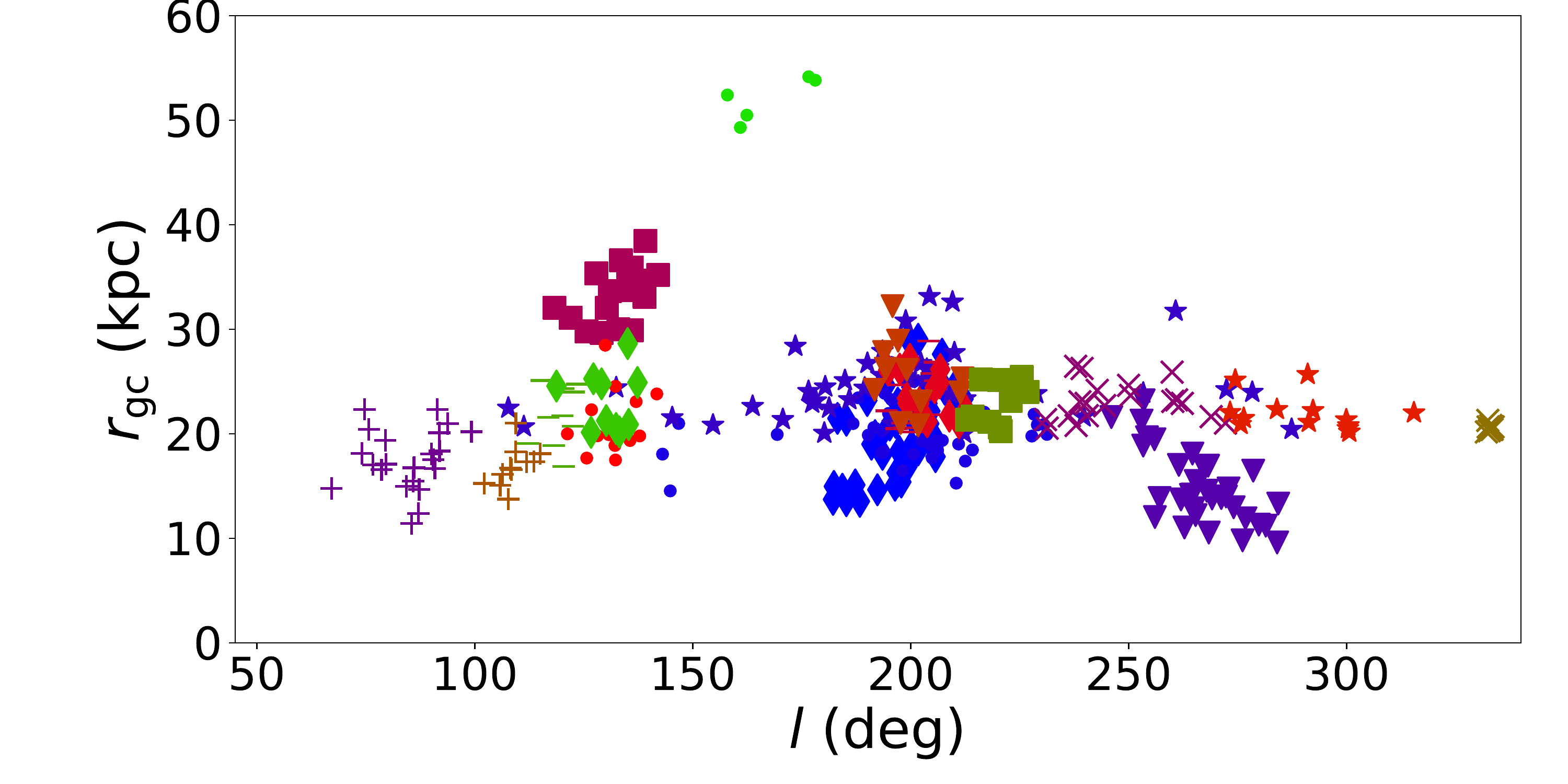} &\\
    \includegraphics[width=.45\textwidth]{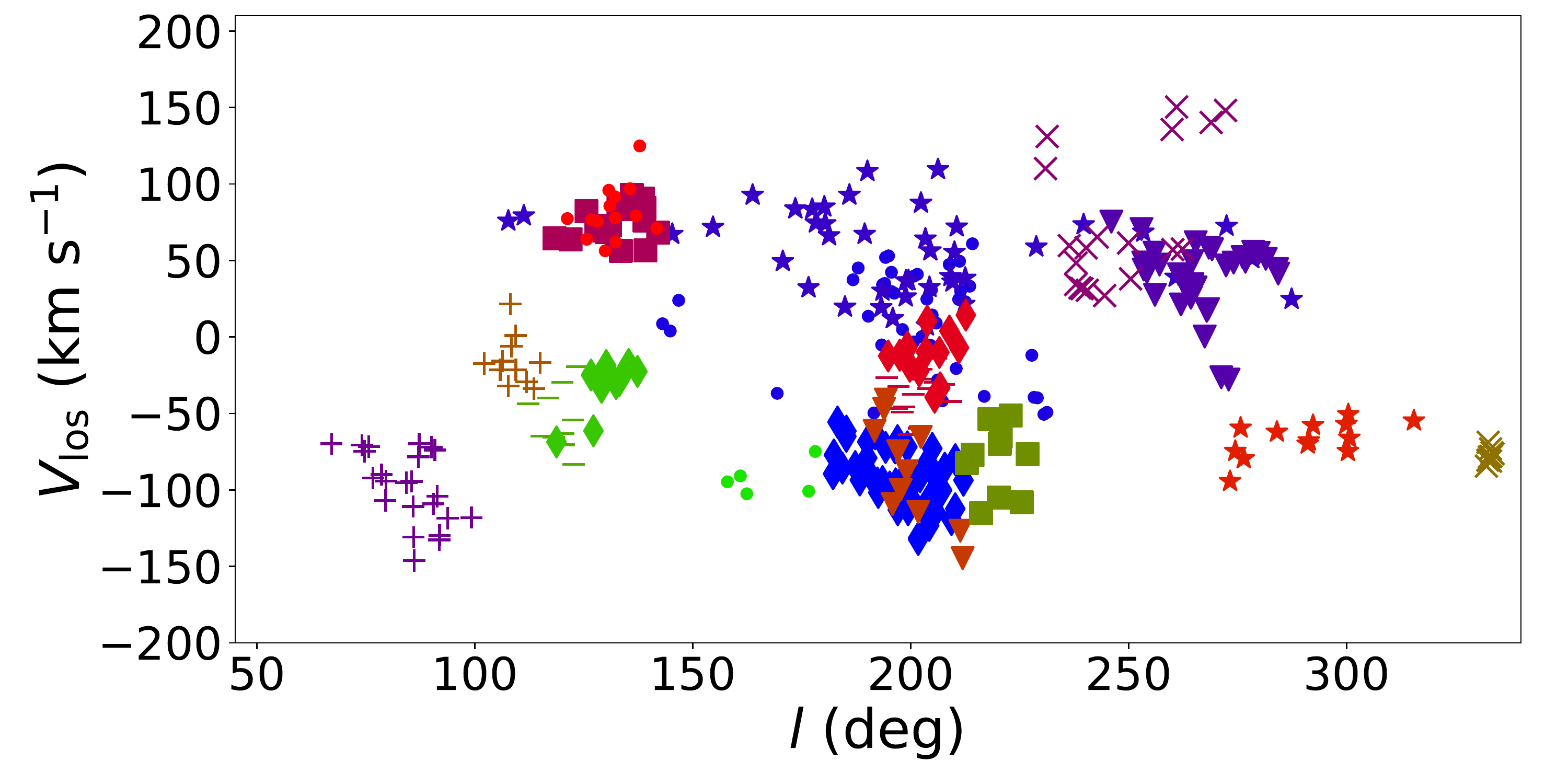} &
    \includegraphics[width=.45\textwidth]{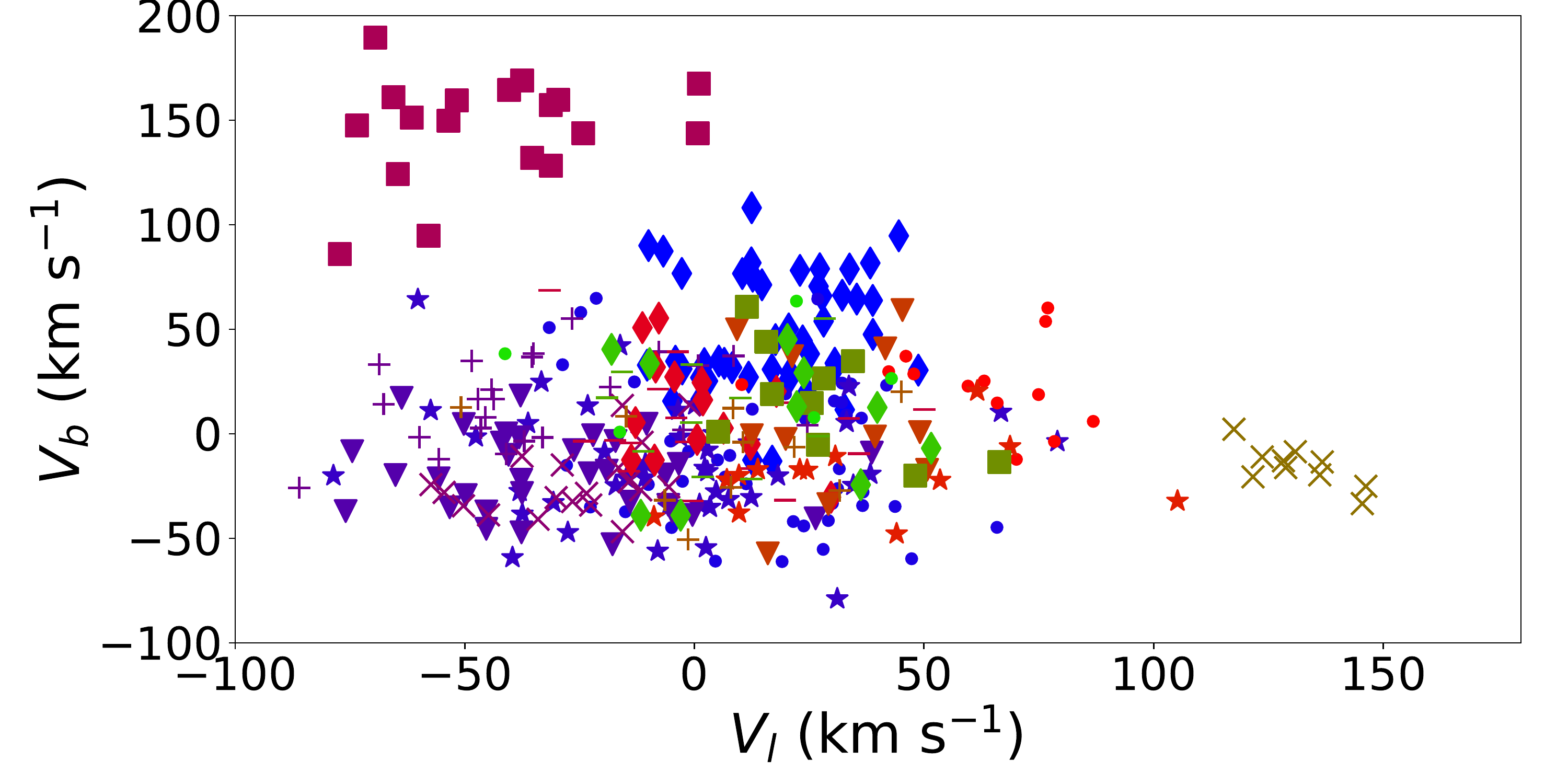} &\\
    \includegraphics[width=.45\textwidth]{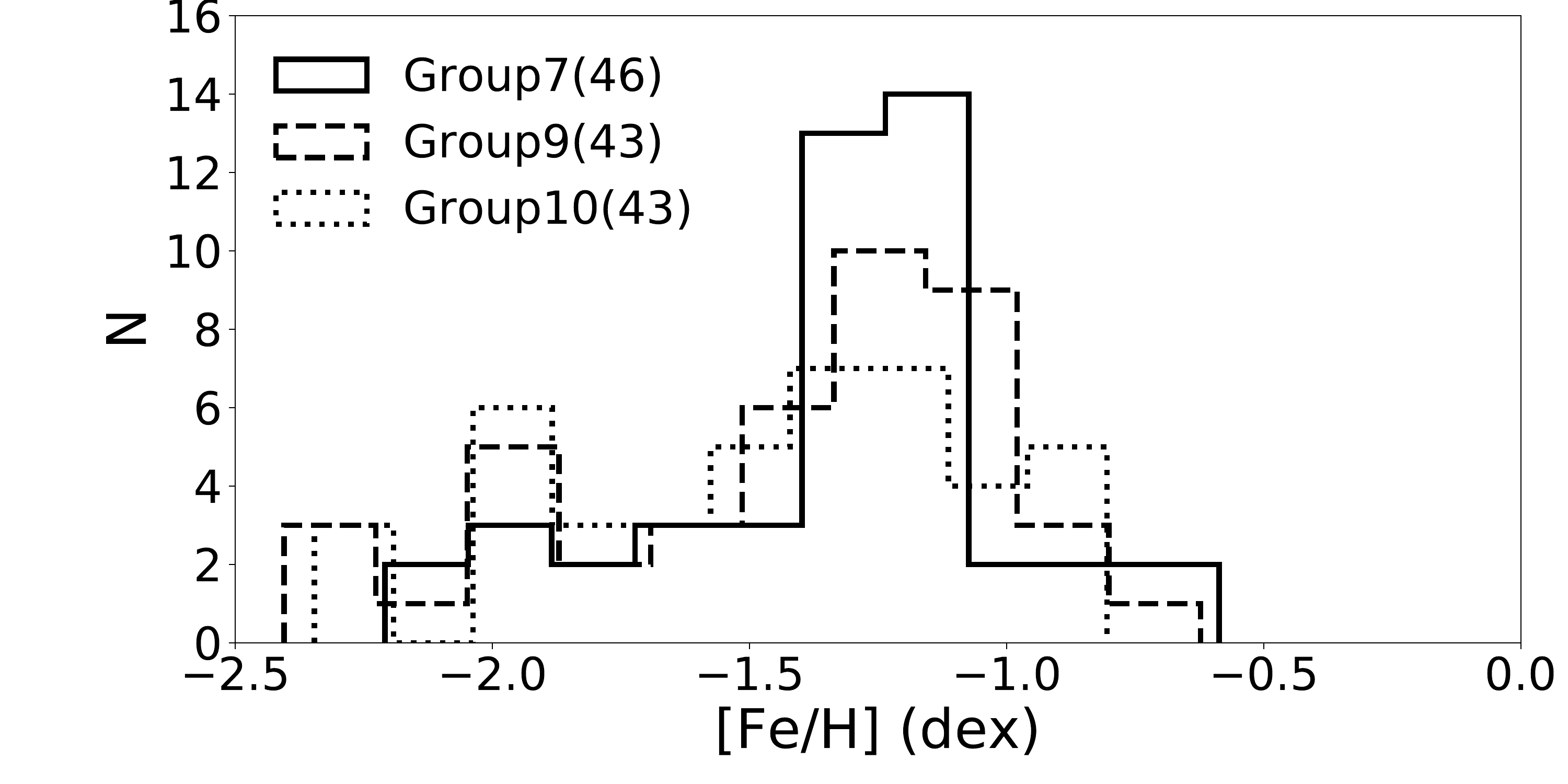} &
    \includegraphics[width=.45\textwidth]{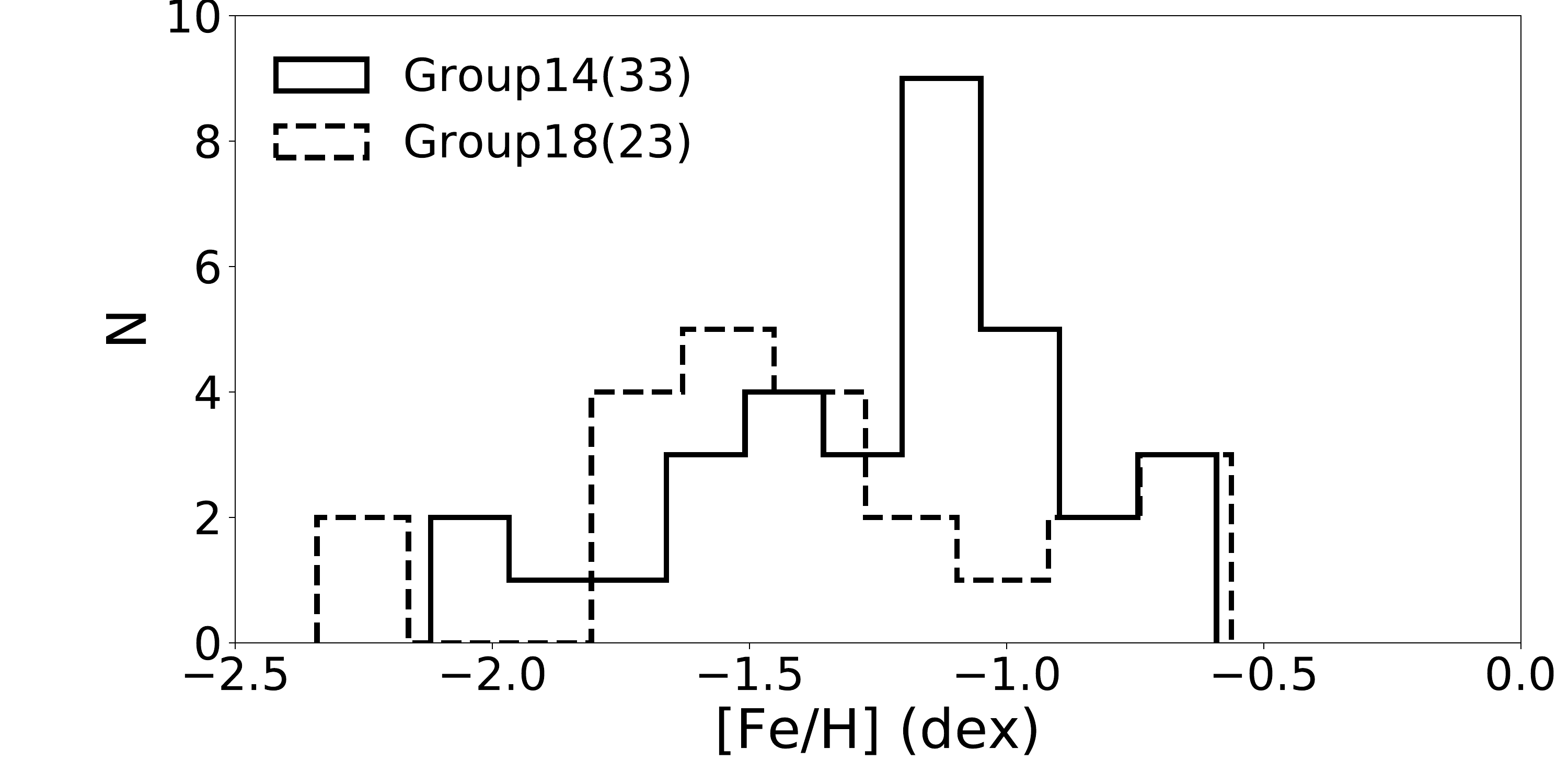} &
  \end{tabular}
\caption{The position and velocity distributions of 18 unknown groups (350 group members), and metallicity distributions of the groups with more than 20 members. The detail properties of these unknown groups are listed in Table \ref{t_unknown}.
} \label{unknown}
\end{figure}

Besides the groups that can be attributed to known streams, there are 18 remaining groups likely relate to unknown substructure. The unknown groups and their velocity-position distribution are shown in Figure \ref{unknown} and Table \ref{t_unknown}. In addition, we plot the [Fe/H] distribution for groups with more than 20 members in the last two panels of Figure \ref{unknown}. The full catalog of K giants with more than 5 members are published online, and a sample is shown in Table \ref{t_catalog}.

\section{Summary}\label{summary}
The stellar halo of our Milky Way are expected to be comprised largely of debris from disrupted satellite galaxies. The debris may appear as coherent streams for some time, but will phase-mix until they become difficult to recognize. Several prominent substructures in Galactic stellar halo have been found in the published literature, such as the famous Sgr streams. Some studies have attempted to quantify the position-velocity substructure of the stellar halo using K giants and BHB stars \citep{Starkenburg09,Xue11,Cooper11,Janesh16}. \citet{Janesh16} even tried to identify members of substructures using SEGUE K giants. However, all previous studies used only 3D positions and 1D radial velocities because of the lack of proper motions at that time. Now, Gaia DR2 can provide useful proper motions for distant halo stars. LAMOST combining with Gaia enables to construct a large sample of $\mathrm{13,554}$ halo K giants with distances up to 100 kpc, radial velocities, metallicities, and proper motions. This paper presents the first attempt to quantify and identify the substructure of the Milky Way’s stellar halo in 6D phase-space.

Based on 4Distance used in previous studies \citep{Starkenburg09,Xue11,Cooper11,Janesh16}, we develop 6Distance to define the distance of two stars in phase-space. By comparing the number of close pairs between observed sample and the diffuse halo constructed by shuffling distances and velocities of the observed sample, we can quantify the amount of substructure in the sample. We find that the substructure increases from inner halo to outer halo, and from metal-poor population to metal-rich population, in agreement with the results of \citet{Xue11,Janesh16}.

Besides quantifying substructures in stellar halo, identifying members of substructure is of particular importance to explore their the origin. We combine 6Distance with FoF algorithm, and manually assign a sky-distance-dependent linking length to identify the substructures. Finally, we find 43 FoF groups (1867 group members), in which 25 groups belong to 5 known substructures: Sgr stream (13 groups), Monoceros Ring (4 groups), Virgo Overdenstiy (4 groups), Hercules-Aquila Cloud (3 groups), and Orphan Stream (1 group); and 18 remaining groups are likely related to unknown substructures. It is worth to point out that for the first time we identify the spectroscopic members of Monoceros Ring both in northern and southern hemispheres, which demonstrates the advantage of LAMOST.

The members of Sgr streams locate in distant halo, and are more metal-rich than other halo stars, we conclude that the Sgr stream dominates both trends of substructure versus metallicity and distance. In addition, we analyze the kinematics and metellicities of the Monoceros Ring, Hercules-Aqulia Cloud, Virgo Overdensity, and the unknown groups with more than 20 grouped members. Monoceros Ring shows more metal-rich than the typical halo stars, and its mean rotation velocity are closer to the thick disk, which may reflect the contamination from other Galactic components, the thick disk in particular.

\acknowledgments
This study is supported by the National Natural Science Foundation of China undergrants (NSFC) Nos. 11873052, 11890694, 11390371/2, 11573032 and 11773033. X.-X.X. thanks the ``Recruitment Program of Global Youth Experts" of China. J.L. acknowledges the NSFC under grants 11703019. L.Z. acknowledges supports from NSFC grants 11703038. This project was developed in part at the 2018 Gaia-LAMOST Sprint workshop supported by the NSFC under grants 11333003 and 11390372.

Guoshoujing Telescope (the Large Sky Area Multi-Object Fiber Spectroscopic Telescope, LAMOST) is a National Major Scientific Project built by the Chinese Academy of Sciences. Funding for the project has been provided by the National Development and Reform Commission.

This work has made use of data from the European Space Agency (ESA) mission {\it Gaia} (\url{https://www.cosmos.esa.int/gaia}), processed by the {\it Gaia} Data Processing and Analysis Consortium (DPAC, \url{https://www.cosmos.esa.int/web/gaia/dpac/consortium}). Funding for the DPAC has been provided by national institutions, in particular the institutions participating in the {\it Gaia} Multilateral Agreement.

\newpage
\bibliographystyle{aasjournal}
\bibliography{Bibtex}
\clearpage

\newpage
\vspace*{\fill}
\vfill
\clearpage

\newpage
\vspace*{\fill}
\vfill
\clearpage

\newpage
\vspace*{\fill}
\vspace*{\fill}
\clearpage

\newpage
\vspace*{\fill}
\vspace*{\fill}
\clearpage

\newpage
\vspace*{\fill}
\vspace*{\fill}
\clearpage

\newpage
\vspace*{\fill}
\clearpage

\newpage
\vspace*{\fill}
\vspace*{\fill}
\clearpage

\newpage
\vspace*{\fill}
\vspace*{\fill}
\clearpage

\newpage
\vspace*{\fill}
s
\vspace*{\fill}
\clearpage

\newpage
\vspace*{\fill}
\vspace*{\fill}
\clearpage

\newpage
\vspace*{\fill}
\vspace*{\fill}
\clearpage

\newpage
\vspace*{\fill}
\vspace*{\fill}
%\clearpage

%\newpage
\vspace*{\fill}
\vfill
\clearpage

\newpage
\clearpage

\newpage
\clearpage

\newpage
\vspace*{\fill}
\vspace*{\fill}
\clearpage

\newpage
\vspace*{\fill}
\vspace*{\fill}
\clearpage

%%%%%%%%%%%%%%%%%%%%%%%%%%%%%%%%%%%%%%%%%%Table%%%%%%%%%%%%%%%%%%%%%%%%%%%%%%%%%%%%%%%%

\newpage
\vspace*{\fill}
\begin{table}[h]
\centering
\caption{Properties of Each sub-sample} \label{parts}
\scriptsize%fontsize
\begin{tabular}{ccccccc}
\tablewidth{0pt}
\hline
\hline
%\\
\multicolumn{2}{c}{Range} & \colhead{Size\textsuperscript{a}} & \colhead{Linking Length} & \colhead{Minimum Group Size \textsuperscript{b}} & \colhead{Number of Groups \textsuperscript{c}} & \colhead{Group Members \textsuperscript{d}}
\\
\multicolumn{2}{c}{(kpc)}
\\
\hline
\multicolumn{2}{c}{$z >$ 5}            & 10304 &  0.045   &  20   &    8  &   543\\
{$z >$ 5}         & $ r_{\rm{gc}} >20$ & 3705  &  0.045   &  10   &   14  &   629\\
{$z >$ 5}         & $ r_{\rm{gc}} >40$ & 497   &  0.060   &   5   &    3  &    70\\
{$z >$ 5}         & $ r_{\rm{gc}} >60$ & 113   &  0.080   &   5   &    2  &    17\\
\hline
\multicolumn{2}{c}{$z<$-5}             & 3250  &  0.045   &  10   &    9  &   672\\
{$z<$ -5}         & $ r_{\rm{gc}} >20$ & 1879  &  0.050   &  10   &    5  &   680\\
{$z<$ -5}         & $ r_{\rm{gc}} >40$ & 205   &  0.080   &   5   &    2  &    66\\
\hline
%\multicolumn{4}{l}{\scriptsize{NOTE. -}}\\
\multicolumn{4}{l}{\textsuperscript{a}\scriptsize{Number of halo K giants.}}\\
\multicolumn{4}{l}{\textsuperscript{b}\scriptsize{Minimum group size.}}\\
\multicolumn{4}{l}{\textsuperscript{c}\scriptsize{Number of groups.}}\\
\multicolumn{4}{l}{\textsuperscript{d}\scriptsize{Total group members.}}
\end{tabular}
\end{table}
\vspace*{\fill}
\clearpage

\newpage
\vspace*{\fill}
\begin{table}[h]
\centering
\caption{Maximum Physical Component Size for Different Linking Length} \label{lk}
\scriptsize%fontsize
\begin{tabular}{cccc}
%\tablewidth{5pt}
\hline
\hline
%\\
\colhead{Linking length} & \colhead{$\theta(l,b)$ (deg)} & \colhead{$\Delta d$ (kpc)} & \colhead{$\Delta V_{*}$\textsuperscript{a} (km s$^{-1}$)}
\\
\hline
0.045   & $  8.1 $ & 4.45 &  22.45\\
0.050   & $  9.0 $ & 5.0 &  25.0\\
0.060   & $ 10.8 $ & 6.0 &  30.0\\
0.080   & $ 14.4 $ & 8.0 &  40.0\\
\hline
%\muflticolumn{4}{l}{\scriptsize{NOTE. -}}\\
\multicolumn{4}{l}{\textsuperscript{a}{$V_*$ stands for $V_l, V_b$, or $V_{\rm{los}}$.}}
\end{tabular}
\end{table}
\vspace*{\fill}

\begin{sidewaystable}[h]
\centering
\caption{Known Substructures in LAMOST Halo K Giants} \label{t_known}
\scriptsize %fontsize
%\tiny %fontsize
\begin{tabular}{cccccccccc}
\tablewidth{0pt}
\hline
\hline
\colhead{Substructure} &
\colhead{R.A.[min,max]\tablenotemark{a}} &
\colhead{Decl.[min,max]\tablenotemark{a}} &
\colhead{$d$\tablenotemark{b}[min,max]\tablenotemark{a}} &
\colhead{$hrv$\tablenotemark{c}[min,max]\tablenotemark{a}} &
\colhead{pmra[min,max]\tablenotemark{a}} &
\colhead{pmdec[min,max]\tablenotemark{a}} &
\colhead{[Fe/H][min,max]\tablenotemark{a}}
\\
\colhead{} &
\colhead{(deg)} &
\colhead{(deg)} &
\colhead{(kpc)} &
\colhead{(km s$^{-1})$} &
\colhead{(mas yr$^{-1}$)} &
\colhead{(mas yr$^{-1}$)} &
\colhead{(dex)}
\\
\hline
Sgr trailing arm  & [11.01,   78.08]  &[-4.33, 27.64] &[12.31, 53.61] &[-244.57, -86.19] &[-1.37,  1.0] &[-4.47, -0.81] &[-2.44, -0.11]\\
Sgr leading arm   & [114.27,  227.0]  &[-5.15, 33.46] &[10.2,  54.69] &[-105.58,  92.99] &[-2.8, -0.58] &[-3.57, -0.12] &[-2.42, -0.35]\\
Monoceros Ring    & [-0.03,   143.98]  &[0.72,  60.15] &[7.73,   22.1] &[-187.2,  118.74] &[-2.19, 1.62] &[-2.34,  0.46] &[-1.67, -0.21]\\
Virgo Overdensity & [160.09, 199.42]  &[-5.65, 24.22] &[10.53, 26.82] &[-19.88,  266.65] &[-2.4, -0.37] &[-3.11,  -0.9] &[-2.29, -0.52]\\
Hercules-Aquila   & [322.06, 349.05]  &[-3.76, 18.77] &[8.07,  29.33] &[-210.57, -55.81] &[0.14,  4.17] &[-3.23,  -0.8] &[-2.42, -0.39]\\
Sgr Debris        & [105.01, 152.71]  &[23.78, 38.86] &[77.77, 95.93] &[-17.38,  133.73] &[-0.54, 0.17] &[-0.85, -0.23] &[-2.44, -1.02]\\
Orphan Stream     & [137.88, 149.12]  &[33.81, 47.85] &[36.33, 48.83] &[89.05,   162.72] &[-1.0,  -0.2] &[-0.85, -0.25] &[-2.25,  -1.3]\\
\hline
\multicolumn{4}{l}{\textsuperscript{\hspace{0.45em} a}[min, max] minimum, maximum.}\\
\multicolumn{4}{l}{\textsuperscript{\hspace{0.45em} b}{Heliocentric distance.}}\\
\multicolumn{4}{l}{\textsuperscript{\hspace{0.45em} c}{Heliocentric radial velocity.}}
\end{tabular}
\end{sidewaystable}

\begin{sidewaystable}[h]
\centering
\caption{Unknown Groups in LAMOST Halo K Giants} \label{t_unknown}
\scriptsize %fontsize
%\tiny %fontsize
\begin{tabular}{cccccccccc}
\tablewidth{0pt}
\hline
\hline
\colhead{GroupID} &
\colhead{R.A.[min,max]\tablenotemark{a}} &
\colhead{Decl.[min,max]\tablenotemark{a}} &
\colhead{$d$\tablenotemark{b}[min,max]\tablenotemark{a}} &
\colhead{$hrv$\tablenotemark{c}[min,max]\tablenotemark{a}} &
\colhead{pmra[min,max]\tablenotemark{a}} &
\colhead{pmdec[min,max]\tablenotemark{a}} &
\colhead{[Fe/H][min,max]\tablenotemark{a}}
\\
\colhead{} &
\colhead{(deg)} &
\colhead{(deg)} &
\colhead{(kpc)} &
\colhead{(km s$^{-1})$} &
\colhead{(mas yr$^{-1}$)} &
\colhead{(mas yr$^{-1}$)} &
\colhead{(dex)}
\\
\hline
7  &[124.97, 153.64] &[16.07, 39.79] &[6.87,  23.19] &[-83.66,    19.55] &[-0.95,  1.82] &[-7.43, -2.04] &[-2.21, -0.59]\\
9  &[154.48, 188.79] &[20.63, 39.43] &[11.15, 22.22] &[-47.93,   133.05] &[-3.04, -0.48] &[-4.13, -1.67] &[-2.4,  -0.62]\\
10 &[151.72, 196.16] &[14.67, 43.55] &[17.18, 30.48] &[18.76,    131.85] &[-2.39,  -0.1] &[-2.76, -0.99] &[-2.35, -0.8,]\\
14 &[159.18, 183.68] &[-5.04,   8.3] &[6.15,  20.28] &[81.52,    208.75] &[-6.57, -0.88] &[-5.58, -1.71] &[-2.12, -0.59]\\
18 &[335.94, 352.55] &[0.97,  22.16] &[8.62,  22.29] &[-329.45, -202.93] &[-0.01,  1.28] &[-3.61, -1.14] &[-2.34, -0.56]\\
20 &[159.11, 175.63] &[-4.23,  16.0] &[16.68, 23.75] &[117.6,    288.24] &[-1.75, -0.84] &[-2.38,  -1.5] &[-2.36, -0.94]\\
21 &[6.93,    28.71] &[16.99, 41.91] &[24.25, 33.11] &[-114.03,  -12.88] &[0.26,   1.23] &[-0.18,   0.9] &[-2.44, -1.41]\\
22 &[118.29,  145.5] &[11.59, 30.23] &[13.79, 21.76] &[-12.56,     76.7] &[-0.47,  0.84] &[-3.79,  -2.1] &[-2.44, -0.81]\\
24 &[142.68, 156.31] &[19.06, 32.16] &[15.42, 21.63] &[20.28,    100.72] &[-0.94,  0.17] &[-3.29, -1.93] &[-2.18,  -0.8]\\
25 &[11.11,   32.38] &[27.66, 37.75] &[11.49, 23.09] &[-91.38,     9.04] &[1.4,    3.86] &[-1.92,  -0.7] &[-2.28, -0.54]\\
26 &[183.72, 197.48] &[2.32,  14.81] &[19.6,  25.26] &[-47.16,    10.15] &[-1.44, -0.32] &[-2.6,  -1.57] &[-2.34, -1.08]\\
28 &[119.65, 143.63] &[16.42, 33.37] &[13.71, 25.85] &[-59.42,    13.73] &[-0.22,  1.26] &[-3.93,  -1.9] &[-1.54, -0.23]\\
30 &[-7.89,  5.41] &[22.56, 30.57] &[9.32,  17.38] &[-206.86, -150.04] &[0.78,   2.93] &[-2.43, -1.09] &[-1.85, -0.46]\\
33 &[198.07, 198.32] &[18.01, 18.25] &[19.83,  21.0] &[-74.28,   -59.05] &[-0.36,  0.07] &[-1.43,  -1.3] &[-2.14,  -1.8]\\
34 &[154.2,  165.53] &[16.17, 22.56] &[15.53, 21.36] &[-57.65,    28.76] &[-0.79, -0.14] &[-3.56, -1.98] &[-1.86, -0.88]\\
35 &[0.52,    13.52] &[34.22,  44.0] &[11.73, 20.92] &[-251.78, -189.63] &[1.25,   2.53] &[-2.01, -0.13] &[-2.16, -0.55]\\
36 &[8.77,    26.62] &[27.3,   40.8] &[14.38, 22.83] &[-226.67, -138.96] &[1.33,   2.46] &[-2.18, -0.49] &[-2.34, -1.23]\\
41 &[33.95,   46.83] &[-2.28,  12.3] &[44.01, 48.91] &[-144.29,   -67.9] &[0.61,   0.94] &[-0.81, -0.52] &[-1.6,   -1.1]\\
\hline
\multicolumn{4}{l}{\textsuperscript{\hspace{0.45em} a}[min, max] minimum, maximum}\\
\multicolumn{4}{l}{\textsuperscript{\hspace{0.45em} b}{Heliocentric distance.}}\\
\multicolumn{4}{l}{\textsuperscript{\hspace{0.45em} c}{Heliocentric radial velocity.}}
\end{tabular}
\end{sidewaystable}

\begin{deluxetable*}{cccccccc}
\tabletypesize{\scriptsize}
%\tablewidth{500pt}
%\tablenum{5}
\tablecaption{Group Catalog: Groups with Five or More Members \label{t_catalog}}
\tablehead{
\\
\colhead{obsid\tablenotemark{a}} &
\colhead{R.A.} &
\colhead{Decl.} &
\colhead{$d$}&
\colhead{$V_{\rm{hel}}$} &
\colhead{pmra} &
\colhead{pmdec} &
\colhead{GroupID\tablenotemark{b}}
\\
\colhead{} &
\colhead{(deg)} &
\colhead{(deg)} &
\colhead{(kpc)} &
\colhead{(km s$^{-1})$} &
\colhead{(mas yr$^{-1}$)} &
\colhead{(mas yr$^{-1}$)} &
\colhead{}
}
%\colnumbers
\startdata
77201039  &15.597092  & 5.310736 &13.6997 &-215.36 & 0.265   &-4.238  &0 \\
20603144  &19.739155  &-0.7779   &13.9554 &-173.39 &-1.366   &-3.4597 &0 \\
185709128 &39.33012   & 3.758879 &12.3098 &-179.17 &-0.346   &-2.451  &0 \\
187012195 &23.141436  &-0.383224 &13.8292 &-188.84 &-0.805   &-3.233  &0 \\
381006238 &36.810744  &-1.672415 &12.8698 &-148.22 &-0.31    &-2.639  &0 \\
21110160  &38.951621  & 1.017566 &12.9395 &-156.68 &-0.15    &-2.698  &0 \\
381014061 &33.997727  &-0.539124 &13.6571 &-174.38 &-0.4502  &-3.003  &0 \\
203913110 &39.6415    & 4.362719 &13.0332 &-199.16 &-0.194   &-2.99   &0 \\
187009063 &22.400078  &-1.022162 &14.6656 &-172.4  &-0.861   &-3.43   &0 \\
496704119 &18.590844  & 8.725264 &14.7694 &-216.16 &-1.173   &-3.4581 &0 \\
\enddata
\vspace{0.45em}
{\textsuperscript{\hspace{0.45em} a}{Unique identifier in LAMOST.}}\\
{\textsuperscript{b}{0-Sgr trailing arm; 1-Sgr leading arm; 2-Monoceros Ring; 3-Virgo Overdensity; 4-Hercules-Aquila; 5-Sgr Debris; 6-Orphan Stream; $>6$-unknown.}}\\
{(This table is available in its entirety in a machine-readable form in the online journal. A portion is shown here for guidance regarding its form and content.)}
\end{deluxetable*}

\end{document}